\documentclass[twocolumn]{aastex631}

\usepackage{color, hyperref}
\usepackage[normalem]{ulem}
\usepackage{amsmath}
\usepackage[caption=false]{subfig}
\usepackage{graphicx}
\usepackage{xspace}
\usepackage{longtable}
\usepackage{hyperref}
\usepackage[all]{hypcap}
\usepackage{chngcntr}

\newcommand{\lx}{$L_{\rm X}$\xspace}

\newcommand{\ergs}{erg s$^{-1}$\xspace}

\newcommand{\mstar}{$M_{0}$\xspace}
\newcommand{\msun}{$M_{\odot}$}
\newcommand{\Zsun}{$Z_{\odot}$\xspace}
\newcommand{\msunyr}{$M_{\odot}$ yr$^{-1}$\xspace}

\newcommand{\Galex}{\textit{GALEX}\xspace}

\newcommand{\HST}{\textit{HST}\xspace}

\newcommand{\chandra}{\textit{Chandra}\xspace}

\newcommand{\flux}{erg s$^{-1}$ cm$^{-2}$\xspace}

\newcommand{\sxpulx}{SXP$_{\rm ULX}$\xspace}
\newcommand{\sedulx}{SED$_{\rm ULX}$\xspace}

\newcommand{\sxpulxtz}{SXP$_{\rm ULX}$($t_{\rm burst}$, $Z$)\xspace}
\newcommand{\compSP}{[SXP$_{\rm ULX}$ + SSP]($t_{\rm burst}$, $Z$)\xspace}
\newcommand{\compSPU}{[SXP$_{\rm ULX}$ + SSP]($t_{\rm burst}$, $Z$, \logU)\xspace}
\newcommand{\logU}{$\log~\mathcal{U}$\xspace}

\renewcommand{\det}[1]{||{#1}||}

\newcommand{\fsps}{\textsc{\tt FSPS}\xspace}
\newcommand{\cloudyfsps}{\textsc{\tt CloudyFSPS}\xspace}
\newcommand{\cloudy}{\textsc{\tt Cloudy}\xspace}

\newcommand*\wideoline[1]{\hbox{\vbox{\hrule height 0.5pt \kern0.5ex\hbox{\kern-0.01em\ensuremath{#1}\kern-0.01em}}}}

\received{29 June, 2023}
\accepted{3 November, 2023}

\shorttitle{High-Energy Ionization from SXPs}
\shortauthors{Garofali et al.}
\graphicspath{{./}{figures/}}

\begin{document}

\title{Modeling the High-Energy Ionizing Output from Simple Stellar and X-ray Binary Populations}

\correspondingauthor{Kristen Garofali}
\email{kristen.garofali@nasa.gov}

\author[0000-0002-9202-8689]{Kristen Garofali}
\affiliation{NASA Goddard Space Flight Center, Code 662, Greenbelt, MD 20771, USA}
\affiliation{William H. Miller III Department of Physics and Astronomy, Johns Hopkins University, Baltimore, MD 21218, USA}
\author[0000-0001-8525-4920]{Antara R. Basu-Zych}
\affiliation{NASA Goddard Space Flight Center, Code 662, Greenbelt, MD 20771, USA}
\affiliation{Center for Space Science and Technology, University of Maryland, Baltimore County, Baltimore, MD 21250, USA}
\affiliation{Center for Research and Exploration in Space Science and Technology, NASA Goddard Space Flight Center, Greenbelt, MD 20771, USA}
\author{Benjamin D. Johnson}
\affiliation{Center for Astrophysics $|$ Harvard \& Smithsonian, 60 Garden Street, Cambridge, MA 02138, USA}
\author{Panayiotis Tzanavaris}
\affiliation{Center for Space Science and Technology, University of Maryland, Baltimore County, Baltimore, MD 21250, USA}
\affiliation{Center for Research and Exploration in Space Science and Technology, NASA Goddard Space Flight Center, Greenbelt, MD 20771, USA}
\affiliation{American Physical Society, Hauppauge, New York, NY 11788, USA}
\author[0000-0002-6790-5125]{Anne Jaskot}
\affiliation{Department of Astronomy, Williams College, Williamstown, MA 01267, USA}
\author[0000-0002-3703-0719]{Chris T. Richardson}
\affiliation{Elon University, Elon, NC 27278, USA}
\author[0000-0003-2192-3296]{Bret D. Lehmer}
\affiliation{Department of Physics, University of Arkansas, Fayetteville, AR 72701, USA}
\author[0000-0001-6366-3459]{Mihoko Yukita}
\affiliation{William H. Miller III Department of Physics and Astronomy, Johns Hopkins University, Baltimore, MD 21218, USA}
\author[0000-0002-2397-206X]{Edmund Hodges-Kluck}
\affiliation{NASA Goddard Space Flight Center, Code 662, Greenbelt, MD 20771, USA}
\author[0000-0001-8667-2681]{Ann Hornschemeier}
\affiliation{NASA Goddard Space Flight Center, Code 662, Greenbelt, MD 20771, USA}
\author[0000-0001-5655-1440]{Andrew Ptak}
\affiliation{NASA Goddard Space Flight Center, Code 662, Greenbelt, MD 20771, USA}
\author[0000-0001-7855-8336]{Neven Vulic}
\affiliation{Eureka Scientific, Inc., Oakland, CA 94602-3017, USA}

\begin{abstract}

We present a methodology for modeling the joint ionizing impact due to a ``simple X-ray population" (SXP) and its corresponding simple stellar population (SSP), where ``simple" refers to a single age and metallicity population. We construct composite spectral energy distributions (SEDs) including contributions from ultra-luminous X-ray sources (ULXs) and stars, with physically meaningful and consistent consideration of the relative contributions of each component as a function of instantaneous burst age and stellar metallicity.
These composite SEDs are used as input for photoionization modeling with \cloudy, from which we produce a grid for the time- and metallicity-dependent nebular emission from these composite populations. We make the results from the photoionization simulations publicly available. We find that the addition of the SXP prolongs the high-energy ionizing output from the population, and correspondingly increases the intensity of nebular lines such as \ion{He}{II}~$\lambda$1640,4686, [\ion{Ne}{V}]~$\lambda$3426,14.3$\mu$m, and [\ion{O}{IV}]~25.9$\mu$m by factors of at least two relative to models without an SXP spectral component. This effect is most pronounced for instantaneous bursts of star formation on timescales $>$ 10~Myr and at low metallicities ($\sim$ 0.1 \Zsun), due to the imposed time- and metallicity-dependent behavior of the SXP relative to the SSP. We propose nebular emission line diagnostics accessible with {\it JWST} suitable for inferring the presence of a composite SXP + SSP, and discuss how the ionization signatures compare to models for sources such as intermediate mass black holes.

\end{abstract}

\section{Introduction} \label{sec:intro}

Over the past decade, spectroscopic observations of high redshift ($z >$ 6) galaxies and their nearby analogs have revealed the presence of strong high-ionization nebular emission lines (e.g., \ion{He}{II}, [\ion{Ar}{IV}], [\ion{Ne}{V}]) and lines with high equivalent widths, indicating the presence of relatively hard ionizing spectra, recent ($\lesssim$ 30 Myr) star formation, and relatively low metallicities ($\sim$ 0.1 \Zsun) \citep[e.g.,][]{garnett1991, Schaerer1996, Guseva2000, thuan2005, izotov2012, shirazi2012, jaskot2013, Stasinska2015, Stark2016,Berg2018,Berg2019,Senchyna2017,Senchyna2019, Olivier2022}. Attempts to model the observed nebular features of such ``extreme emission line galaxies" (EELGs) have revealed that the spectral energy distributions (SEDs) from young, low-metallicity stellar populations have difficulty reproducing the observed strengths of a number of high-ionization emission line species, a problem that has been referred to as the ``high-energy ionizing photon production problem" \citep{Berg2021}. 

Addressing this problem is key to understanding how the heating and reionization of the Universe proceeds at $z > 6$, a regime now becoming more accessible spectroscopically thanks to {\it JWST}. In the coming decade, second generation interferometers such as the Hydrogen Epoch of Reionization Array (HERA) will probe even earlier cosmic epochs (6 $< z <$ 50) through the cosmic 21-cm signal. This high-redshift 21-cm signal is sensitive to the UV to soft X-ray ($<$ 2 keV) radiation field produced by the first galaxies' dominant ionizing populations. Initial results from HERA already suggest that the emergent 0.5-2.0~keV X-ray luminosity per galaxy star formation rate (\lx(0.5--2~keV)/SFR) from galaxies at $z > 6$ is at least an order of magnitude higher than measured for local galaxies \citep{hera2023}. These results point to increased production efficiency of photons with energies $\gtrsim$ 500~eV from galaxies at high redshift, possibly due to increased formation efficiency of accreting compact objects at the lower metallicities and/or younger stellar population ages characteristic of early galaxies \citep[e.g.,][]{Fragos2013b}. Developing a comprehensive accounting of the major sources of high-energy ionizing photons in EELGs, both near and far, therefore requires a framework for modeling how {\it additional} ionizing sources---such as fast shocks, superbubbles, and X-ray binaries (XRBs)---evolve as a function of redshift-dependent properties such as metallicity and stellar population age \citep[e.g.,][]{Allen2008, oskinova2022, Simmonds2021}.

One potential alternative source of high-energy ionizing photons is radiative shocks, which may be produced as a consequence of massive star evolution through winds and supernovae explosions \citep[e.g.,][]{izotov2012,izotov2021}. For fast shocks, there are publicly available grids of emission line ratios for a range of shock properties \citep{Allen2008}. Based on such models, shocks appear capable of sufficient high-energy ionizing photon production to reproduce select observed nebular line intensities for some shock velocities  \citep[e.g.,][]{thuan2005}; however, it is not yet clear whether fast shocks are capable of simultaneously reproducing suites of high-ionization nebular lines (i.e., from UV to optical), particularly if there is a metallicity-dependence to the observed line ratios \citep[e.g.,][]{Senchyna2017}.

Numerous works have also investigated the need for changes to prescriptions used in {\it stellar} spectral population synthesis and photoionization modeling, such as changes to assumed abundance patterns \citep[e.g.,][]{Berg2021, Grasha2021}, the initial mass function (IMF) and associated products of binary evolution \citep[e.g.,][]{Gotberg2019,Senchyna2021}, and/or assumptions about stellar rotation and line-driven winds for massive stars at very low metallicities \citep[e.g.,][]{Telford2021}. There are publicly available and well-vetted libraries for stellar atmosphere models and isochrones that allow for testing how these different prescriptions affect the ionizing output from stellar populations \citep[see e.g.,][for a review of models]{Conroy2013}. With stellar spectral population synthesis, these ingredients can be flexibly combined for a given IMF to produce the SEDs for simple stellar populations (SSPs)---populations at a single burst age and metallicity---for use in photoionization modeling and spectrophotometric fitting \citep[e.g.,][]{Conroy2009,Conroy2010,Byler2017,prospector21}. These tools enable investigation of how tweaks to stellar population models can address the high-energy ionizing photon production problem \citep[e.g.,][]{Berg2019,Senchyna2021}. However, presently available stellar population models, including those that model some of the products of binary evolution, still have difficulty reproducing the observed intensities of select high-ionization nebular emission lines \citep[e.g., \ion{He}{II};][]{Stanway2019}. This highlights the need for more comprehensive accounting of the products of binary evolution that may contribute to the ionizing photon budget. 

Accreting stellar mass compact objects (i.e., XRBs) are one such product of binary evolution, and have long been considered as a potential source of high-energy ionizing photons \citep[e.g.,][]{garnett1991,schaerer2019,Kovlakas2022}. Recent photoionization modeling that combines XRBs with stellar populations hints that XRBs may be sufficient to reproduce observed high-ionization emission line strengths {\it in some cases} \citep[e.g.,][]{Senchyna2020,Simmonds2021}. However, drawing a cohesive picture of the efficacy of XRBs to addressing the high-energy ionizing photon production problem remains challenging due to the following issues: (1) differing prescriptions for intrinsic XRB SEDs and; (2) minimal consideration towards how to scale XRB power output relative to the stellar population as a function of star formation history (SFH) and metallicity. Critically, for XRBs we currently lack many of the ingredients necessary for truly flexible spectral population synthesis, and therefore a means to understand the XRB contribution to high-energy ionizing photon production. In this work, we take steps to address this via the development of a flexible framework for including XRBs alongside SSPs for use in photoionization modeling. 

XRBs, and specifically those sources at the extreme bright-end of the luminosity function known as ultra-luminous X-ray sources (ULXs), are a compelling option for producing high-energy ionizing photons in high redshift galaxies and their local analogs for a few key reasons. Specifically, such sources (1) form swiftly ($\gtrsim$ 3--5~Myr) in multiple generations following a burst of star formation \citep[e.g.,][]{Linden2010,Wiktorowicz2017,West2023}, and are therefore capable of producing ionizing photons on longer timescales than single massive stars; (2) dominate the X-ray power output from normal star-forming galaxies (i.e., those without an actively accreting supermassive black hole) at high specific SFRs \citep[e.g.,][]{Lehmer2016}; and (3) have integrated luminosities directly proportional to SFR that increase with decreasing gas-phase metallicity, resulting in increased contribution to the photon budget in low-metallicity, highly star-forming galaxies \citep[e.g.,][]{Prestwich2013,Basu-Zych2013,Basu-Zych2016, Brorby2017}. For ULXs specifically, there is also increasing evidence for the presence of powerful outflows \citep[e.g.,][]{Pinto2016}, which may additionally produce shocks \citep{Lopez2019,Gurpide2022}. ULXs may therefore be important sources of both radiative and mechanical feedback at the low metallicities and high specific SFRs  characteristic of high redshift galaxies and their local analogs. 

A key consideration for modeling the ionizing contribution from accreting stellar mass compact objects such as ULXs is that they should form as a consequence of the evolution of the stellar population. To address this, we present a methodology for constructing a simple X-ray population (SXP), analogous to an SSP (i.e., single age and metallicity), and an approach to a physically consistent coupling of the SXP to the corresponding SSP for use in photoionization modeling. This approach is aimed at providing a more comprehensive framework for investigating the XRB contribution to the high-energy ionizing photon production problem. 

This paper is organized as follows: in Section~\ref{sec:sxp} we present the main elements used in constructing the SXP and in Section~\ref{sec:sxp_ssp} we outline how the SXP is scaled and added to the corresponding SSP to create the SED for a composite population. In Section~\ref{sec:cloudy} we describe our grid-based photoionization modeling set-up using the composite SXP + SSP SEDs. In Section~\ref{sec:results} we present key results from the photoionization modeling for the nebular emission from the composite populations, and describe how to access our simulation results. In Section~\ref{sec:discuss} we discuss the SXP model and photoionization simulation results in the context of X-ray detectability, recent literature results for the contribution of X-ray sources to the production of high-ionization emission lines, and alternative ionizing sources. Finally, in Section~\ref{sec:summary}, we summarize our key findings.    

\section{Constructing a Simple X-ray Population}\label{sec:sxp}

To construct the inputs for the photoionization modeling that follows, we apply the formalism of SSPs to accreting stellar mass compact objects. In what follows, we refer to models for single age and metallicity accreting compact object ``populations" as ``SXPs." In constructing these SXPs we consider selection of the following three key ingredients: (1) power output for the population as a function of stellar population age and metallicity; (2) the dominant source type in the population; and (3) physically motivated {\it intrinsic} source spectra. In this section, we describe the SXP used in this work based on selections for these three ingredients. 

\subsection{Normalization as a Function of Metallicity and Instantaneous Burst Age}\label{sec:sxp_scal}

Constructing a true SXP requires a prescription for the normalization of X-ray emitting population as a function of burst age and metallicity. To describe this behavior, we use the theoretical predictions for the X-ray power output per stellar mass from XRBs as a function of age and stellar metallicity from the binary population synthesis models presented in \citet{Fragos2013a}. We opt to use these theoretical predictions, as opposed to empirical measurements, for two key reasons. First, the evolution of XRB power output---particularly on finely resolved timescales $\lesssim$ 100~Myr that are important for luminous XRB formation---is not yet well-constrained observationally \citep[e.g.,][]{Lehmer2017}. In addition, empirical constraints on XRB radiative power as a function of time or metallicity are typically reported in terms of \lx/SFR, where the SFR is a quantity averaged over some extended timescale (i.e., 10 or 100~Myr). In this way, available empirical measurements are a time-averaged \lx for the {\it observed} XRB population. Such measurements are therefore not ideally suited to simulating the ionizing output from the {\it intrinsic} XRB population formed from an instantaneous burst of star formation, as we aim to do here. 

A critical component of the selected theoretical population synthesis models from \citet{Fragos2013a} is therefore that the predictions are provided in terms of radiative power (\lx) from the intrinsic XRB population per initial mass formed in a burst of star formation (\mstar). Nearly all other theoretical binary population synthesis models currently in the literature that focus on XRBs provide predictions in terms of {\it number counts} of sources per stellar mass as function of time and metallicity \citep[e.g.,][]{Linden2010,Wiktorowicz2019}. For the purposes of the photoionization modeling that follows, \lx/\mstar~is the preferred quantity since transforming number counts to total radiative power is non-trivial, and requires additional assumptions about how the luminosities of different sources should be calculated from theoretical mass supply rates. The models from \citet{Fragos2013a} track the mass supply rate to the compact object population---both black holes (BHs) and neutron stars (NS)---as a function of time. Following prescriptions in \citet{Fragos2008,Fragos2009}, the bolometric \lx is calculated directly for persistent XRBs from the mass supply rate along with the mass and radius of the accretor, assuming different conversion efficiencies for BHs and NS. For transient XRBs, a modified version of this formalism is applied, considering typical duty cycles and outburst luminosities. 

\begin{figure}[t!]
    \centering
    \includegraphics[width=9cm]{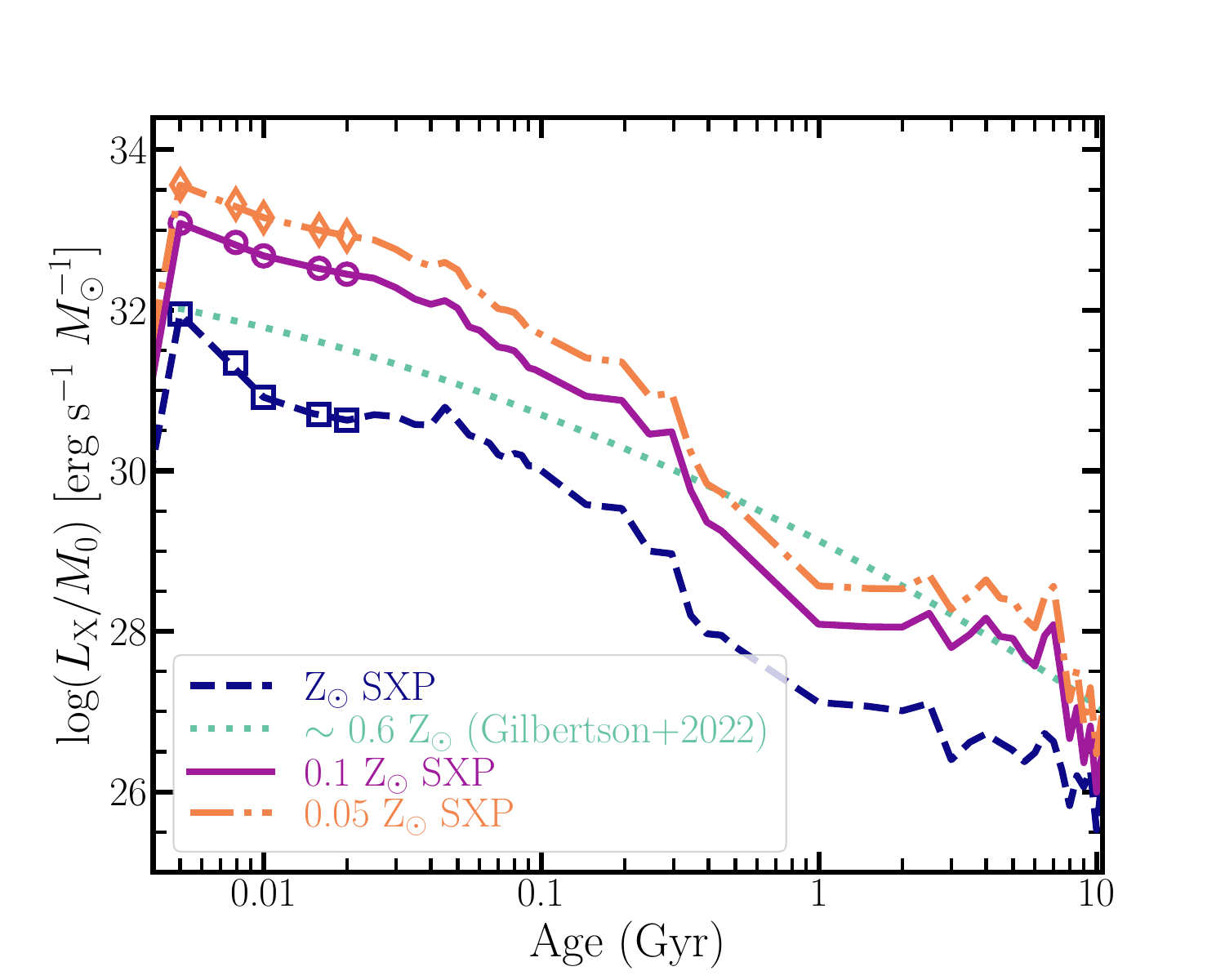}
    \caption{The adopted scaling of \lx/\mstar for the SXP as a function of instantaneous burst age and metallicity, where \mstar~corresponds to the initial stellar mass (\Zsun: dark blue dashed line, 0.1 \Zsun: purple solid line, and 0.05 \Zsun: orange dash-dot line) from theoretical binary population synthesis models \citep{Fragos2013a}. The instantaneous burst ages at which we model the SXP contribution in the photoionization simulations are marked with dark blue squares, purple circles, and orange diamonds for the \Zsun, 0.1 \Zsun, and 0.05 \Zsun cases, respectively. The SXP component in our models turns on for instantaneous bursts $\geq 5$~Myr. Empirical results for \lx/\mstar~from star-forming galaxy stacks in the \chandra~Deep Fields, for which the median metallicity ($\sim$ 0.6 \Zsun) is intermediate to the theoretical scalings adopted here, are shown as a green dotted line for reference.}
    \label{fig:lxm_age}
\end{figure}

The \citet{Fragos2013a} binary population synthesis models used here provide published values of \lx/\mstar~for XRBs at three different stellar metallicities\footnote{Throughout, we assume $Z$ = 0.02 corresponds to \Zsun for stellar metallicities. For abundances from \citet{Nicholls2017}, this corresponds to a gas-phase metallicity 12 + log(O/H) = 8.76.}: $Z$ = 0.002 (0.1 \Zsun), $Z$ = 0.02 (\Zsun), and $Z$ = 0.03 (1.5 \Zsun), and burst ages from 0--10~Gyr. To calculate the normalization of the SXP as a function of age and metallicity, we select a set of seven discrete burst ages and three metallicities from the reference model in \citealt{Fragos2013a} (i.e., their Figure~2, data provided via private communication).

The seven discrete time steps ($t_{\rm burst} \sim$ \{1, 3, 5, 8, 10, 16, 20\}~Myr) are selected to maximize \lx/\mstar (i.e., capture a bounding case for XRB ionization), and to cover relevant timescales for important stages of massive star evolution. On these short burst timescales, the XRB population power output is dominated by high-mass XRBs (HMXBs), sources descended from the most massive stars ($\gtrsim$ 8~\msun). As such, we calculate the normalization of the SXP to reproduce the theoretical \lx(0.5--8~keV)/\mstar~from the HMXB component of the reference model in \citet{Fragos2013a}. We additionally elect to turn on the SXP with a delay time that accounts for the minimum time for the first BHs to form and begin accreting \cite[e.g.,][]{Belczyski2008,Linden2010}. The first time step in the simulations with SXP contribution is 5~Myr. 

Two of the three metallicities selected for our models include those available from the published theoretical population synthesis results, namely $Z = 0.002$ (0.1 \Zsun) and $Z = 0.02$ (\Zsun), where the 0.1 \Zsun scaling for \lx/\mstar~is relevant for nearby EELGs and sources at $z \gtrsim 3$ \citep[e.g.,][]{shirazi2012,Madau2014}. We additionally add a third metallicity set at 0.05 \Zsun, though this metallicity was not originally included in the theoretical population synthesis models. To create normalizations for the SXP for this extremely metal-poor case, we assume the same dependence of \lx/\mstar~on post-starburst timescale as the 0.1 \Zsun model, but allow the absolute normalization of \lx/\mstar~to be a factor of 3$\times$ higher than for the 0.1 \Zsun at each time step. This scaling corresponds roughly to the upper bound on the observed $\langle$\lx/SFR$\rangle$ from galaxies with 12 + log(O/H) $\sim$ 7.4 from \citet{Lehmer2021}. Though speculative, we include this case for reference, noting that it will likely need to be revised either theoretically or observationally going forward. 

In Figure~\ref{fig:lxm_age}, we show the adopted theoretical \lx/\mstar~as a function of instantaneous burst age for the three metallicities employed in the photoionization simulations that follow, where the symbols denote burst ages for which the SXP contribution is non-zero and therefore modeled (i.e., we do not display the 1 and 3 Myr time steps in this figure, as we assume there is no SXP contribution on these timescales.\footnote{In Section~\ref{sec:sxp_nebAge}, we discuss relaxing the assumption that there is a delay-time for SXP formation. In the Appendix we show select line diagnostic results from simulations with no delay-time dependence, where we allow the immediate formation of the SXP alonside the SSP. The results from these no delay-time simulations at 1--3 Myr are similar to the results for the SXP contribution at 5 Myr shown throughout.} For reference, we also plot in this figure empirical results from \citet{Gilbertson2022} for \lx/\mstar~for a sample of star-forming galaxy stacks from \chandra~Deep Fields. The galaxy stacks from the \chandra~Deep Fields have median metallicity $\sim$ 0.6 \Zsun, intermediate to our selected metallicity range, illustrating that the empirical results are in qualitative agreement with the behavior of the adopted theoretical relations on the timescales of interest for these simulations. 

The adopted theoretical values for \lx/\mstar~from Figure~\ref{fig:lxm_age} can be considered as good approximations for the {\it average} radiative output from a population. For the purposes of our simulations, this is appropriate as we seek to simulate a well-sampled population; however, for observed samples 
the X-ray luminosity function (XLF) may not be as well-populated. Stochastic sampling of the XLF is particularly important to consider at very low SFRs ($\lesssim$ 0.1 \msunyr), where observed values for \lx/$M_{\star}$~or \lx/SFR can be subject to scatter of order $\sim$ 1 dex, especially for sources at the bright end of the XLF \citep{Lehmer2021}. We return to this point in Section~\ref{sec:discuss_bestDetect} in discussing how results from these simulations should be compared against observations.  

\subsection{Source Type: Ultra-luminous X-ray Sources}\label{sec:sxp_ulx}

With the adopted scaling for \lx/\mstar, we next consider the source type that dominates the power output from the population. In star-forming galaxies without a central accreting supermassive black hole, XRBs dominate the galaxy-integrated X-ray point source emission. For predominantly young stellar populations ($\lesssim$ 100~Myr) HMXBs are the dominant accretor population, while in galaxies with older stellar populations low-mass XRBs (i.e., sources with low-mass donor stars) dominate the radiative output \citep[e.g.,][]{garofali2018,Lehmer2019,Lehmer2021}.  

Regardless of stellar population age, the 
total radiative power output from a population will be dominated by the sources that populate the bright end of the XLF. Such sources are ULXs, off-nuclear point sources with \lx $>$ 10$^{39}$ erg s$^{-1}$. These luminosities, under the assumption of isotropic emission, exceed the Eddington limit for a 10 \msun~BH or 1.4 \msun~NS: 

\begin{equation}\label{eq:ledd}
L_{\rm Edd} = \frac{4 \pi GMc}{\kappa_{T}} =  1.5 \times 10^{38}~m~\frac{1.7}{1+X}~{\rm erg s^{-1}}
\end{equation}

\noindent where $M$ is the mass of the compact object (such that $m\equiv M/M_{\odot}$), $X$ is the hydrogen mass fraction (assumed to be $X = 0.73$, corresponding to solar composition), $G$ is the gravitational constant, $c$ is the speed of light, and $\kappa_{T}$ = 0.2(1 + $X$) is the Thomson scattering opacity.

Models to explain the extreme apparent luminosities from ULXs fall broadly into two categories: (1) isotropic emission from sub-Eddington accretion onto intermediate mass black holes \citep[IMBHs; e.g.,][]{Colbert1999} or (2) super-Eddington mass transfer via modified accretion disks onto either stellar mass BHs or strongly magnetized NS \citep[e.g.,][]{King2001,Begelman2006,Poutanen2007,King2009}. Recent results from X-ray timing studies favor the second scenario, as the detection of pulsations in an increasing number of ULXs indicates the presence of a NS for at least some fraction of the ULX population \citep[e.g.,][]{Bachetti2014,Israel2017, Brightman2018}. Broad-band and high-resolution X-ray spectroscopy of ULXs also support the picture of stellar mass accretors with high mass transfer rates. In particular, the observed spectral turnover for ULXs at energies $\gtrsim$ 3--8 keV is incompatible with models for intermediate mass or supermassive BHs accreting at sub-Eddington rates \citep{Gladstone2008}, and the detection of absorption lines for some ULXs provide evidence for the presence of fast outflows \citep[Section~\ref{sec:sxp_sedAcc};][]{Pinto2016,Kosec2018a,Kosec2018b,Kosec2021}, and potentially strongly magnetized NS accretors \citep{Walton2018, Brightman2022}.

These findings suggest ULXs (\lx $\sim$ 10$^{39}-10^{42}$ \ergs, i.e., below the hyper-luminous regime) are a distinct class of stellar mass accretors with super-Eddington mass supply rates\footnote{As discussed in \citet{King2023}, the phrase ``super-Eddington accretion" can lead to ambiguity, as stellar mass BHs and NS in ULXs do not actually accrete matter at highly super-Eddington rates; instead, these authors suggest that for ULXs it would be more appropriate to refer to the mass supply or transfer rate as the super-Eddington quantity in question. We adhere to this convention throughout.}. This in turn implies that ULXs are indeed the high luminosity extension of the XRB luminosity function. In this work, we treat ULXs as the bright extension of the HMXB luminosity function specifically, as we consider timescales relevant to formation and evolution of the most massive stars ($\leq$ 20~Myr, Section~\ref{sec:sxp_scal}). 

Given that they occupy the bright end of the XLF, ULXs in star-forming galaxies can surpass by at least an order of magnitude the combined output of various HMXB sub-populations filling out the lower luminosity---and more highly populated---portion of the luminosity function \citep[e.g.,][]{Lehmer2021}. {\it In this work, we therefore choose to model the SXP as a population of ULXs.} Assuming a well-sampled XLF (i.e., high SFR) as we model here, a ULX-based SXP is a reasonable approximation for the total XRB population. Indeed, many highly star-forming galaxies (SFRs $\geq$ 3 \msunyr) show ULX-like spectra globally \citep{Garofali2020}. Because various HMXB sub-populations likely have different accretion state distributions and scaling relations with metallicity and stellar population age, approximating the SXP using ULXs considerably simplifies the modeling that follows. Hereafter, we refer to the SXP as ``\sxpulx", to indicate the source type modeled. This selection captures a bounding case for HMXB\footnote{There is no explicit ULX population included in the theoretical models from \citet{Fragos2013a}, and we therefore use the HMXB component, corresponding to the bright end of the XLF, for normalizing the SXP; however, the theoretical models do allow for XRBs undergoing outbursts to exceed Eddington. We also note that empirical values for \lx/\mstar~from galaxy-integrated studies, where the populations can be assumed to be dominated by sources at the brightest end of the XLF, i.e., ULXs, appear relatively consistent with the theoretical scaling relations ostensibly for HMXBs employed here, as shown in Figure~\ref{fig:lxm_age}.} contribution to production of high-energy ionizing photons in terms absolute normalization as emphasized above, as well as hardness of spectral shape (Section~\ref{sec:sxp_sed}). 

\subsection{Model for Ultra-luminous X-ray Source SED}\label{sec:sxp_sed}

The efficacy of the \sxpulx in the  production of high-energy ionizing photons is highly dependent on the shape of the {\it intrinsic} SED and normalization relative to the corresponding SSP \citep[e.g.,][]{Simmonds2021}. Unfortunately, this intrinsic shape for ULXs is difficult to measure observationally and therefore remains highly uncertain. At the typical extragalactic distances ($\gtrsim$ 3 Mpc) to ULXs, fluxes are $<$ 10$^{-12}$ \flux, such that high-resolution X-ray spectroscopy is currently feasible for a limited number ($\sim$ 10) of the brightest sources \citep{Kosec2018a}. At CCD-resolution, many more sources are available for spectral fitting, but these fits are often performed using phenomenological models with no obvious physical basis, which cannot be reliably extrapolated into wavelength regimes outside the observed bandpass. The lack of comprehensive multiwavelength catalogs for ULXs (from IR--to--X-ray) is a further limitation to measuring their intrinsic SEDs, resulting in large uncertainties especially in the extreme ultraviolet (EUV) to very soft X-ray regime ($\sim$ 54--200 eV). Critically, this is the wavelength regime responsible for setting the intensity of high-ionization nebular emission lines. 

We therefore opt to model our SED (hereafter \sedulx) using analytic prescriptions from theoretical models for stellar mass compact objects with super-Eddington mass supply rates. In the sections that follow, we outline the key components of the \sedulx, and motivate our selections. The \sedulx employed throughout this work is qualitatively similar to empirical models derived from analyses of ULXs detected within their own optical nebulae, where high-ionization emission lines are also detected \citep{Berghea2010,Berghea2012,Kaaret2009,lebouteiller2017, Simmonds2021}. In such analyses, both the broad-band X-ray data and the observed emission line strengths provide constraints on the form of the ULX SED, where the line emission is particularly useful for constraining the shape of the unseen EUV portion. We note that such a multi-wavelength approach has promise for constraining the shape of the intrinsic SED; however, there is as yet no comprehensive approach in this respect for ULXs given the lack of collated multiwavelength data sets.

\subsubsection{Accretor Type}\label{sec:sxp_sedCO}

As a starting point to constructing the \sedulx, we must select the mass of the accretor as well as the type (BH or NS), as the SED shape will depend not only on accretor mass and mass supply rate, but also on whether the accretion flow encounters a hard surface and/or a strong magnetic field (see Section~\ref{sec:discuss_altIon_NS}). Because the selected \lx/\mstar normalizations from Section~\ref{sec:sxp_scal} span burst ages $\leq$ 20~Myr, we choose a stellar mass BH ($<$ 100 \msun) as the accretor for the \sedulx. Theoretical binary population synthesis models show that the fraction of NS accretors contributing to the observed ULX population is still low on timescales $<$ 20~Myr post-starburst, and that accreting BHs may have ages $\sim$ 4-40~Myr during their ULX phase \citep{Middleton2017,Wiktorowicz2017,Wiktorowicz2019}. 

Given that we likewise elect to model the population at three discrete stellar metallicities (\Zsun, 0.1 \Zsun, and 0.05 \Zsun), we construct the \sedulx model as a function of approximate typical BH masses for ULXs at each of these metallicities: 8~\msun~for $Z \geq$ 0.1 \Zsun and 16~\msun~for $Z <$ 0.1 \Zsun, following theoretical population synthesis results for ULXs \citep{Wiktorowicz2017}.

At present it is not clear theoretically or observationally what the typical BH mass is in a ULX as a function of burst age. Thus, we consider it a reasonable simplifying assumption at present to use approximate BH masses as a function of {\it metallicity only}. It is important to note that while these selected BH masses factor into the parametrization of the \sedulx shape (Section~\ref{sec:sxp_sedParam}), they do not strongly affect the {\it normalization} of the \sedulx, which is instead set by the theoretical scaling relations from Section~\ref{sec:sxp_scal}, which are also key to setting the simulated nebular line intensities. 

\subsubsection{Accretion Flow Geometry \& Components}\label{sec:sxp_sedAcc}

\begin{figure}[t!]
    \centering
    \includegraphics[width=9cm]{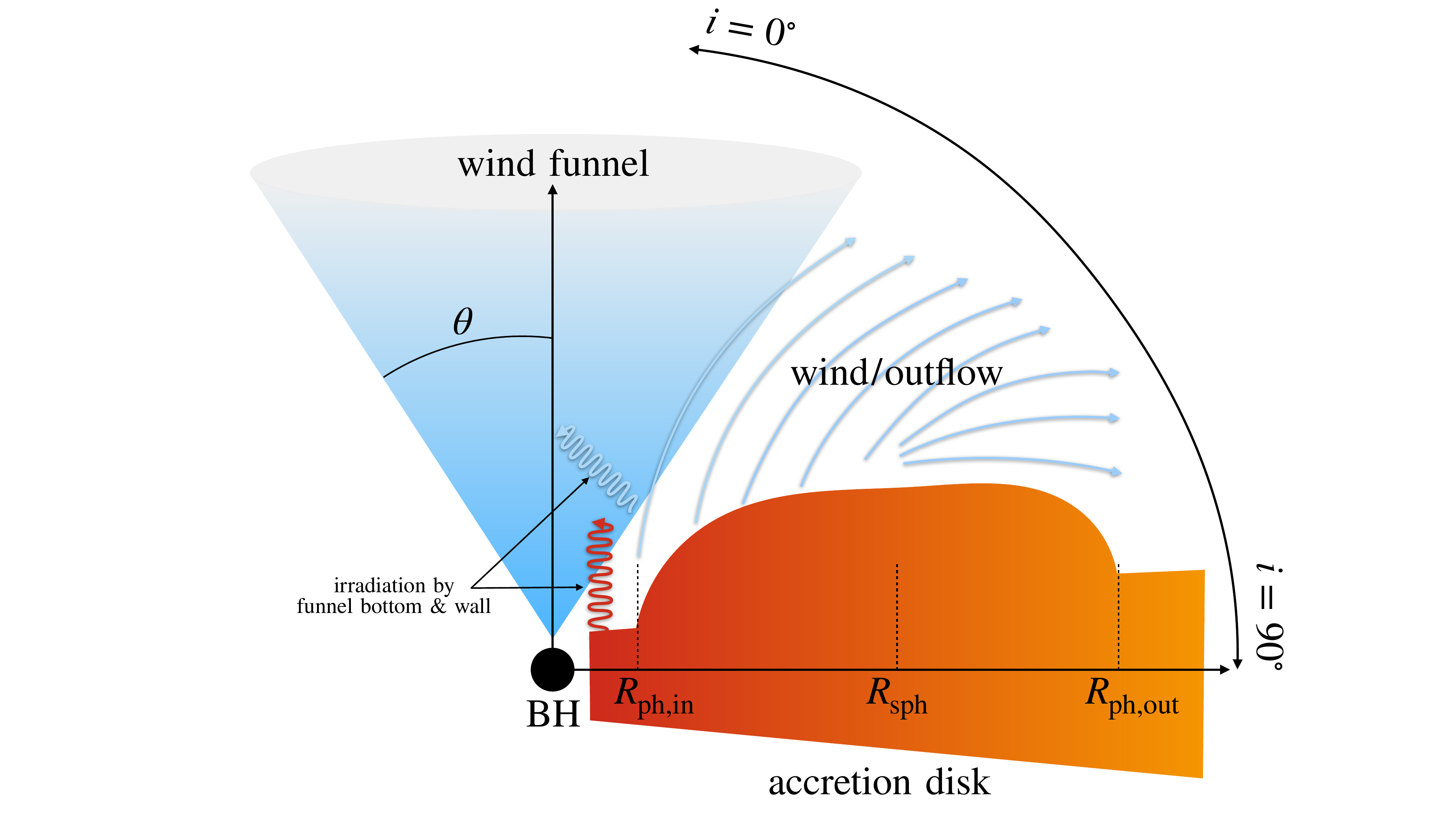}
   
    \caption{Cartoon depiction of the accretion flow geometry for the \sedulx model. Key components and parameters are labeled, including the accretor (BH) and accretion disk, evacuated wind funnel and associated funnel opening angle ($\theta$), the inner and outer photospheric radii ($R_{\rm ph,in}$, $R_{\rm ph,out}$, respectively), the spherization radius ($R_{\rm sph}$), and the inclination angle ($i$) with respect to an observer. The wind or outflow, which is launched from the accretion disk and assumed to be a quasi-spherical component, is shown by the blue arrows. Irradiating photons emanate from the hot inner accretion flow and funnel walls, as labeled.}\label{fig:flow_cartoon}
\end{figure}

Under the selection of a stellar mass accretor with super-Eddington mass supply rate for the \sedulx, theory and observation alike suggest the presence of a strong outflow, which is a key component in defining the SED shape. In the supercritical regime the accretion flow consists of distinct zones delineated by the wind/outflow component \citep[e.g.,][]{ss73,lipunova1999,Poutanen2007,Abolmasov2009}. We illustrate these zones in Figure~\ref{fig:flow_cartoon}, as follows:  
\begin{itemize}
    \item The hot inner region of the accretion flow before the wind launching zone ($R < R_{\rm ph,~in}$). 
    \item The region between the wind launching zone and so-called the spherization radius ($R_{\rm ph,~in} < R < R_{\rm sph}$), where the wind is optically thick. The spherization radius ($R_{\rm sph}$) can be thought of as the region where the disk is unstable and super-Eddington (i.e., the actual accretion rate is Eddington within this radius, and limited through the expulsion of matter from the disk as a wind).  
    \item The region beyond the spherization radius where the wind becomes optically thin ($R > R_{\rm sph}$). The outer extent of the wind is given by the effective photospheric radius of the outflow, $R_{\rm ph,~out}$.
\end{itemize}

Under such a model, the launching of the disk wind leads to an evacuated ``wind funnel," wherein the hard inner disk emission is geometrically collimated and scattered by the wind. At much larger disk radii ($R > R_{\rm sph}$), the wind becomes optically thin and can be approximated as a pseudo-photosphere (i.e., a  quasi-spherical component with a blackbody spectrum). This picture of the accretion flow is broadly consistent with CCD-resolution spectra of ULXs, which typically require multi-component models to fit their observed spectra \citep[e.g.,][]{Gladstone2009,sutton2013,Middleton2015a}.

In defining our \sedulx model, we therefore include a component representing the hot inner disk and wind funnel, as well as a component for the outflow. In addition, we consider the effects of Compton scattering and irradiation on the resultant SED shape. Irradiation---where one component of the accretion flow absorbs and re-emits radiation from another component---may be critically important in the context of \sxpulx contribution to high-energy ionizing photon production. This is because reprocessed emission due to irradiation can produce a spectrum with a more substantial EUV component, where photon energies exceed ionization potentials for lines such as \ion{He}{II}~$\lambda$4686 and [\ion{Ne}{V}]~$\lambda$3426. Given the geometry of the supercritical accretion flow described above, it is unlikely that there is direct irradiation of the outer regions of the accretion disk by hot inner disk photons \citep[e.g.,][]{gierlinski2009}. In the case of a ULX with an outflow, the wind likely blocks and reprocesses the hot inner disk emission. Irradiation may therefore occur via the wind irradiating the outer disk \citep{Vinokurov2013,Yao2019}, or via ``self-irradiation" of inner disk and funnel wall emission within the wind funnel itself \citep{Abolmasov2009}. 

For the \sedulx in this work, we choose to model the irradiated component as being due to self-irradiation within the wind funnel. This is because there are uncertainties in how effectively soft photons from the wind thermalize in the outer disk \citep{Kaaret2009,Grise2012}, and additional complexities in handling the geometry and radiative transfer for wind irradiation. To represent the supercritical accretion flow with self-irradiation, we opt to use the {\tt sirf}, or ``self-irradiated funnel", template as implemented in {\tt XSPEC} \citep{Abolmasov2009}. This particular template provides a physical model coupling all components of the supercritical accretion flow, including the hot inner disk, outflow, and the effects of irradiation consistent with the previously described flow geometry. To additionally account for the effects of Compton scattering, we include the {\tt simpl} (SIMple Power Law) convolution component, which Comptonizes flux output from the {\tt sirf} model. Thus, our \sedulx is implemented as {\tt simpl(sirf)} in {\tt XSPEC}, with relevant parameters listed in Table~\ref{tab:mod_params}. Under this accretion flow model, the mass supply rate relative to Eddington sets the physical parameters (i.e., characteristic radii and temperatures) that define the shape of the {\it intrinsic} spectrum, while the viewing angle ($i$ in Figure~\ref{fig:flow_cartoon}) determines the shape of the {\it observed} spectrum. In this respect, the {\tt sirf} template is ideal, as it allows for changing the inclination angle to simulate changes to the apparent emergent spectrum for an observer, a point we return to in more detail in Section~\ref{sec:discuss_bestDetect}. 

\begin{deluxetable}{ccccc}
\tabletypesize{\small}
\centering
\tablewidth{\textwidth}
\tablecaption{Summary of calculated and selected parameters for the \sedulx model, as implemented via the {\tt simpl(sirf)} templates in {\tt XSPEC}. Parameter names follow nomenclature used in {\tt XSPEC}, as described in the text. \label{tab:mod_params}}
\tablehead{
\colhead{Component} & \colhead{} & \colhead{Name} & \colhead{Value} & \colhead{Units} \\
\colhead{(1)} & \colhead{} & \colhead{(2)} & \colhead{(3)} & \colhead{(4)}
}

\startdata 
\hline
& \vline & $T_{\rm funnel,~in}$ & Eq.~\ref{eq:tfunnel} & keV \\
& \vline & $r_{\rm ph,~in}$ & Eq.~\ref{eq:rphin} & $R_{\rm sph}$ \\
& \vline & $r_{\rm ph,~out}$ & Eq.~\ref{eq:rphout} & $R_{\rm sph}$ \\
& \vline & $\theta$ & Eqs.~\ref{eq:btheta}--\ref{eq:bmdot} & deg \\
{\tt sirf} & \vline & $i$ & 0 & deg \\
& \vline & $\alpha$ & 0 & \dots \\
& \vline & $\gamma$ & 4/3 & \dots \\ 
& \vline & $\dot{m}$ & Eq.~\ref{eq:mdot_approx} & \dots \\ 
& \vline & irrad. & 4 & \dots \\
& \vline & norm. & \citet{Fragos2013a} & \dots \\
\hline
& \vline & $\Gamma$ & 2.5 & \dots \\
{\tt simpl} & \vline & Frac. Sctr & 0.1 & \dots \\
& \vline & Up Sctr. Only & 0 & \dots \\
\hline 
\enddata 
\end{deluxetable}

\subsubsection{Parametrization of SED Shape}\label{sec:sxp_sedParam}

We aim to construct the \sedulx model with as few {\it selected} parameter values as possible. Therefore, to the extent possible, we calculate parameter values for the {\tt sirf} template analytically from a few key selected physical properties \citep[see][for prescriptions]{ss73,Poutanen2007,Abolmasov2009}. 

The primary selected properties influencing the \sedulx shape are the accretor mass and the shape of XLF, both of which are assumed to depend on metallicity. For the former, our model selections are described in Section~\ref{sec:sxp_sedCO}. For the latter, we employ a model for the metallicity-dependent XLF. Below we describe the form of the XLF model, and outline how the physical properties (i.e., Table~\ref{tab:mod_params}) defining the shape of the \sedulx are determined analytically from this model.

The XLF describes the number count of sources in bins of luminosity that contribute to the \sxpulx. To calculate this distribution of sources as a function of metallicity, we employ the empirical metallicity-dependent XLF from \citet{Lehmer2021}: 

\begin{equation}\label{eq:xlf}
\begin{split}
\frac{dN_{\rm HMXB}}{dL} = {\rm SFR}~A_{\rm HMXB}~{\rm exp}[-L/L_{\rm c}(Z)] \\
\times \begin{cases}
L^{\gamma_{1}} & (L < L_{b})\\    
L_{b}^{\gamma_{2}(Z) - \gamma_{1}}L^{-\gamma_{2}(Z)} & (L > L_{b})
\end{cases}
\end{split}
\end{equation}

\noindent In this model, $\gamma_{2}(Z)$ is a metallicity-dependent bright-end slope (see Equation~2 from \citealt{Lehmer2021}), and log$L_{\rm c}(Z)$ is a metallicity-dependent exponential cutoff luminosity (see Equation~3 from \citealt{Lehmer2021}). For the remaining model parameters that do not depend on metallicity ($A_{\rm HMXB}$, $\gamma_{1}$, and $L_{b}$), we use the best-fit values from Table~2 in \citet{Lehmer2021}. 

The best-fit form of this XLF model is derived from a sample of 55 local ($<$ 30~Mpc) star-forming galaxies spanning gas-phase metallicities 7.0 $\leq$ 12 + log (O/H) $\leq$ 9.2. To populate the \sxpulx using this model, we restrict ourselves to sampling bins of luminosity between 10$^{39}$ $\leq$ \lx $\leq$ 10$^{41}$ \ergs. The lower limit corresponds roughly to the Eddington limit for stellar mass accreting compact objects, and the upper limit corresponds to the potential upper limit on \lx for accreting stellar mass compact objects, i.e., below the regime of hyper-luminous sources \citep{Tranin2023}. 

Because Equation~\ref{eq:xlf} is an empirical model, sampling from this XLF results in {\it observed} ULX number counts in bins of {\it observed} luminosity ($L_{\rm obs}$). By contrast, the shape of the accretion flow model from Section~\ref{sec:sxp_sedAcc} is set by {\it intrinsic} physical properties, which are often difficult to measure directly for a given ULX or a population of sources. As such, we must transform the observed XLF quantities (number counts, or relative contribution to the total population, and $L_{\rm obs}$) into the intrinsic physical properties from Table~\ref{tab:mod_params}. To do so, we follow a methodology similar to the approach outlined in \citet{Kovlakas2022}. 

Taking this approach ($L_{\rm obs}$ $\xrightarrow{}$ intrinsic physical parameters) requires defining the intrinsic physical properties of interest. To begin with, the mass transfer rate is a parameter critical to defining the spectral shape. By definition, a source in an ultra-luminous accretion state will have a mass supply rate exceeding the Eddington rate:

\begin{equation}\label{eq:mcr}
\begin{split}
    \dot{M}_{\rm Edd} = \frac{48 \pi G M}{c \kappa_{T}} \simeq \\
    & 3 \times 10^{-8}~m~~M_{\odot}~{\rm yr}^{-1}
\end{split}
\end{equation}

\noindent where $m$ is the mass of the accretor in units of \msun. It is often convenient to define a dimensionless mass transfer rate $\dot{m}$, as follows: 

\begin{equation}\label{eq:mdot}
\dot{m} = \dot{M}/\dot{M}_{\rm Edd}
\end{equation}

\noindent A source in the ultra-luminous state would therefore be expected to have $\dot{m}$ $\gg$ 1. The {\it intrinsic} luminosity then depends logarithmically on the dimensionless mass transfer rate \citep{Poutanen2007}: 

\begin{equation}\label{eq:lbol}
L^{\rm int}_{\rm bol} = L_{\rm Edd}(1 + 0.6{\rm ln}(\dot{m}))
\end{equation}

Following \citet{King2009}, the {\it observed}, or apparent, luminosity may differ from the intrinsic luminosity due to geometrical beaming, given by beaming factor $b$, such that: 

\begin{equation}\label{eq:lobs}
L_{\rm obs} = L^{\rm int}_{\rm bol} / b
\end{equation}

The combination of Equations~\ref{eq:lbol}--\ref{eq:lobs} connects the XLF-derived property $L_{\rm obs}$ with the intrinsic properties $b$, $\dot{m}$, and $L_{\rm Edd}$, where the Eddington limit is given by Equation~\ref{eq:ledd} for the assumed BH mass $m$. The number of free intrinsic parameters can be further reduced given a reasonable assumption about the physical meaning of the beaming factor $b$. Under the formalism outlined in \citet{King2009}, the beaming factor $b$ is related to the funnel opening angle ($\theta$, as shown in Figure~\ref{fig:flow_cartoon}) by b = $\Omega / 4\pi$, where $\Omega = 2 \times 2\pi (1-{\rm cos}\theta)$ is the combined solid angle from both cones of the wind funnel. In this way, the beaming factor ($b$) also sets the probability that the source is observed:   

\begin{equation}\label{eq:btheta}
b = P_{\rm obs}(\theta) = 1 - {\rm cos}\theta
\end{equation}

For large funnel opening angles ($\theta$ $\sim$ 90$^{\circ}$), the beaming factor approaches unity, and there is a high probability the source is observed. For very small opening angles, by contrast, the beaming factor is small and the observed luminosity of the central disk is boosted, but the probability that the source is observed at X-ray wavelengths is low. \citet{King2009} suggest that the beaming factor itself may depend on the mass transfer rate, and we adopt this view for our model to reduce the number of predetermined parameter values. Specifically, we parameterize $b$ in terms of the mass transfer rate as follows:

\begin{equation}\label{eq:bmdot}
b =
\begin{cases}
1 & \dot{m} \leq 8.5\\    
(8.5/\dot{m})^{2} & \dot{m} > 8.5  
\end{cases}
\end{equation}

\noindent By this definition, high mass transfer rates lead to small funnel opening angles, correspondingly small beaming factors, and therefore low probabilities for observing hard emission from the source. 

Further combining Equation~\ref{eq:bmdot} with Equations~\ref{eq:lbol}--\ref{eq:lobs} enables solving for $\dot{m}$ in terms of $L_{\rm obs}$ and $L_{\rm Edd}$. In particular, this can be accomplished via an analytic approximation as outlined in \cite{Kovlakas2022}: 

\begin{equation}\label{eq:mdot_approx}
\dot{m} = 
\begin{cases}
L_{\rm obs} / L_{\rm Edd} & L_{\rm obs} / L_{\rm Edd} \leq 1 \\ 
e^{(L_{\rm obs} / L_{\rm Edd} - 1)} & 1 < L_{\rm obs} / L_{\rm Edd} \leq 3.14 \\
8.5 \times (\frac{L_{\rm obs} / L_{\rm Edd}}{3.14})^{4/9} & L_{\rm obs} / L_{\rm Edd} > 3.14
\end{cases}
\end{equation}

\noindent We note that $L_{\rm obs}$ is determined from the metallicity-dependent XLF, and $L_{\rm Edd}$ is determined by the accretor mass. Therefore, given {\it only} selections for the form of the XLF and the typical BH mass as a function of metallicity, we can analytically calculate $\dot{m}$, $b$, and $\theta$ using Equations~\ref{eq:btheta}--\ref{eq:mdot_approx}.

\begin{figure}[t!]
    \centering
    \includegraphics[width=9cm]{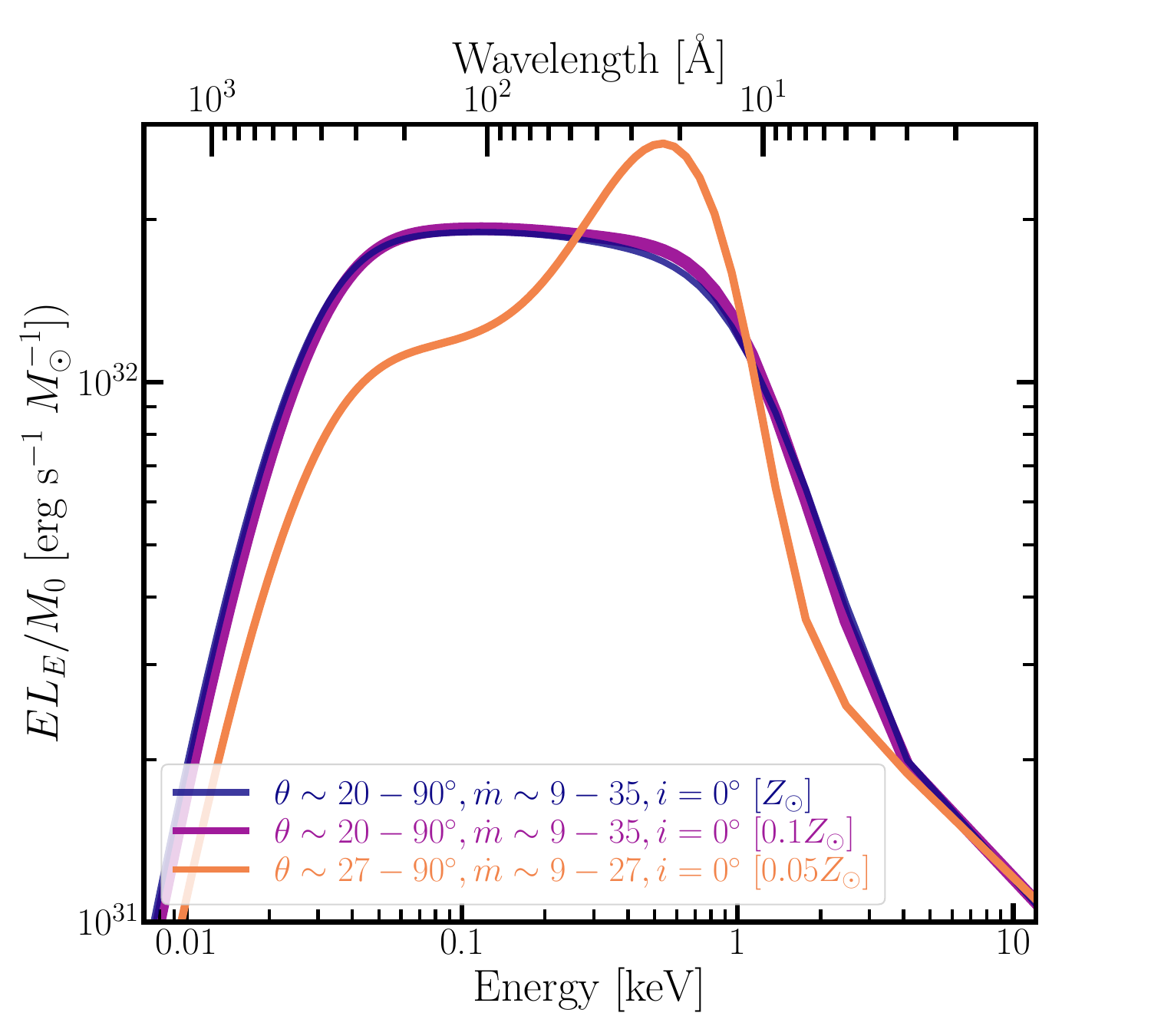}
    \caption{The intrinsic broad-band \sedulx for the \sxpulx (i.e., {\tt simpl(sirf)} in {\tt XSPEC}) for each of the three metallicities simulated in this work (\Zsun: dark blue, 0.1 \Zsun: purple, and 0.05 \Zsun: orange). All SEDs are normalized to the same bolometric luminosity for plotting purposes. The shape of the intrinsic SED is set by the distribution of funnel opening angles ($\theta$) and mass transfer rates  ($\dot{m}$) for the accreting population, where the distribution depends on metallicity. For all simulations in this work, we assume $i$ = 0$^{\circ}$.}\label{fig:fidseds}
\end{figure}

With $\theta$, $\dot{m}$, and $b$ specified as above, and with the selection for $m$ from Section~\ref{sec:sxp_sedCO}, it is relatively straightforward to calculate a number of the remaining critical parameters setting the shape of the {\tt sirf} component of the model (see Table~\ref{tab:mod_params}). The spherization radius where the outflow is launched is calculated as $R_{\rm sph}$ = $\frac{18}{cos\theta} R_{\rm S} \dot{m}$, where $R_{\rm S} = 2 \frac{GM}{c^{2}}$ is the Schwarzschild radius\footnote{This is proportional to the dimensionless spherization radius from \citet{Poutanen2007}, given as $r_{\rm sph} = 5/3 \times \dot{m}$, but critically also depends on the geometry of the funnel.}. The minimum inner radius for the photosphere ($r_{\rm ph, in}$) in units of this spherization radius ($R_{\rm sph}$) can be calculated as as the innermost stable circular orbit ($R_{\rm isco} = 6 \frac{GM}{c^{2}}$) divided by $\theta$: 

\begin{equation} \label{eq:rphin}
r_{\rm ph,in} = \frac{1}{6\dot{m}}
\end{equation}

\noindent The corresponding temperature at the inner edge of the wind funnel ($T_{\rm funnel,~in}$), and the outer radius of the photopshere ($r_{\rm ph, out}$), also in units of the spherization radius, can be calculated analytically following \citet{Abolmasov2009}:

\begin{equation} \label{eq:tfunnel}
\begin{split}
T_{\rm funnel,~in} & = 0.038(\Omega'{\rm tan}^{2}\theta~r_{\rm ph, in})^{-1/4}\\  
& \times~(2.8 + 0.6 {\rm ln}~\dot{m_{3}})^{1/4} \dot{m_{3}}^{-1/2} M_{10}^{1/4}~keV
\end{split}
\end{equation}

\begin{equation}\label{eq:rphout}
r_{\rm ph,out} \simeq \left[ \frac{2\Omega'\sqrt{\dot{m}}}{(1+\alpha)\sqrt{cos\theta}}\right]^{1/(1+\alpha)}     
\end{equation}

\noindent where $\dot{m_{3}} = \dot{M}$ in units of 10$^{3}$ $\dot{M}_{\rm Edd}$, $\Omega' = -8\pi~\rm{ln}~\rm{sin}\theta$, $M_{10}$ is the mass of the BH in units of 10 \msun, and $\alpha$ is the exponent describing the velocity law for the wind. Our Equation~\ref{eq:tfunnel} corresponds to Equation 35 from \citet{Abolmasov2009}, where we have assumed $T_{\rm funnel,~in}$ is equal to the temperature at the bottom of the funnel ($T_{\rm bot}$) for the {\tt sirf} template. 

The remainder of the parameters for the {\tt sirf} component of the model are selected, as summarized in Table~\ref{tab:mod_params}, where notation corresponds to symbols used in {\tt XSPEC}. In particular, the model has $i$ = 0$^{\circ}$ (such that the cloud in the photoionization modeling described in Section~\ref{sec:cloudy} sees all components of the inflow/outflow), $\alpha = 0$ (no acceleration for the wind), $\gamma = 4/3$ (default value, the adiabatic index for the inner parts of the accretion flow), and irrad. = 4 (number of iterations for irradiation).

Finally, we convolve the {\tt sirf} component with a Comptonization component ({\tt simpl}) in order to reproduce the typical high-energy ($>$ 2 keV) shape of observed ULX spectra. All parameters values of the {\tt simpl} component are selected, namely the photon power-law index $\Gamma = 2.5$, and the scattering fraction (10\%). For observed ULXs fit with the {\tt simpl} convolution component, the recovered Comptonization parameters are broadly consistent with those listed in Table~\ref{tab:mod_params}; however, such parameters are typically not tightly constrained from observations \citep[e.g.][]{Walton2013,walton2014}. As such, we consider the \sedulx shape at energies $\gtrsim$ 2~keV, which is set by the {\tt simpl} component,  critical for comparisons to apparent observed \lx, but not particularly physically informative in terms of shape. 

The procedure for parametrizing the SED shape is performed on a {\it per luminosity bin} basis. The XLF from Equation~\ref{eq:xlf} provides the characteristic $L_{\rm obs}$, and therefore corresponding $\dot{m}$, $b$, and $\theta$, for each bin, as well as the number of sources in each luminosity bin (i.e., relative contribution to the total). The \sedulx for the total population is then the weighted sum of the SEDs from each luminosity bin. In this way, the resultant population-integrated \sedulx represents sources with a range of mass transfer rates, and beaming factors/funnel opening angles. In Figure~\ref{fig:fidseds} we show the population-integrated \sedulx for each of the three metallicities used in this work. 

As is evident from Figure~\ref{fig:fidseds}, the procedure outlined above results in a metallicity-dependent shape for the \sedulx. In particular, for $Z \gtrsim 0.1$~\Zsun, sources with the largest funnel opening angles ($\theta \sim 90^{\circ}$) contribute $\sim$ 35--45\% of the total luminosity (going from \Zsun to 0.1 \Zsun, respectively), while for $Z \lesssim 0.05$ \Zsun sources with $\theta \sim 90^{\circ}$ contribute 99\% of the total population-integrated luminosity. This change in the fractional contribution of sources with different characteristic funnel opening angles/mass transfer rates is a consequence of the change to the shape of the XLF as a function of metallicity {\it and} the change to the typical BH (accretor) mass with metallicity. The former alters the number distribution of sources with luminosity, as the bright-end slope of the XLF becomes increasingly more flat with decreasing metallicity. The latter affects the Eddington limit (Equation~\ref{eq:ledd}), and therefore the threshold for a source with a given $L_{\rm obs}$ being mildly versus highly super-Eddington. For the typical accretor mass of 8~\msun~for the $Z \geq$ 0.1~\Zsun models, the Eddington limit is $\sim$ 1.2 $\times$ 10$^{39}$ \ergs. Taking this limit in conjunction with Equations~\ref{eq:btheta}--\ref{eq:mdot_approx}, only sources with \lx $\gtrsim$ 3.8 $\times$ 10$^{39}$ \ergs will have $\theta <$ 90$^{\circ}$. By contrast, for the typical BH mass of 16~\msun~for the $Z \leq$ 0.05 \Zsun models, the Eddington limit increases to $\sim$ 2.4 $\times$ 10$^{39}$ \ergs, implying that only sources with \lx $\gtrsim$ 7.5 $\times$ 10$^{39}$ \ergs will have $\theta <$ 90$^{\circ}$. Sources at these very high luminosities are quite rare at all metallicities for the adopted XLF model. 

As a consequence of these model selections, highly super-Eddington sources---those with smaller funnel opening angles and therefore stronger EUV-emitting quasi-spherical outflow components---appear less frequently in the lowest metallicity model relative to the higher metallicity models. The \sedulx models for $Z \geq 0.1$~\Zsun are therefore flatter through the EUV than the \sedulx model at 0.05~\Zsun. As a final note, we reiterate that this procedure sets only the metallicity-dependent shape of the \sedulx. The overall normalization for the model is calculated following the theoretical scaling of \lx/\mstar with starburst age and metallicity from \citet{Fragos2013a} as described in Section~\ref{sec:sxp_scal}.

We make the \sedulx models at each metallicity available as part of this work (Section~\ref{sec:sxp_avail}), noting they represent {\it only} the accretion flow, and therefore do not account for any emission related to the putative donor star. We have implemented the \sedulx without selecting a unique donor star in part because the distribution of such sources for ULXs, though the subject of intense study, is still not well-constrained theoretically or observationally \citep[e.g.,][]{Liu2004,Copperwheat2005,Tao2011,Motch2011,Soria2012,Grise2012donor,Gladstone2013,Heida2014,Heida2016,Heida2019,Heida2019n300,Lau2019,Wiktorowicz2017,Wiktorowicz2021}. Identifying unique optical, UV, or IR counterparts at the extragalactic distances to most ULXs remains challenging. Constraining donor stars SEDs through photometry or spectroscopy is likewise difficult given uncertainties such as the contribution of emission from the accretion flow itself, and effects of irradiation of the donor star. In the photoionization simulations that follow, the \sedulx is always coupled to the SED for a stellar population, thereby making a range of potential donor stars available to the accretor population. We consider this a sensible choice for modeling the spectra or nebular emission from simple {\it populations}; however, comparison of this model to broad-band observations for an {\it individual} ULX would warrant matching the \sedulx with the spectrum for an appropriate donor star, and therefore further customization of the \sedulx model presented here. 
 
\section{Constructing a Physically Consistent Composite Population}\label{sec:sxp_ssp}

In constructing a composite population with contributions from both stars and ULXs, we start from the assumption that the \sxpulx can be specified using the same parameters governing the evolution of the stellar population. Ideally, the \sxpulx contribution would be determined directly from stellar population synthesis models themselves; however, there is currently no publicly available set of models for which properties of both the stellar and accreting compact object population are easily recoverable. Absent such models, our method aims to stitch together the \sxpulx model with publicly available models for SSPs in a physically consistent way.

To specify the stellar component (i.e., SSPs) of the composite population, we opt to use Flexible Stellar Population Synthesis (\fsps) and the associated {\tt Python-FSPS} bindings \citep{Conroy2009, Conroy2010, newpyfsps}. For the SSPs, we use the Binary Population and Spectral Synthesis \citep[BPASSv2.2;][]{eldridge2017} models available in \fsps, which set the prescriptions for the IMF, as well as metallicity- and time-dependent properties of the spectra for the stellar population. We use the spectra from BPASS v2.2 that include products of binary evolution, noting that this version does not explicitly include accreting compact objects as part of the population. As a consequence of using the default BPASS SSPs included in \fsps, the IMF has an upper mass cutoff of 100~\msun, a low-mass (0.1--0.5 \msun) IMF slope of $-1.30$, and a high-mass (0.5--100~\msun) IMF slope of $-2.35$ (matching the IMF used in the theoretical models from \citealt{Fragos2013a}). For the BPASS models $Z = 0.02$ corresponds to \Zsun. By convention in~\fsps, each SSP is normalized by stellar mass.

\begin{figure*}
    \centering
    \includegraphics[width=17cm]{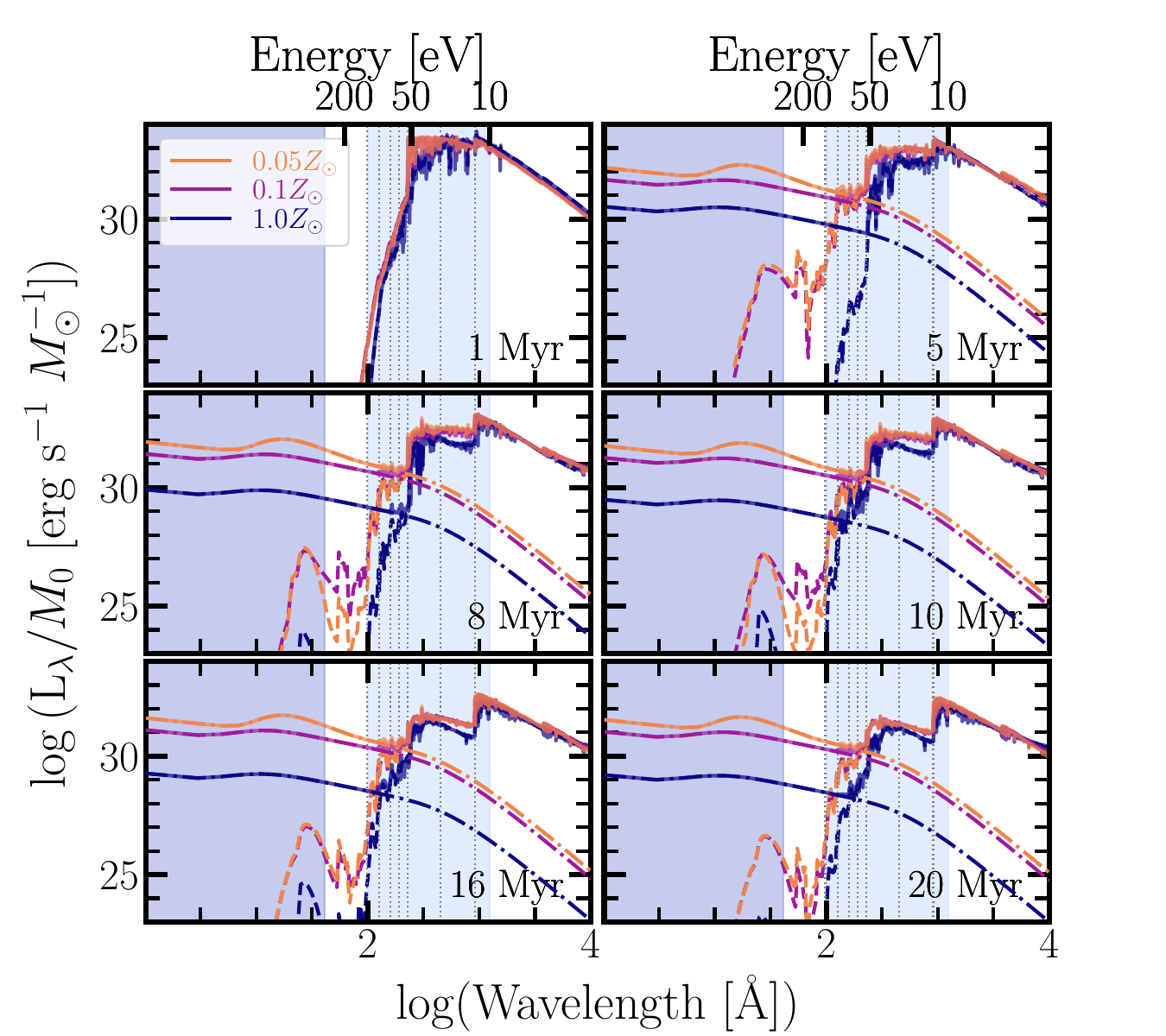}
    \caption{The composite \sxpulx + SSP SEDs (solid lines) for the sampled metallicities (dark blue: \Zsun, purple: 0.1 \Zsun, and orange: 0.05 \Zsun) normalized to the initial stellar mass (in this case 1~\msun~formed in an instantaneous burst), where the burst age is annotated in each panel. In each panel, the dash-dot line shows the \sxpulx component of the composite, while the dashed line denotes the SSP contribution. The dark blue shaded region indicates the 0.5--12 keV range, while the light blue shaded region shows the EUV regime. Grey dotted vertical lines mark the ionization potentials for select lines, from high to low energies (left to right): [\ion{Ne}{V}], [\ion{Ne}{IV}], [\ion{O}{IV}], \ion{C}{IV}, \ion{He}{II}, [\ion{Ar}{II}], and H. The addition of the \sxpulx to the corresponding SSP substantially adds to the intensity at energies $\gtrsim$ 54~eV (log(Wavelength \AA) = 2.36), particularly for the low-metallicity models with $t_{\rm burst} >$ 10~Myr.}
    \label{fig:sxp_ssp_age}
\end{figure*}

We construct a grid of BPASS SSPs in age and metallicity corresponding to the selected burst ages and metallicities at which we model the \sxpulx ($t_{\rm burst} \sim$ \{1, 3, 5, 8, 10, 16, 20\}~Myr, and $Z$ = \{0.001, 0.002, 0.02\}). Because the \sxpulxtz models we produce are already in terms of 1~\msun~stellar mass, we simply add them to the corresponding SSP to produce the composite \compSP. This ensures some measure of physical consistency between the SSP and SXP, for example, accounting for the delay between formation of high mass stars and the appearance of the first X-ray emitting accreting BHs. This procedure also incorporates metallicity dependent effects on the SSP and SXP evolution. For example, the increase in radiative output from the \sxpulx with decreasing metallicity shown in Figure~\ref{fig:lxm_age} is likely the consequence of weaker line-driven winds for massive stars at low metallicities, which result in less mass and angular momentum loss from the binary. This in turn results in XRBs with more massive compact objects and tighter orbits, which facilitate more efficient mass transfer, thereby producing more luminous sources \citep{Linden2010,Mapelli2010}. Pairing a low metallicity \sxpulx (high \lx/\mstar) with a high metallicity SSP would therefore be inconsistent, i.e., would assume that massive stars with strong line-driven winds typically produce XRBs with tight orbits and massive accreting compact objects. 

In Figure~\ref{fig:sxp_ssp_age} we show the \compSP models normalized to 1~\msun~of initial stellar mass (solid lines), where each panel corresponds to the annotated instantaneous burst age. The \sxpulx and SSP components are shown as dash-dot and dashed lines, respectively. We do not display the 3~Myr burst, as it is very similar to the 1~Myr panel, prior to \sxpulx formation. In all panels, the 0.5--12~keV bandpass is highlighted via the dark blue shaded region, while the EUV range is shown in light blue. These \compSP models, which span wavelengths 1--10$^{8}$ \AA, are used as the spectral input for photoionization modeling with \cloudy.

\section{Photoionization Simulations with \cloudy}\label{sec:cloudy}

We perform photoionization simulations using the \compSP models as input to \cloudy~v17.02 \citep{cloudy17}. To construct the \cloudy-specific input files for our photoionization simulation grid and organize the \cloudy~output, we employ a modified version of the \cloudyfsps~code \citep{cloudyfsps}. Below we describe the \cloudy~simulation set-up, relevant grid parameters, and saved output from the simulations. 

For the cloud geometry in all simulations, we assume a closed spherical shell with a fixed inner radius $R$ = 10$^{19}$~cm and a constant density $n_{\rm H} = 100$ cm$^{-3}$. The cloud is ionized by a central source, which we set as the \compSP SED, implying coincident mixing between the \sxpulx and SSP components. This cloud set-up is most appropriate for radiation-bounded regions, while the assumed density is appropriate to H{\sc ii} regions. The chosen value for the inner radius ($R$) corresponds to $\sim$ 0.3--3 $R_{\rm S}$ over the entire simulation grid, where $R_{\rm S}$ is the Str\"{o}mgren\footnote{$R_{\rm S} = (\frac{3 \mathcal{Q}_{\rm H}}{4\pi~\alpha(T_{\rm e})n_{\rm e}n_{\rm H}})^{1/3}$, where $\mathcal{Q}_{\rm H}$ is the hydrogen ionizing photon rate, $\alpha(T_{\rm e})$ is the recombination rate, $n_{\rm e}$ is the electron density, and $n_{\rm H}$ is the hydrogen density. To obtain the range of $R_{\rm S}$ from our simulations, we calculate $\mathcal{Q}_{\rm H}$ directly from the input SEDs, use $\alpha(T_{\rm e})$ = 2.6 $\times$ 10$^{-13}$~cm$^{3}$~s$^{-1}$ (corresponding to $T_{\rm e}$ = 10$^{4}$~K, as in \citealt{Jaskot2016}), and assume $n_{\rm e}$ = $n_{\rm H}$ = 100 cm$^{-3}$.} radius. 

We set $i$ = 0 for the \sedulx component of the \compSP, which is equivalent to assuming an isotropic distribution of photons from the composite spectrum is incident on the cloud. This is a simplifying assumption, but reasonable for the production of high-energy ionizing photons as considered here. The outflow component of the \sedulx is quasi-spherical, and it is this component in particular that emits strongly in the EUV. We consider more complex geometrical effects beyond the scope of the present work, but note that although the cloud in all simulations is subject to an isotropic distribution of photons (effective $i$ = 0), simulation results can be post-processed with $i > 0$ to simulate different viewing angles for an observer. 

In setting the chemical composition of the cloud, we follow the prescriptions outlined in \citealt{Richardson2022}. In brief, we employ reference abundances and scaling factors with metallicity for these abundances from \citet{Nicholls2017}. We include both Orion grains and polycyclic aromatic hydrocarbons (PAHs) in the cloud \citep{Baldwin1991,Abel2008}. We adopt a metallicity-dependent prescription for the gas-to-dust ratio, such that the abundances of all grains are scaled from their default values to satisfy a broken power-law form for the gas-to-dust ratio \citep[Equation 5 in][]{Richardson2022}. Finally, we include depletion on to dust grains using a custom set of depletion factors \citep[Table~1][]{Richardson2022}.

The \compSP models specify the SED of the central ionizing source normalized by initial stellar mass. In order to set the intensity of the composite population for the \cloudy~simulations, we normalize each model by the dimensionless ionization parameter $\mathcal{U}$: 

\begin{equation} \label{eq:logU}
    \mathcal{U} \equiv \frac{\mathcal{Q}_{\rm H}}{4 \pi R^{2}n_{\rm H} c}
\end{equation}

\noindent where $\mathcal{Q}_{\rm H} \equiv \int_{\nu_{0}}^{\infty} \frac{f_{\nu}}{h\nu}d\nu$ is the rate of emitted photons capable of ionizing hydrogen (i.e., $h\nu_{0} = 13.6$~eV), $R$ is the inner cloud radius (hydrogen ionized region) in cm, $n_{\rm H}$ is the hydrogen number density in cm$^{-3}$, and $c$ is the speed of light. By this definition, the dimensionless ionization parameter can be thought of as the density of photons relative to the density of atoms. For the simulations presented here we select seven values of \logU = [$-4.0, -3.5, -3.0, -2.5, -2.0, -1.5, -1.0$], a range that is typical for modeling H{\sc ii} regions or starbursts \citep[e.g.,][]{Dopita2000,kewley2013}. Each model is therefore specified as \compSPU before being run through \cloudy. 

Because $R$ and $n_{\rm H}$ remain fixed\footnote{We note that fixing $R$ and varying only ionizing flux means certain geometric effects are not accounted for in these models; such geometrical effects can substantially change both high-ionization and neutral line emission \citep[e.g.,][]{Ramambason2022}.} in our models, changing \logU~amounts to changing $\mathcal{Q}_{\rm H}$ and therefore the effective stellar mass, as more or fewer sources are needed for older and younger bursts, respectively, to produce sufficient ionizing photons to achieve the intensity specified by \logU. Following the same convention as \citet{Byler2017}, we define the rate of ionizing photons {\it per solar mass} as $\hat{\mathcal{Q}}_{\rm H} \equiv \mathcal{Q}_{\rm H}$ / \msun, where $\mathcal{Q}_{\rm H}$ is the rate of ionizing photons corresponding to the \logU~that sets the intensity for a given model in the \cloudy~simulations. With this convention, output quantities can be recovered in terms of the initial stellar mass formed in the burst (\mstar) through multiplication by $\hat{\mathcal{Q}}_{\rm H}$/$\mathcal{Q}_{\rm H}$. Normalization of input or output intensities by \mstar~produces the intensity per 1~\msun~initial stellar mass formed in a burst. We specify this normalization throughout, where appropriate.

\begin{figure}[t!]
    \centering
    \includegraphics[width=9cm]{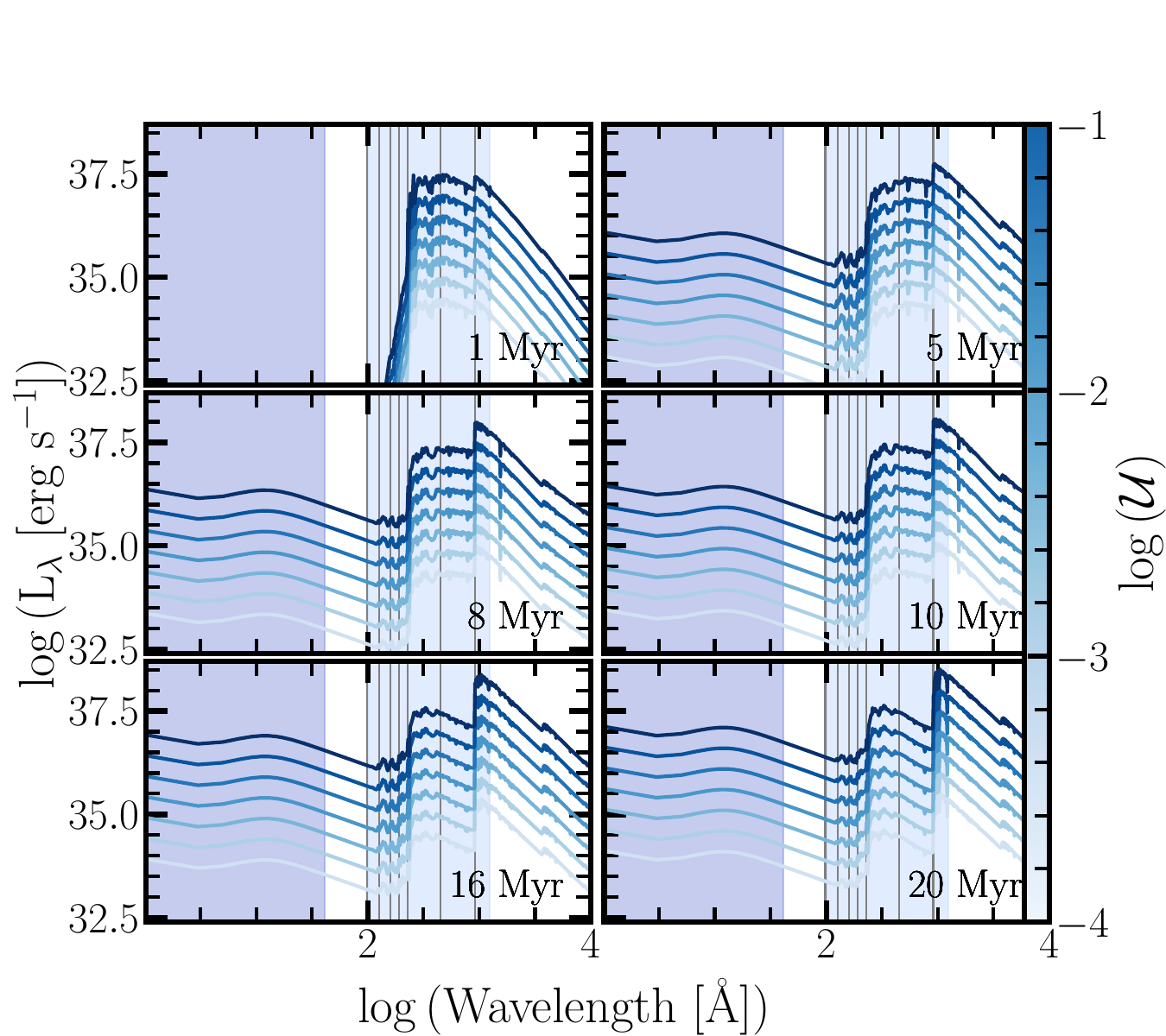}
    \caption{The composite \sxpulx + SSP SEDs at 0.1 \Zsun for a selection of instantaneous burst ages, normalized by a range of \logU~values, as noted in the colorbar. Models with high \logU have the highest intensities, and correspond to the largest stellar mass formed in a burst. As in Figure~\ref{fig:sxp_ssp_age}, the darker blue shaded region shows the X-ray bandpass, the lighter blue shaded region encompasses the EUV regime, and the grey vertical lines denote a selection of relevant ionization potentials (13.6--126~eV, from right to left). Each composite SED is used as input for the \cloudy~photoionization simulations.}
    \label{fig:sxp_ssp_age_logU}
\end{figure}

With these parameters ($t_{\rm burst}$, $Z$, and \logU), the overall grid input to \cloudy~consists of 147 separate models, corresponding to different combinations of the seven values for $t_{\rm burst}$, three values for $Z$, and seven values for \logU. In Figure~\ref{fig:sxp_ssp_age_logU}, we show the \compSPU models for select burst ages and the full range for \logU~for the 0.1 \Zsun case. The models with \logU~= $-$1 have the highest intensities, or correspondingly, the largest stellar mass formed in a burst ($\mathcal{Q}_{\rm H}$/$\hat{\mathcal{Q}}_{\rm H}$ $\sim 10^{4}-10^{6}$ \msun~for \logU~= $-$1 and $t_{\rm burst}$ = 1--20~Myr). 

We again employ a modified version of \cloudyfsps~to create the relevant \cloudy~input files for each model in the grid, execute \cloudy~simulations, and organize the \cloudy~output files. For each grid point, we allow the \cloudy~simulations to iterate to convergence, up to a maximum number of five iterations, and set the stopping criteria for the calculations to when the cloud temperature falls below 100 K or the ionized fraction falls to 1\%. Building on the nebular emission line lists presented in \citet{Byler2017} and \citet{Byler2018}, we record intensities from 406 emission lines, spanning the far IR to the near UV. The full line list is included in the Appendix in Table~\ref{tab:line_list}, including vacuum wavelength (\AA), line name, and \cloudy~specific line ID.

\section{Results}\label{sec:results}

\begin{figure*}
    \centering
    \includegraphics[width=17cm]{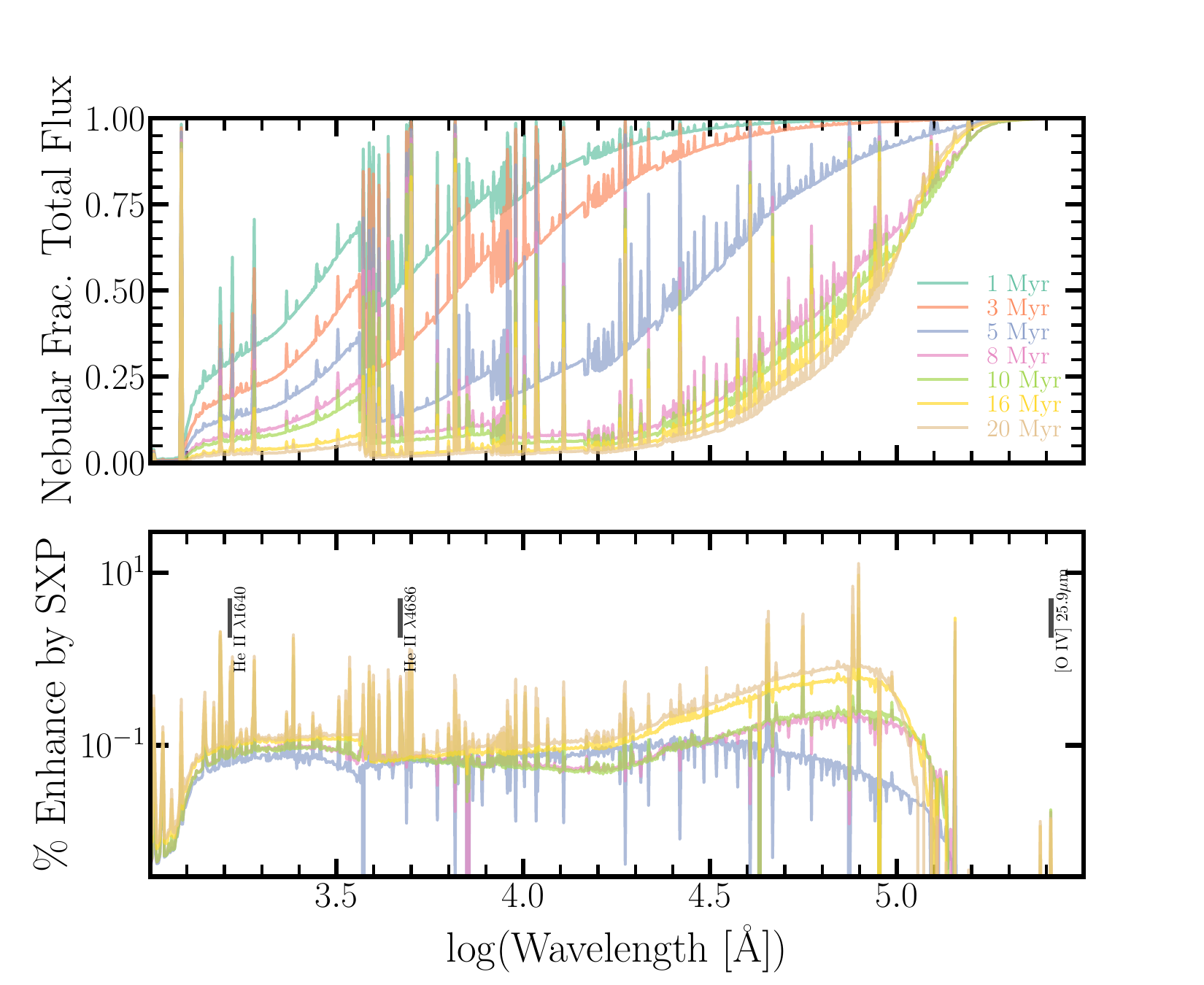}
    \caption{{\it Top}: The fractional contribution of the nebular line and continuum emission to the total flux (stellar and nebular) as a function of wavelength and instantaneous burst age for models with $Z$ = 0.1 \Zsun and \logU~= $-$1. {\it Bottom}: The percentage enhancement in nebular line and continuum flux due to the addition of the \sxpulx. For all burst timescales simulated, the \sxpulx-driven enhancement in the total nebular emission is typically small ($<$ 5\%). The black labels denote high-ionization species whose nebular {\it line} intensity is particularly enhanced due to the inclusion of the \sxpulx component.}
    \label{fig:nebflu_en}
\end{figure*}

We highlight select results illustrating the importance of the time- and metallicity-dependence of the \sxpulx implementation, and provide potential diagnostics for investigating a \sxpulx ionizing contribution. We further describe the availability of our models and simulation results.

\subsection{Characteristics of Nebular Emission Due to \texorpdfstring{\sxpulx}{SXP} and Potential \texorpdfstring{\sxpulx}{SXP} Diagnostics}\label{sec:sxp_neb}

A key result from the photoionization simulations is that the inclusion of the \sxpulx component does not significantly alter the {\it broad-band} colors in the UV--to--FIR relative to the case of the BPASS SSPs alone. This can easily be seen by looking at the enhancement in total nebular (line and continuum) emission due to the addition of the \sxpulx. In Figure~\ref{fig:nebflu_en} we show the fractional contribution of the nebular emission to the total flux as a function of wavelength and burst age in the top panel, and the percentage {\it enhancement} in this nebular flux due to the addition of the \sxpulx in the bottom panel. The nebular contribution to the total flux is high ($>$ 50\%) at all wavelengths on short burst timescales ($\leq$ 5 Myr, before the \sxpulx has turned on), and in the MIR for all burst timescales simulated here. However, as the bottom panel illustrates, the enhancement in nebular emission due to the inclusion of the \sxpulx is typically very small ($<$ 5\%). In the broad-band sense, the major effect of the \sxpulx is therefore to slightly prolong the ionizing output relative to the SSP alone, a consequence of the imposed age-dependence of our models relative to the BPASS SSPs. 

Due to the shape of the \sedulx as extended through the EUV, a key outcome from the addition of the \sxpulx is a change to the intensity of lines with excitation potentials $\geq$ 54 eV. The \cloudy~line list for our simulations (Table~\ref{tab:line_list}) includes 63 lines redward of Ly$\alpha$ with ionization potential $\geq$ 54 eV \citep{NIST}, of which 28 have ionization potentials in excess of 90~eV, including a number of so-called ``coronal lines" (i.e., forbidden transitions with very high ionization potentials). In order to establish which of these high-excitation lines may be used as diagnostics for \sxpulx ionization, we consider lines redward of Ly$\alpha$ that satisfy the following criteria: (1) are relatively strong in the models with \sxpulx contribution, as measured in terms of median line flux relative to median flux in H$\beta$ or Pa$\beta$\footnote{We use H$\beta$ as the reference line for the high-ionization lines, which are mostly in the UV and optical, and Pa$\beta$ for coronal lines in the IR.} (i.e., log($\widetilde{f}_{\rm line}/\widetilde{f}_{{\rm H}\beta}$) $\geq -3$ or log($\widetilde{f}_{\rm line}/\widetilde{f}_{{\rm Pa}\beta}$) $\geq -3$); and (2) have a median enhancement in line flux by a factor of $\sim$ 2$\times$ relative to the SSP-only models (log($\widetilde{f}_{\rm line, SXP + SSP}/\widetilde{f}_{\rm line, SSP}$) $\gtrsim$ 0.3). There are only a select few emission lines satisfying both of these criteria: \ion{He}{II}~$\lambda$1640,4686, and [\ion{O}{IV}]~25.8832$\mu$m. We annotate these lines in Figure~\ref{fig:nebflu_en}, and note them in Table~\ref{tab:line_list} ({\rm *} flag). 

\begin{figure}[t!]
    \centering
    \includegraphics[width=9cm]{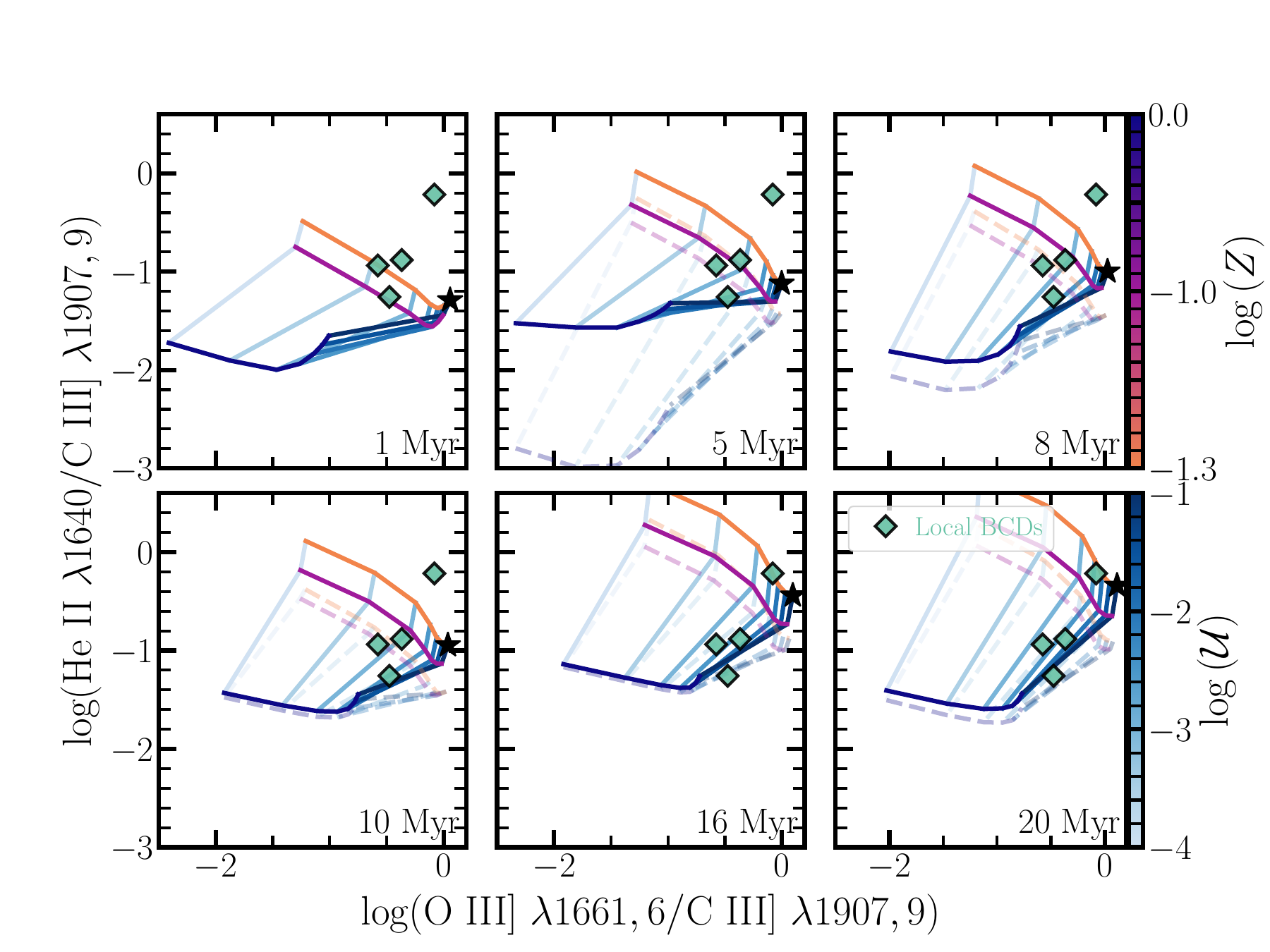}
    \caption{Potential UV diagnostic diagram for \sxpulx ionization, as a function of instantaneous burst age. The \compSPU grid is shown as solid lines, where orange to dark blue lines represent metallicities 0.05 \Zsun--\Zsun and light blue to dark blue lines represent \logU~= $-4$ to $-1$, as illustrated by the colorbars. In all panels, the star represents the 0.05 \Zsun and \logU~= $-1$ grid point. The dashed transparent lines in the background show the corresponding grid with SSP ionizing contribution only. A sample of local BCDs for which these UV lines have been measured are shown as cyan diamonds, where the uncertainties in line ratios are smaller than the plotted points \citep{Senchyna2017}. The addition of the \sxpulx hardens the ionizing spectrum at a given \logU, thereby increasing the intensity of \ion{He}{II}~$\lambda$1640 relative to \ion{C}{III}]~$\lambda$1907,9. This effect is more pronounced at solar metallicity for $t_{\rm burst}$ = 5~Myr, following depletion of WR stars, and low metallicities with $t_{\rm burst}$ $>$ 10~Myr (bottom panels), following depletion of the most massive stars.}
    \label{fig:he2u}
\end{figure}

\begin{figure}[t!]
    \centering
    \includegraphics[width=9cm]{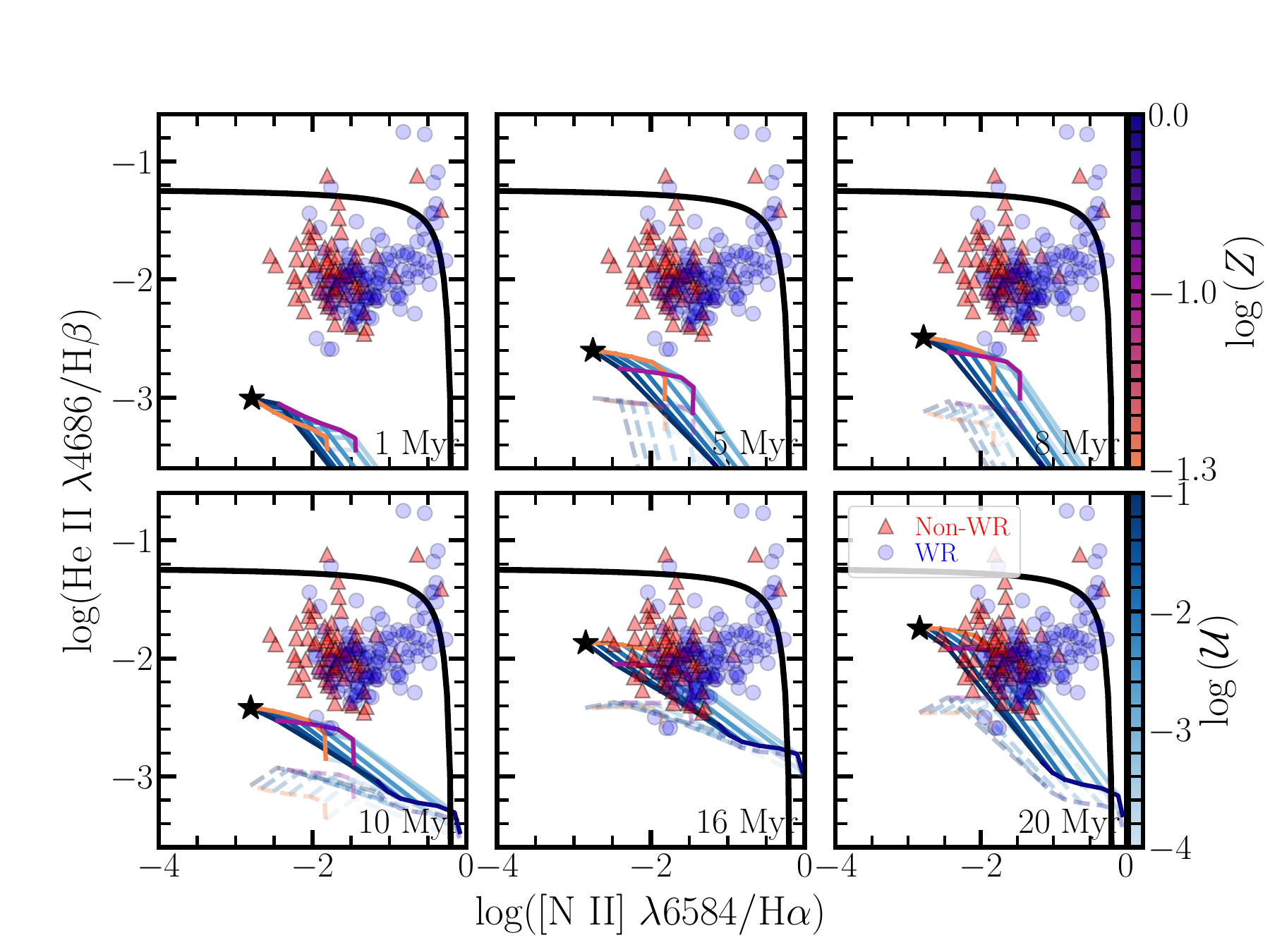}
    \caption{Same as Figure~\ref{fig:he2u}, but for a potential optical diagnostic diagram accounting for the inclusion of \sxpulx ionization. The data points are a sample of star-forming galaxies from \citet{shirazi2012} with strong nebular \ion{He}{II}~$\lambda$4686 emission (red triangles: galaxies without WR star features, blue circles: galaxies with WR star features). The black solid line is the empirically derived relation from \citet{shirazi2012} to separate ionization due to star formation (below the line) versus ionization due to active galactic nuclei or composite populations (above the line). The models with \sxpulx contribution overlap with the observed intensities for some of the Non-WR feature galaxies, particularly for $t_{\rm burst}$ $>$ 10~Myr and $Z$ $\leq$ 0.1 \Zsun; however, neither the \compSPU nor the SSP-only grids are capable of reproducing some of the more extreme galaxies, particularly those with detected WR features and higher \ion{N}{II}/H$\alpha$ near the empirical maximum starburst line.
}
    \label{fig:n2ha_v_he2hb}
\end{figure}

\begin{figure}[t!]
    \centering
    \includegraphics[width=9cm]{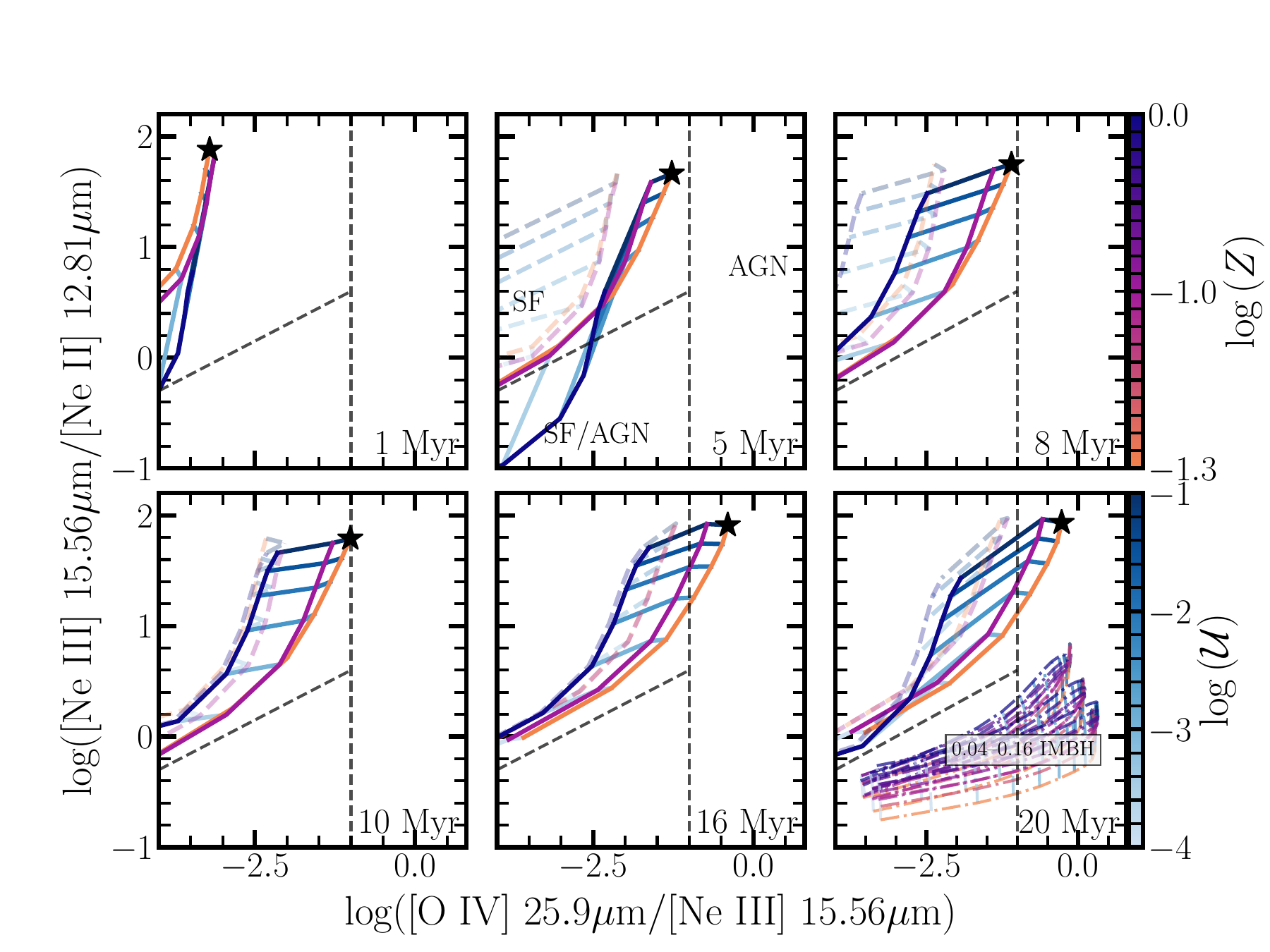}
    \caption{Same as Figure~\ref{fig:he2u}, but for a potential IR diagnostic diagram for \sxpulx ionization. The dot-dash grids are photoionization simulations of a 10$^{3}$~\msun~IMBH at different fractional contributions relative to a 20~Myr SSP from \citet{Richardson2022}, where the color scheme is the same as for the \compSPU grid for values \logU and $Z$ in common between the grids. The dashed lines show potential classification regions for pure star-formation ionization, composite star-formation and AGN ionization, and pure AGN ionization, also from \citet{Richardson2022}, as annotated in the center top panel.}
    \label{fig:ir_bpt}
\end{figure}

In Figures~\ref{fig:he2u}--\ref{fig:ir_bpt} we show emission line diagnostic diagrams in the UV, optical, and IR, respectively, to highlight the potential for uncovering ionization due to an \sxpulx using these lines. In general, diagnostic diagrams such as these are constructed using lines relatively close in wavelength to reduce the effects of reddening and make the lines used in the diagnostic accessible with a single instrument, reducing uncertainties due to cross-calibration. In all line diagnostic diagrams that follow, the \compSPU models (i.e., models {\it with} \sxpulx contribution) are shown as the solid lines, where orange to dark blue lines represent metallicities from $Z$ = 0.05--1 \Zsun and light blue to dark blue lines represent ionization parameters from \logU~= $-4$ to $-1$. For all grids in the diagnostic plots, the star denotes the model with lowest metallicity (0.05 \Zsun) and highest ionization parameter (\logU~= $-1$). The BPASS SSP-only models are shown as slightly transparent, dashed lines in the background with the same color scheme. In cases where the models with \sxpulx contribution are indistinguishable from SSP-only models (e.g., burst ages $<$ 5 Myr), only solid lines are shown.  

For the UV diagnostic in Figure~\ref{fig:he2u}, the (\ion{O}{III}]~$\lambda$1661,6)/(\ion{C}{III}]~$\lambda$1907,9) ratio is sensitive to the ionization parameter, C/O ratio (i.e., abundance pattern, see discussion in Section~\ref{sec:sxp_nebMetal}), and uses lines that are relatively strong in the UV \citep{Byler2018}. The \ion{He}{II}~$\lambda$1640/\ion{C}{III}]~$\lambda$1907,9 ratio in this same diagram is sensitive to the hardness of the ionizing spectrum. For burst timescales where the \sxpulx contributes ($\geq$ 5 Myr), this diagnostic illustrates that the ionizing spectrum is harder at all metallicities relative to the SSP-only models due to the inclusion of the \sxpulx, resulting in stronger intensity of \ion{He}{II}~$\lambda$1640 relative to \ion{C}{III}]~$\lambda$1907,9 at the same ionization parameter.

For this set of line diagnostics, we compare the \compSPU grid to a sample of local blue compact dwarf (BCD) galaxies from \citet{Senchyna2017} for which these same lines have been measured. While most of the data points from the BCDs (cyan diamonds) are consistent with both the SSP-only and \compSPU grid for ionization parameters \logU~$\geq$ $-3.0$, the inclusion of the \sxpulx increases the intensity of \ion{He}{II}~$\lambda$1640 relative to \ion{C}{III}]~$\lambda$1907,9, particularly at low metallicities and older burst ages. The composite grid is therefore consistent even with the most extreme data point, but only at later times ($>$ 10~Myr). Given the relative sparsity of the metallicity sampling in our grids and in the absence of a more detailed age analysis---which we consider beyond the scope of this work---it is not clear from this qualitative comparison whether the SSP-only or \sxpulx grid is preferred for such galaxies. Nonetheless, the comparison serves to illustrate how the addition of the \sxpulx alters the parameter space spanned by the grid, therefore changing the range of physical properties consistent with the observed population. 
In the case of Figure~\ref{fig:he2u}, the addition of the \sxpulx hardens the ionizing spectrum at a given \logU, which in effect mimics an SSP with slightly lower metallicity. 

For an optical emission line diagnostic we present Figure~\ref{fig:n2ha_v_he2hb}, where the [\ion{N}{II}]~$\lambda$6584/H$\alpha$ ratio is sensitive to metallicity and \logU (among other parameters, e.g., \citealt{kewley2013}), and the \ion{He}{II}~$\lambda$4686/H$\beta$ ratio again probes the hardness of the radiation field. Here we compare to a sample of star-forming galaxies with strong nebular \ion{He}{II} emission from SDSS identified by \citet{shirazi2012}. A little under half of this sample lack Wolf-Rayet (WR) star features in their spectra (``Non-WR"; red triangles), while the remaining galaxies show evidence for broad emission line features due to the presence of WR stars (``WR"; blue circles). In general, SSP-only models have a difficult time reproducing the observed range of \ion{He}{II}~$\lambda$4686/H$\beta$, as the dashed grids in Figure~\ref{fig:n2ha_v_he2hb} illustrate. The \compSPU grid is capable of reproducing the observed intensities for a subset of these data points, notably those without clear WR features in their spectra and those with WR features at lower metallicities, but only for select grid points. In this case, the ability of the \compSPU grid to explain a subset of the galaxies with strong \ion{He}{II}/H$\beta$ where SSP-only models cannot is again a consequence of a harder radiation field due to the addition of the \sxpulx. However, the magnitude of the increase to the ionizing photon rate due to the \sxpulx is strongly dependent on SFH. In particular, the \sxpulx contribution for this line ratio reaches a maximum for $t_{\rm burst} >$ 10~Myr. This has important consequences for the compatibility of \sxpulx models with the observed properties, such as age as inferred from equivalent widths (EWs), of some EELGs, a point we discuss in more detail in Section~\ref{sec:sxp_nebAge}.

Interestingly, the observed WR feature galaxies at higher metallicities are difficult to reproduce using either the models with \sxpulx ionizing contribution or SSP-only photoionization. This is primarily because such galaxies have large ratios of both \ion{He}{II}~$\lambda$4686/H$\beta$, implying hard ionizing spectra, and [\ion{N}{II}]~$\lambda$6584/H$\alpha$. We explore stellar population age in the context of this diagnostic in Section~\ref{sec:sxp_nebAge}, metallicity in Section~\ref{sec:sxp_nebMetal}, and alternative ionizing sources in Section~\ref{sec:discuss_altIon_IMBH}. Here we instead comment on the effect of stellar wind contamination of nebular emission. Lines such as \ion{He}{II} can be produced either via stellar winds (broad emission line component) or photoionization (narrow emission line). Moderately good spectral resolution is therefore needed to distinguish the broad stellar from narrow nebular components for such lines \citep{brinchmann2008}. If not fully resolvable, residual emission from the broader stellar wind component could enhance the line intensity assumed to be purely nebular. It is likely that stellar wind contamination, if present, operates preferentially to inflate line intensities at higher metallicities, where line driven winds from massive stars are stronger \citep{Vink2001}. It is therefore important to consider the presence and magnitude of potential stellar wind contamination in observations of such nebular line species, particularly at higher metallicities, when comparing to the purely nebular line emission determined from photoionization simulations \citep[e.g.,][]{Byler2018}. 

In the IR, we consider the diagnostic presented in Figure~\ref{fig:ir_bpt} for lines accessible with {\it JWST} \citep[e.g.,][]{Weaver2010}. Here the [\ion{O}{IV}]~25.9$\mu$m/[\ion{Ne}{III}]~15.66$\mu$m ratio is sensitive to hardness of the ionizing spectrum, while [\ion{Ne}{III}]~15.66$\mu$m/[\ion{Ne}{II}]~12.81$\mu$m is primarily sensitive to the ionization parameter. For this diagnostic, we include the classification regions from \citet{Richardson2022} for ionization due to pure star-formation (SF), SF/active galactic nuclei (AGN), and pure AGN, as well as the results from photoionization simulations for a 10$^{3}$ \msun~IMBH with fractional contribution 4--16\% relative to a 20~Myr BPASS SSP (described in more detail in Section~\ref{sec:discuss_altIon_IMBH}). The \compSPU grid is largely consistent with the pure SF region in this diagnostic, though offset from the SSP-only grid, again demonstrating how the addition of the \sxpulx hardens the ionizing spectrum. The models including \sxpulx contribution veer into the composite region for some of the solar metallicity $t_{\rm burst}$ = 5~Myr models, and into the AGN ionization region only for lowest metallicities ($\lesssim$ 0.1 \Zsun) and highest ionization parameters (\logU~$>$ -3) for bursts with $t_{\rm burst}$ $>$ 10~Myr. This indicates, not surprisingly, that the \compSP photoionization signature looks less like pure SF when the fractional contribution of the \sxpulx component relative to the SSP is most pronounced. As in Figure~\ref{fig:he2u}, this occurs when the strongest ionizing components of the stellar population are depleted, namely WR stars ($\sim$ 5 Myr at roughly solar metallicities) and the most massive stars that will explode as supernovae or collapse to a BH ($>$ 10~Myr). We note that the \sxpulx grid does {\it not} significantly overlap with IMBH grids, suggesting some power in this emission line diagnostic for distinguishing between ionization due to BHs separated by orders of magnitude in mass; however, this too depends on how the accreting BH (at any mass) is scaled relative to the stellar population, a point that we return to in more detail in Section~\ref{sec:discuss_altIon_IMBH}. 

Finally, we comment on high-ionization lines that do not satisfy both criteria in terms of strength relative to H$\beta$ or Pa$\beta$ and enhancement relative to the SSP-only case. These lines fall broadly into three categories: (1) lines that are strong relative to H$\beta$ or Pa$\beta$, but not significantly enhanced relative to the SSP-only models (flagged with \textdagger~in Table~\ref{tab:line_list}); (2) lines that are weak relative to H$\beta$ or Pa$\beta$, but are significantly enhanced relative to the SSP-only models (flagged with \textdaggerdbl~in Table~\ref{tab:line_list}); or (3) lines that are weak relative to H$\beta$ or Pa$\beta$ (including those with zero flux), and are not significantly enhanced relative to the SSP-only models (flagged with {\rm \textbar\textbar}~in Table~\ref{tab:line_list}). Lines flagged with \textdagger~in Table~\ref{tab:line_list} have a median log($\widetilde{f}_{\rm line}/\widetilde{f}_{{\rm H}\beta}$) $\sim -1.6$ and median enhancement of 1.007$\times$ in flux relative to the SSP-only models. By contrast, lines flagged \textdaggerdbl~in Table~\ref{tab:line_list} have log($\widetilde{f}_{\rm line}/\widetilde{f}_{{\rm Pa}\beta}$) $\sim -6$, but median enhancement of 300,000$\times$ in line flux relative to the SSP-only models. The latter case encompasses many coronal lines with ionization potentials $>$ 90~eV, i.e., the regime where the SSP provides relatively little flux, but the \sxpulx component is substantial. As such, there is a higher diagnostic potential for discerning the \sxpulx ionization contribution using coronal lines, many of which are in the IR, but only if the fractional contribution of the \sxpulx relative to the SSP is high enough to boost the line strengths into an easily detectable range. 

\begin{figure}[t!]
    \centering
    \includegraphics[width=9cm]{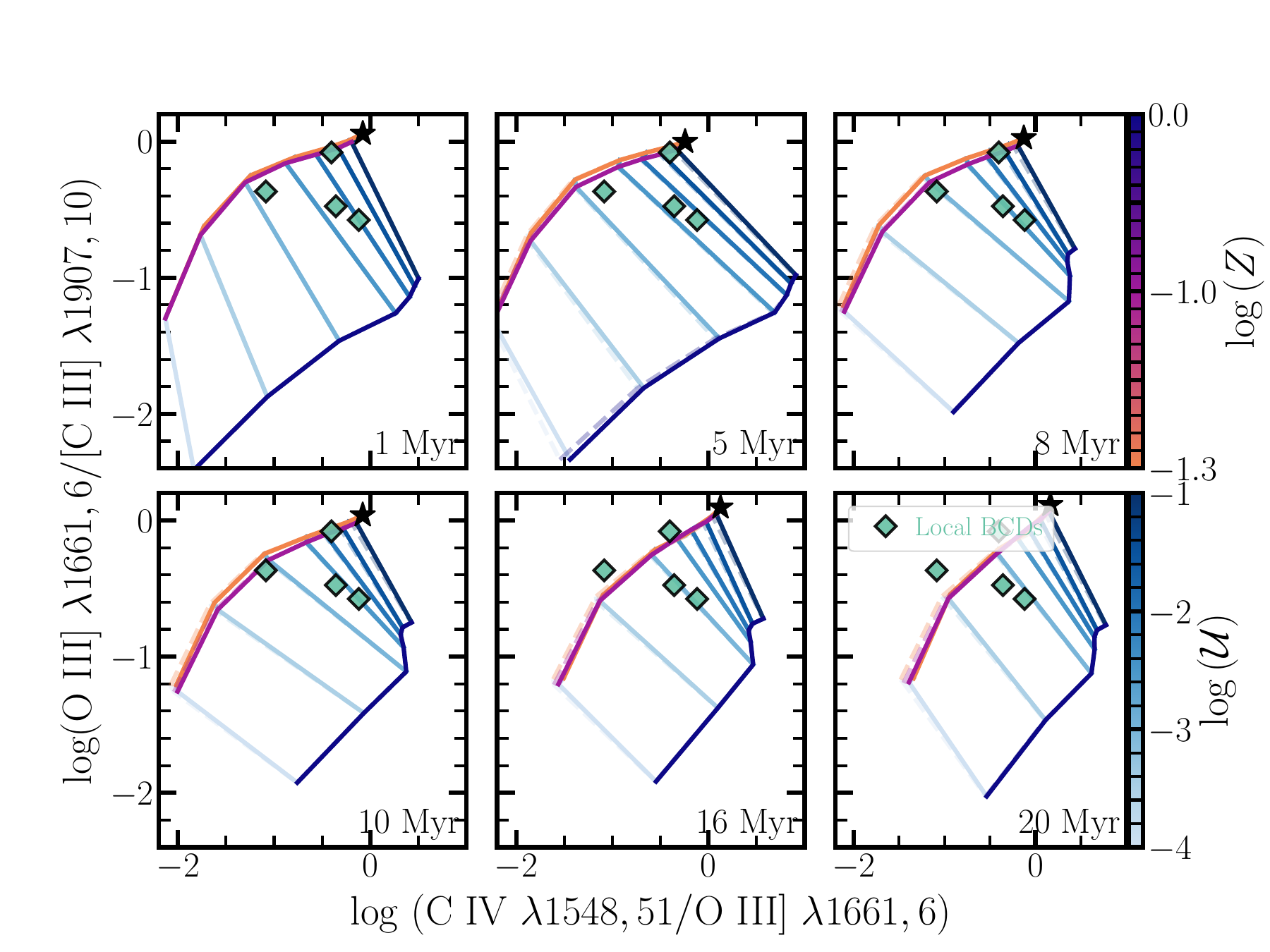}
    \caption{A UV emission line diagnostic where the \sxpulx contribution does not change the high-ionization line species appreciably relative to the SSP-only grid. The cyan diamonds are again a sample of local BCDs for which these UV lines have been measured \citep{Senchyna2017}.}
    \label{fig:c4uv}
\end{figure}

\begin{figure}[t!]
    \centering
    \includegraphics[width=9cm]{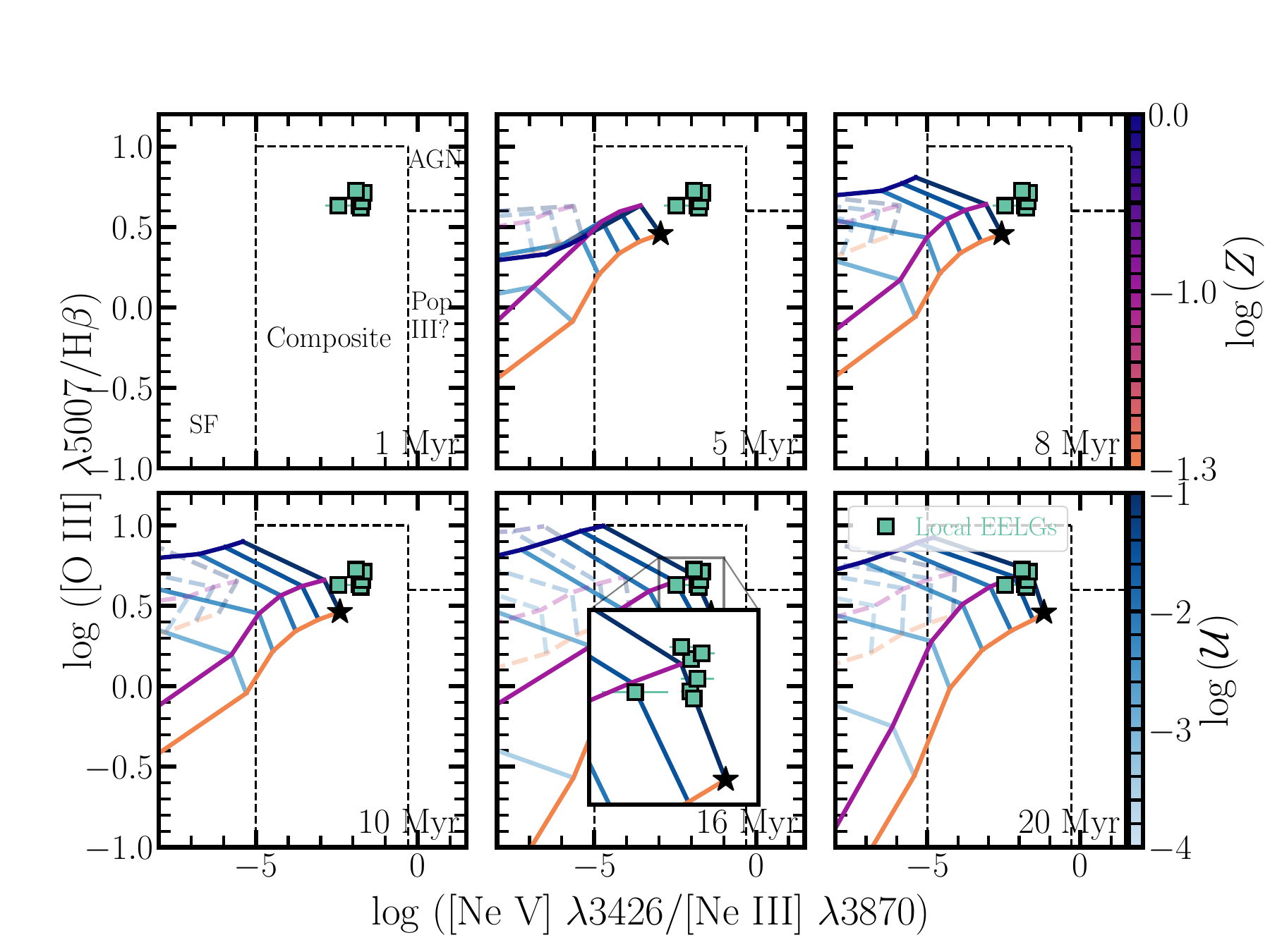}
    \caption{Same as Figure~\ref{fig:he2u}, but for a potential UV/optical diagnostic  diagram for \sxpulx contribution. For these high-ionization lines, the \sxpulx contribution is substantial relative to the SSP-only case, but the high-ionization line intensities are generally weak except for select grid points (low metallicities and high \logU). The dashed lines illustrate classification regions for pure star-formation ionization, ionization from composite populations, ionization due to AGN, and potential ionization due to Pop III stars from \citet{Cleri2023}. The cyan squares are the observed line ratios from a sample of local EELGs \citep{izotov2012,Berg2021,Olivier2022} that are consistent with select \compSPU models in a very narrow range of grid space for older $t_{\rm burst}$ (e.g., see 16~Myr inset).}
    \label{fig:ne5ne3opt}
\end{figure}

\begin{figure}[t!]
    \centering
    \includegraphics[width=9cm]{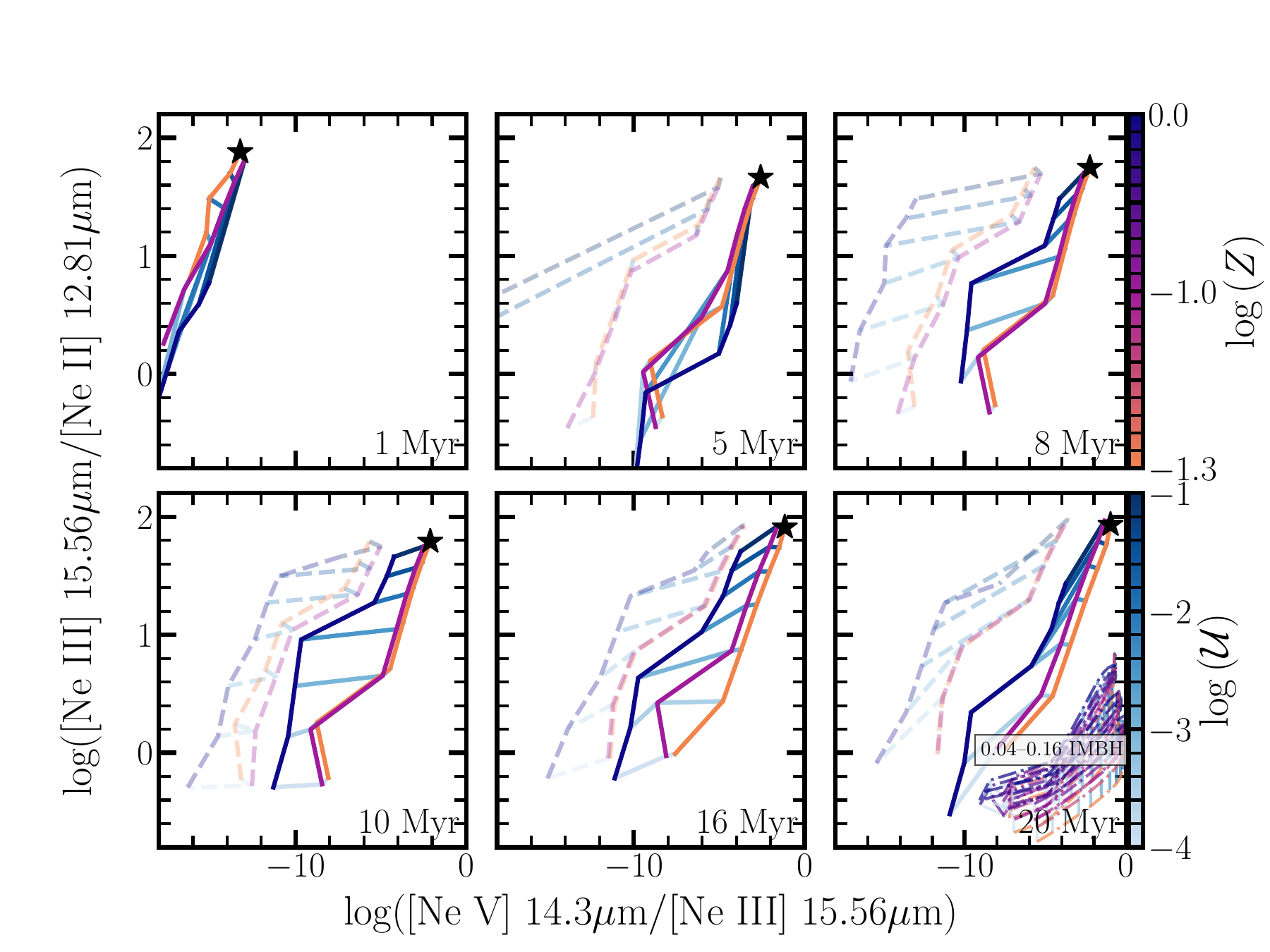}
    \caption{Same as Figure~\ref{fig:ne5ne3opt}, but for a potential IR diagnostic, albeit using a weak coronal line that is highly sensitive to hardness of the ratio field. As in Figure~\ref{fig:ir_bpt}, the IMBH models from \citet{Richardson2022} for a selection of IMBH fractional contributions are shown for reference.}
    \label{fig:ne5ne3ir}
\end{figure}

We illustrate these two cases---moderately strong lines only marginally enhanced by the \sxpulx and weak lines heavily enhanced due to the addition of the \sxpulx---via additional emission line diagnostics in Figures~\ref{fig:c4uv}--\ref{fig:ne5ne3ir}. In Figure~\ref{fig:c4uv}, we show the \compSPU grid relative to the same sample of local BCD galaxies from as in Figure~\ref{fig:he2u}. Here, the grid with \sxpulx contribution is only slightly offset to higher values of \ion{C}{IV}~$\lambda$1548,51/\ion{O}{III}]~$\lambda$~1661,6 compared to the SSP-only grid, reflecting the very modest enhancement in \ion{C}{IV}~$\lambda$1548,51 emission due to the \sxpulx. The lack of overlap between some of the galaxies and the grids in this diagnostic at later burst ages is likely a function of the adopted C/O ratio for the simulations in this work (see Section~\ref{sec:sxp_nebMetal}). 

In Figures~\ref{fig:ne5ne3opt}--\ref{fig:ne5ne3ir}, we show emission line diagnostics in the optical and IR that include the high-ionization line [\ion{Ne}{V}]. For these emission line diagnostics, the addition of the \sxpulx has a profound effect on the grid relative to the SSP-only case, 
given the ratio of [\ion{Ne}{V}]/[\ion{Ne}{III}] compares lines with ionization potentials of $\sim$ 126~eV and 64~eV, respectively. In Figure~\ref{fig:ne5ne3opt}, the SSP-only models rarely reach log([\ion{Ne}{V}]~$\lambda$3426/[\ion{Ne}{III}]~$\lambda$3870) $> -5$, while the grid with \sxpulx contribution begins to populate the composite region for low metallicities and high-ionization parameters \citep[classification regions from][]{Cleri2023}. In Figure~\ref{fig:ne5ne3ir}, the \compSPU grid traces out a relatively narrow vertical track, distinct from the SSP-only and IMBH photoionization cases. However, for both of these diagnostics, the potential power for discerning \sxpulx contribution to the ionizing photon budget comes at the expense of line strength. The ratio of [\ion{Ne}{V}]~14.3$\mu$m/[\ion{Ne}{III}]~15.56$\mu$m traced out by the grid with \sxpulx contribution spans roughly 10 orders of magnitude. Such lines are therefore undetectable for pure star-formation ionization and only potentially detectable (e.g., with {\it JWST}) with an \sxpulx ionizing contribution for a narrow range in \logU, $t_{\rm burst}$, and $Z$. This implies that very high-ionization species such as [\ion{Ne}{V}] or [\ion{Ar}{V}] may be more reliable diagnostics of AGN or IMBH photoionization rather than \sxpulx ionizing contribution; however, the potential selection biases inherent in using particular lines should be considered in assessing the reliability of any such classification diagnostic \citep[e.g.,][]{Richardson2022}. 

\subsection{Consequences of Time Dependence of the \texorpdfstring{\sxpulx}{SXP} Relative to the SSP}\label{sec:sxp_nebAge}

The key parameters affecting the time-dependence of the nebular emission from the \compSPU grid are (1) the selection to impose a delay time for the \sxpulx to turn on and; (2) the combination thereafter with SSPs of {\it corresponding} $t_{\rm burst}$. Imposing a delay time creates a scenario in which the \sxpulx prolongs the ionizing output of the population beyond that of the products of binary evolution already included in BPASS, as would be expected if the \sxpulx component is indeed descended from the stellar component, allowing time for compact object formation and accretion to begin. In this way, the \sxpulx, or accreting compact objects in general, are another way of rejuvenating the ionizing power from a population. 

In spite of this, the magnitude of the {\it additional} contribution of the \sxpulx to the emergent nebular emission has time-dependent limits by virtue of being coupled to the time evolution of the SSPs.  Although the absolute \lx/\mstar scaling for the \sxpulx is largest for younger instantaneous bursts at all metallicities (Figure~\ref{fig:lxm_age}), the \sxpulx generally hardens the incident ionizing spectrum more profoundly {\it relative to the BPASS SSPs} at older burst ages. That is, the fractional contribution of the \sxpulx to the ionizing photon budget is typically highest for burst ages $>$ 10 Myr in the models considered here. This corresponds to the timescale on which the harder components of the stellar ionizing spectrum have been depleted (e.g., WR and other massive stars), while the \sxpulx component remains appreciable. Given the imposed delay time and coupling with BPASS SSPs, the \sxpulx contribution to the total rate of ionizing photons is $\lesssim$ 2\% for burst ages between 5--10 Myr, and increases only to 5--7\% on timescales $>$ 10~Myr.

The consequences of this time-dependence are particularly evident in Figures~\ref{fig:n2ha_v_he2hb} and~\ref{fig:ne5ne3opt}, where we compare to observed samples of EELGs. In these figures, it is clear that the \sxpulx ionizing contribution can only explain the high line ratios in the observed galaxies {\it for older $t_{\rm burst}$}. This result has already been noted previously for other EELG samples. For example, a detailed investigation of the efficacy of different ionizing sources relative to a reference sample of metal-poor star-forming galaxies from \citet{Plat2019} found that products of binary evolution are capable of producing stronger high-ionization emission lines, but primarily for lower EW(H$\beta$), which corresponds to older ages assuming a burst of star formation.

In addition to their strong high-ionization emission line ratios, many EELGs may in fact be characterized by young bursts of star formation. The sample of star-forming galaxies in \citet{Plat2019}, have median \ion{He}{II}~$\lambda$4686/H$\beta$ $\sim$ -1.9, and median EW(H$\beta$) $\sim$ 160, corresponding to $t_{\rm burst}$ $\sim$ 6~Myr for BPASS burst models. Similarly, a sample of galaxies from SDSS DR7 and DR10 investigated by \citet{Stasinska2015} have median \ion{He}{II}~$\lambda$4686/H$\beta$ $\sim$ -1.9 and median EW(H$\beta$) $\sim$ 110, corresponding to $t_{\rm burst}$ $\sim$ 11~Myr. While the models with \sxpulx contribution can reach these values of \ion{He}{II}~$\lambda$4686/H$\beta$, they only do so for $t_{\rm burst}$ $>$ 10~Myr, as illustrated in Figure~\ref{fig:n2ha_v_he2hb}. This timescale is incompatible with the measured EW(H$\beta$) values, assuming bursts of star formation, for some non-negligible portion of galaxies with strong \ion{He}{II}~$\lambda$4686 detections \citep[e.g.][]{Stasinska2015,Plat2019}. Thus, the \sxpulx is unlikely to be a meaningful source of ionizing photons for the galaxies with very high EWs in such samples.

\begin{figure*}
    \centering
    \includegraphics[width=17cm]{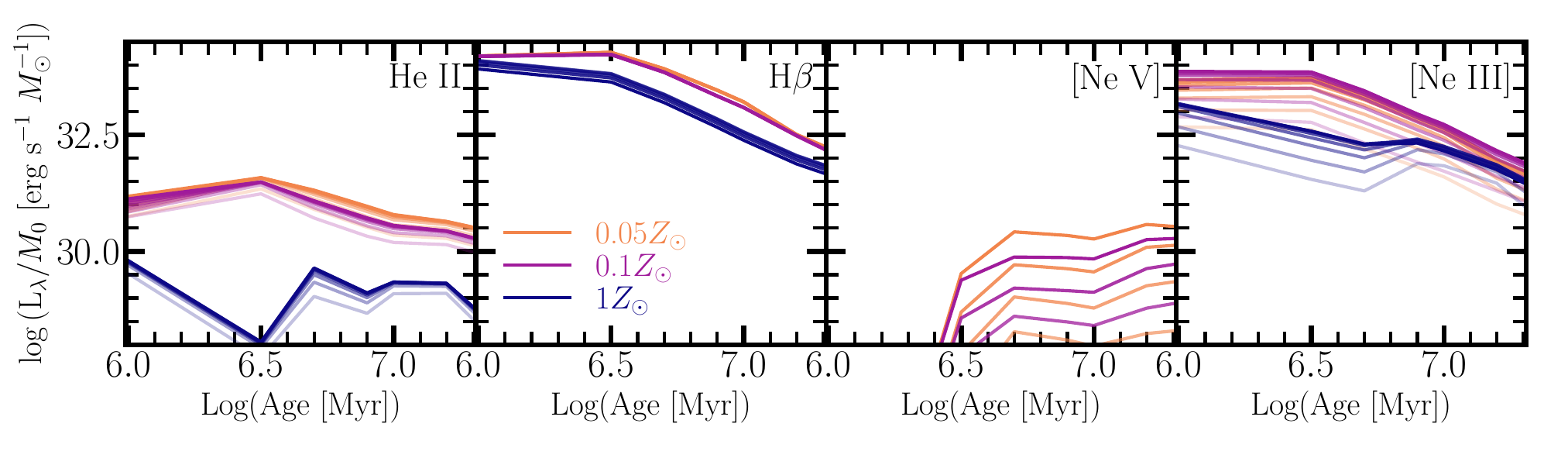}
    \caption{Nebular emission line luminosities normalized by stellar mass for \ion{He}{II}~$\lambda$4686, H$\beta$, [\ion{Ne}{V}]~$\lambda$3426, and [\ion{Ne}{III}]~$\lambda$3870 as a function of burst age. As before, in each panel dark blue corresponds to \Zsun, purple corresponds to 0.1 \Zsun, and orange corresponds to 0.01 \Zsun models. Increasing line opacity denotes increasing \logU. The ratio of a high-ionization line (e.g., \ion{He}{II}, [\ion{Ne}{V}]) with significant \sxpulx contribution relative to a lower ionization line (e.g., H$\beta$, [\ion{Ne}{III}]) is largest at later times ($>$ 10~Myr), particularly for the lower metallicity models. This corresponds to the timescale on which the ionizing contribution from the stellar population has decreased significantly, while the \sxpulx ionizing contribution remains appreciable. 
    }
    \label{fig:tlinelum}
\end{figure*}

Figure~\ref{fig:ne5ne3opt} similarly underscores this point, where the local EELGs are consistent with the models with \sxpulx contribution for a narrow range of the grid corresponding to $\sim$ 0.1 \Zsun, \logU~$> -2$, and $t_{\rm burst}$ $>$ 10~Myr. However, five of the seven galaxies in this figure have measured EW(H$\beta$) $\sim$ 100--260, corresponding to $t_{\rm burst}$ $\sim$ 5--15~Myr, and no clear correlation between EW and the strength of [\ion{Ne}{V}]~$\lambda$3426/[\ion{Ne}{III}]~$\lambda$3870 \citep{izotov2012}. The remaining galaxies presented in the figure have measured ages 1--10~Myr \citep{Olivier2022}. Assuming fidelity in these absolute ages (i.e., bursts of star formation), an \sxpulx contribution may only be important for a small subset of these galaxies, with alternative source(s) required to explain the observed line intensities at high EWs \citep[e.g.,][]{Olivier2022}.

In summary, our models suggest select high-ionization emission line ratios can be produced by ionization due to an \sxpulx, but only for $t_{\rm burst} >$ 10~Myr and low metallicities. For galaxies where the SFH includes a young ($<$ 10 Myr) burst component, the \sxpulx contribution to the ionizing photon budget is not significant enough to reproduce the observed line ratios that include high-ionization lines such as \ion{He}{II} or [\ion{Ne}{V}]. This is because, for the models considered here, the pure stellar ionizing continuum is substantial for $t_{\rm burst} <$ 10~Myr, resulting in strong emission from lines such as H$\beta$. This strong stellar ionizing flux effectively dilutes the \sxpulx contribution, which primarily affects the high-ionization line (e.g., \ion{He}{II}) rather than hydrogen ionization. We further illustrate this point in two ways. First, in Figure~\ref{fig:tlinelum} we show nebular emission line luminosities normalized by stellar mass for \ion{He}{II}~$\lambda$4686, H$\beta$, [\ion{Ne}{V}]~$\lambda$3426, and [\ion{Ne}{III}]~$\lambda$3870 as a function of burst age. In comparing the evolution of the intensities of \ion{He}{II} and H$\beta$, it is clear both lines reach their {\it absolute} maximum in intensity for $t_{\rm burst} \lesssim$ 3--5~Myr; however, while the H$\beta$ intensity drops by almost two orders of magnitude thereafter, the evolution of the \ion{He}{II} intensity is much flatter over the same time span, owing to the \sxpulx contribution, resulting in a higher ratio of \ion{He}{II}~$\lambda$/H$\beta$ at later times. The same is true for [\ion{Ne}{V}]~$\lambda$3426 relative to [\ion{Ne}{III}]~$\lambda$3870. For this ratio, it is also particularly evident when the \sxpulx contribution turns for $t_{\rm burst} \geq$ 5~Myr, as there is no appreciable [\ion{Ne}{V}] emission prior to this. 

To underscore the dominance of the stellar ionizing contribution relative to the \sxpulx at early times, we reran the photoionization simulations without the strict delay-time dependence for the \sxpulx component. For these ``no-delay time" simulations, we extrapolated \lx/\mstar from Figure~\ref{fig:lxm_age} to burst ages 1--3~Myr, allowing the \sxpulx to contribute to the ionizing budget at every time step. As Figure~\ref{fig:n2ha_v_he2hb_nodelay} shows, the immediate formation of the \sxpulx alongside a 1~Myr or 3~Myr burst only weakly increases \ion{He}{II}/H$\beta$, given that the strength of the stellar ionizing continuum is still quite substantial on these timescales. 

In this way, the addition of the \sxpulx may contribute significantly to high-energy ionizing photon production only in some cases (e.g., EELGs characterized by bursts of star formation with EW(H$\beta$) $< 120$). Such a caveat likely holds in general for other products of stellar or binary evolution with hard spectra that form preferentially on timescales $\gtrsim$~5~Myr. The rejuvenating ionizing power from such sources simply comes too late to explain observed line intensities in high EW EELGs \citep[e.g.,][]{jaskot2013}. Other hard ionizing sources that are capable of operating on a range of timescales, including the earliest times post-starburst, {\it and} that have flexibility in their relative scaling with respect to the stellar population are therefore still required for EELGs with very young burst ages \citep[e.g., shocks, very massive stars, IMBHs;][]{Senchyna2021,oskinova2022}. To understand where the \sxpulx ionizing contribution may be most critical to high-energy ionizing photon production, measured SFHs or other proxies for age such as EW(H$\beta$) should be considered alongside emission line diagnostics when comparing observed samples with these models.

\subsection{Consequences of Metallicity-Dependence of the  \texorpdfstring{\sxpulx}{SXP} and Adopted Abundance Patterns}\label{sec:sxp_nebMetal}

The photoionization simulation results are based on the assumption that the metallicity of the \sxpulx corresponds to the stellar metallicity of the SSP, as would be expected if the \sxpulx and SSP form and evolve from the same parent stellar population. Consequently, a low-metallicity SSP, which already has a somewhat harder ionizing spectrum, will always be combined with an \sxpulx with higher formation efficiency (\lx/\mstar) than the \sxpulx combined with a solar-metallicity SSP. 

The imprint of the metallicity-dependent normalization of the \sxpulx is particularly evident in emission line diagnostics that include metallicity-sensitive line ratios (e.g., [\ion{N}{II}]/H$\alpha$). To illustrate the magnitude of this effect, we run simulations where we remove the correlation between the \sxpulx normalization and stellar metallicity. In these ``no stellar metallicity dependence" simulations, we allow an SSP at a given $t_{\rm burst}$ to be coupled with an \sxpulx normalized using \lx/\mstar from the orange curve (i.e., maximum) in Figure~\ref{fig:lxm_age}, regardless of the SSP metallicity. The results are shown in Figure~\ref{fig:n2ha_v_he2hb_nometal}, where higher ratios of \ion{He}{II}~$\lambda$4686/H$\beta$ are produced at a given [\ion{N}{II}]/H$\alpha$, particularly for the solar metallicity case. This results in much more overlap between the grid with the \sxpulx contribution and the observed data points, for a broader range of burst ages; however, we note that such an elevated \lx/\mstar for ULXs at solar metallicities is not supported by observations. Even allowing for variations due to stochastic sampling of the XLF, sources with \lx $\gg$ 10$^{39}$ \ergs become more rare at higher metallicities, due to steepening of the bright end of the XLF with increasing metallicity \citep{Lehmer2021}. We therefore include this ``no stellar metallicity dependence" example simply to illustrate the effects of our model choices. 

While the \sxpulx may not be an efficient additional source of ionizing photons at high metallicity given the relations in Figure~\ref{fig:lxm_age}, other sources of high-energy ionizing photons may operate preferentially in this regime. For example, as pointed out in \citet{shirazi2012}, the sub-sample of galaxies with WR features in Figure~\ref{fig:n2ha_v_he2hb} generally has higher metallicity, which correlates with higher stellar mass and SFR, and therefore potentially stronger stellar feedback. For such galaxies, this stronger stellar feedback (i.e., winds and supernovae) may drive shocks, which could be an additional source of high-energy ionizing photons (e.g., Section~\ref{sec:discuss_altIon_Shock}).  

The choice to couple stellar and gas-phase metallicities and the adopted abundance patterns for these simulations are also of critical importance for select emission line diagnostics. Although \lx/\mstar increases with decreasing metallicity thereby increasing the number of high-energy ionizing photons, at very low metallicities the absolute abundance of some elements will be low enough that this becomes the dominant factor in setting the line intensity \citep[e.g., discussion in][]{Senchyna2020}. This can easily be seen in the shape of the \compSPU grid in Figure~\ref{fig:ne5ne3ir}, where the lowest metallicity models including \sxpulx contribution do not uniformly have the strongest [\ion{Ne}{V}]~14.3$\mu$m/[\ion{Ne}{III}]~15.56$\mu$m ratios, as would be expected if \sxpulx production efficiency (i.e., \lx/\mstar) were the dominant factor in determining line strength.

Likewise, the adopted C/O ratio in the simulations affects the extent of the grids as presented in Figures~\ref{fig:he2u} and \ref{fig:c4uv}. Our abundance pattern corresponds to log(C/O) = $-$0.77 for 0.1 \Zsun, and up to log(C/O) = $-$0.33 for \Zsun. The former is close to the average value (log(C/O) = $-$0.71) measured from metal-poor star-forming dwarf galaxies from \citet{Berg2019}, and within the large (0.17 dex) intrinsic dispersion on that value. Adopting a lower C/O ratio would shift the grids in Figure~\ref{fig:c4uv} up and to the left \citep{Byler2018}, thereby making the the models consistent with the full observational sample at all time steps. 

The abundance ratios of C/O, N/O, and C/N in particular can be strongly affected by chemical enrichment history {\it and} the ability of a given galaxy to retain enriched gas for subsequent generations of star formation. These abundance ratios may therefore be a complex function of SFH, which modifies the dominant channel for enrichment (e.g., core-collapse supernovae versus asymptotic giant branch stars), and properties like the galaxy potential well. Given these complexities, we do not attempt a more detailed investigation of abundance pattern prescriptions here. However, we do note that the simulations presented in this work rest on the assumption that gas-phase and stellar metallicities are tied, whereas in some environments these may be decoupled \citep[e.g.,][]{Steidel2016}. If such a situation arises preferentially at higher redshifts because, for example, SFHs and therefore enrichment histories are substantially different than for local reference samples, the models presented here would not be an appropriate reference point given the adopted abundance patterns. Fortunately, {\it JWST} is now capable of directly probing abundance ratios out to these high redshifts \citep[e.g.,][]{Arellano2022}, which will provide much improved constraints for tuning model abundance patterns going forward.

\subsection{Availability of Models}\label{sec:sxp_avail}

We make our models and simulation results available in two separate formats, as outlined below.

First, via a standalone github repository\footnote{\url{https://github.com/kgarofali/sxp-cloudy/tree/v1.0}} \citep{kgsxp-repo}, we provide the emission line intensities for all points in the grid (7 $t_{\rm burst}$ $\times$ 3 $Z$ $\times$ 7 \logU) corresponding to the \compSPU model presented in this work (i.e., with all parameters specified in Section~\ref{sec:cloudy}.) In this same repository, we provide code examples in {\tt python} for reading and plotting simulation results, allowing users to quickly plot the \compSPU grid for different emission line diagnostics. Users can therefore compare data or other model grids to these simulations including \sxpulx contribution. We likewise provide Table~\ref{tab:line_list} in a machine-readable format, such that lines can be selected on the basis of relevant flags included in the table notes.

We additionally make the models, including both spectra (\sedulx) and nebular and line continuum output from \cloudy, available directly from \fsps\footnote{\url{https://github.com/cconroy20/fsps/commit/4c527a9}}, and {\tt Python-FSPS} \citep{newpyfsps}. Accessing the nebular models with \sxpulx contribution is only supported for BPASS SSPs in this release of {\tt Python-FSPS}. The \sxpulx contribution can be turned on by specifying the {\tt add$\_$xrb$\_$emission = True} flag when initializing a stellar population model. With this flag specified, one can toggle on or off line and continuum emission ({\tt add$\_$neb$\_$emission} and {\tt nebemlineinspec} parameters), switch between models with and without dust grains\footnote{Although we only present models in this work that include dust grains in the cloud, the models available in \fsps include simulations run both with and without dust grains in the cloud.} in the cloud ({\tt cloudy$\_$dust} parameter), and change gas-phase metallicity and ionization parameter ({\tt gas$\_$logz} and {\tt gas$\_$logu} parameters) in the usual way. 

The models included in \fsps are of a larger grid size (10 $t_{\rm burst}$ $\times$ 11 $Z$ $\times$ 7 \logU) than the results presented here, but include a reduced line list (166 nebular emission lines). The larger grid size and reduced line list are required for parity in implementation with existing nebular models already included in \fsps. The expanded grid included in \fsps encompasses all the values of $t_{\rm burst}$, $Z$, and \logU from the grid presented in this work. It includes three supplementary points in $t_{\rm burst}$ = \{2, 4, 12\}~Myr, where the earlier bursts occur before the \sxpulx has turned on, and the later supplementary $t_{\rm burst}$ is intermediate to timescales presented in this work for which the \sxpulx provides some ionizing contribution. The expanded grid also includes 8 supplementary values in $Z$, namely $Z$ = \{0.003, 0.004, 0.006, 0.008, 0.01 , 0.014, 0.03 , 0.04\}. Including these additional $Z$ grid points requires interpolating the relations from \citet{Fragos2013a} shown in Figure~\ref{fig:lxm_age}. We therefore recommend proceeding with caution if employing models in \fsps that include the supplementary $Z$ values quoted above, as the behavior of \lx/\mstar~as a function of $Z$ and $t_{\rm burst}$ for these models is more uncertain. The reduced line list included in \fsps has been carefully selected to include lines that are most significant as \sxpulx ionization diagnostics, as well other important nebular diagnostics. The lines included in the reduced list are bolded in Table~\ref{tab:line_list}.

From \fsps, one can build SEDs for a population with or without \sxpulx contribution, including consideration of nebular line and continuum emission. In this way, users can simulate spectro-photometric data for populations including \sxpulx contribution for use in SED fitting codes \citep[e.g.,][]{prospector21,Doore2023}.  

\section{Discussion}\label{sec:discuss}

We have already presented potential UV, optical and IR emission line diagnostics for \sxpulx ionization in Section~\ref{sec:results}. Here we present best practices for diagnosing \sxpulx ionization when using X-ray observations, and summarize the results from this work relative to recent literature results for photoionization due to high-energy sources. We additionally present a brief discussion of alternative sources of hard ionizing photons and their connection to the \sxpulx model presented here. 

\begin{figure}[t!]
    \centering
    \includegraphics[width=9cm]{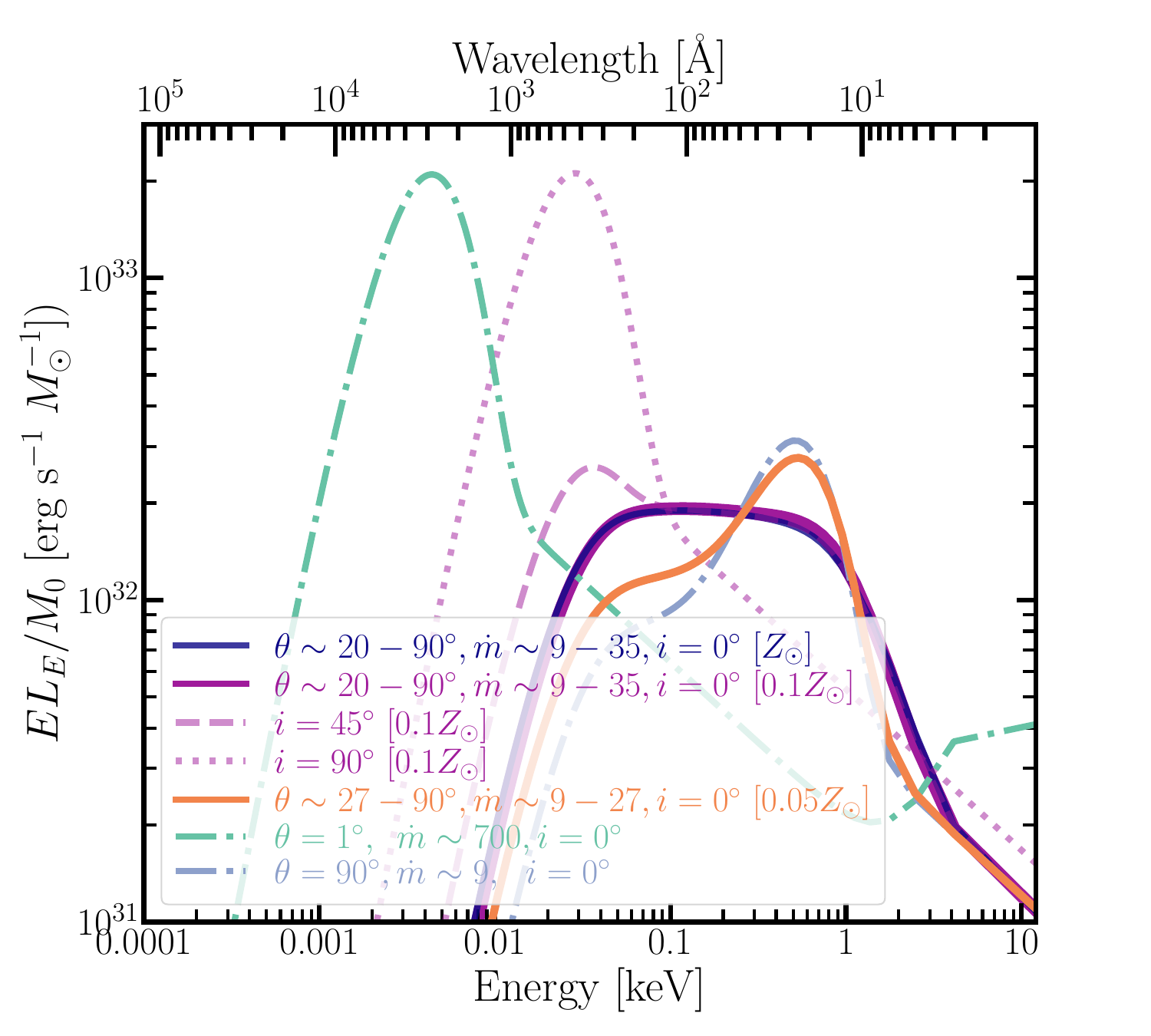}
    \caption{Same as Figure~\ref{fig:fidseds}, but now illustrating the effect on the observed and intrinsic SED from changing the inclination angle ($i$), and the mass transfer rate/funnel opening angle ($\dot{m}$/$\theta$), respectively. The \sedulx, which is the intrinsic model for the photoionization simulations where $\theta$ and $\dot{m}$ have a distribution of values, is shown via the solid lines for each of the three metallicities simulated in this work (\Zsun: dark blue, 0.1 \Zsun: purple, and 0.05 \Zsun: orange). For the 0.1 \Zsun case, we show three viewing angles for observers: $i$ = 0$^{\circ}$ (solid purple line, face-on inclination), $i$ = 45$^{\circ}$ (dashed purple line, intermediate inclination), and $i$ = 90$^{\circ}$ (dotted purple line, edge-on inclination). We additionally show via dash-dot lines two bounding cases for the shape of the intrinsic SED, where the values of $\theta$ and $\dot{m}$ are still tied to one another, but single-valued for the whole population: $\theta = 1^{\circ}$ for all sources (light green), and $\theta = 90^{\circ}$ for all sources (light blue).}\label{fig:varseds}
\end{figure}

\subsection{Caveats For Applying \texorpdfstring{\sxpulx}{SXP} Formalism to X-ray Observations: Model Assumptions}\label{sec:discuss_bestDetect}

The emergent \lx(0.5--8~keV) from the \sxpulx models is very similar to the input \lx (i.e., Figure~\ref{fig:lxm_age}), as the hard photons are attenuated very little by the cloud. However, comparison of the simulation results with X-ray observations likely requires a more careful transformation of the intrinsic \lx from the simulations into a model for observed X-ray counts. We must consider of factors such as line of sight absorption \citep[e.g.,][]{Wilms2000}, viewing angle \citep[e.g.,][]{Abolmasov2009,Kovlakas2022}, source variability \citep[e.g.,][]{Earnshaw2018}, and instrumental effects. Such considerations likely require a full forward-modeling approach \citep[e.g.,][]{Gilbertson2022}, but could allow the use of relatively low count X-ray data in spectro-photometric fitting \citep[e.g.,][]{Doore2023}. Nevertheless, implementation will be non-trivial, and we defer full consideration to a future work. Here, we instead provide caveats relevant for interpreting the simulation results with respect to the observations, given factors affecting X-ray detectability.

A primary consideration in interpreting \lx values from the photoionization simulations is that the \sedulx model assumes a specific geometry appropriate to a supercritical accretion flow. In such a model, the wind or outflow component creates an effective funnel geometry, which collimates and scatters the harder inner disk emission for face-on viewers and may obscure this component for more edge-on viewing angles \citep[e.g.,][]{Begelman2006,Poutanen2007,King2023}. Under this geometry, the observed \lx depends on inclination angle ($i$), and also on the mass supply rate ($\dot{m}$) itself. We discuss how changes to each of these parameter values changes the observed and intrinsic SED for the population, respectively. 

In all simulations, we assume $i$ = 0$^{\circ}$, such that the cloud sees all components of the accretion flow. In Figure~\ref{fig:varseds}, we show the effect on the {\it observed} SED shape from changing the viewing angle for the 0.1 \Zsun case. The intrinsic model ($i$ = 0$^{\circ}$) is shown by the solid purple line, while the SED for an intermediate viewing angle ($i$ = 45$^{\circ}$) is shown with the dashed purple line, and for a perfectly edge-on inclination ($i$ = 90$^{\circ}$) by the dotted purple line. The transition from a face-on to an edge-on viewing angle reduces the apparent (observed) hard-band X-ray emission, and increases the apparent EUV--to--UV emission as the quasi-spherical outflow component begins to dominate the line-of-sight emission. 

In this same figure, we also show the effect on the {\it intrinsic} SED from changing the dominant mass transfer rate ($\dot{m}$) and therefore funnel opening angle ($\theta$) assumed for the population. In the \sedulx model (dark blue, purple, and orange solid lines), the ratio of $L_{\rm obs}$/$L_{\rm Edd}$ sets the mass transfer rate, and therefore beaming factor and funnel opening angle. Because $L_{\rm obs}$ is sampled from the metallicity-dependent XLF, the \sedulx is constructed from a {\it distribution} of values in $\dot{m}$, $b$, and $\theta$. We show the effect of assuming $\dot{m}$ is single-valued for the entire \sxpulx via the light green and blue dash-dot lines in this figure. The light green line corresponds to the case where all sources in the population are highly super-Eddington ($\dot{m}$ $\sim$ 700, $\theta = 1^{\circ}$), while the light blue line corresponds to a population of entirely mildly super-Eddington accretors ($\dot{m}$ $\sim$ 9, $\theta = 90^{\circ}$). For a population with solely highly super-Eddington mass transfer rates, where all sources have very small funnel opening angles, the SED shape is dominated by the outflow component, which is a pseudo-blackbody peaking in the UV. By contrast, for the population composed solely of sources with mildly super-Eddington mass transfer rates, the accretion disk component dominates the emission, resulting in an SED that peaks at soft X-ray wavelengths. The \sedulx model employed in this work is intermediate to these two bounding cases, as it includes contribution from sources ranging from mildly to highly super-Eddington, and it accordingly spans EUV to hard X-ray wavelengths.

The diversity of observed spectral shapes for ultra-luminous sources may therefore be explained by differences in both funnel opening angle and viewing angle. For example, ultra-luminous supersoft sources \citep[e.g.,][]{Soria2016,Urq2016} may represent accreting compact objects with modest mass supply rates and large funnel opening angles viewed edge-on, while ultra-luminous UV sources may be compact objects with higher mass transfer rates likewise viewed edge-on \citep[e.g.,][]{Kaaret2010}.

Unfortunately, the dependence of mass transfer rate (or lack thereof) on funnel opening angle, and corresponding distribution of funnel opening angles is not well-known for either the intrinsic or the observed ULX population. Some binary population synthesis models suggest that the dominant accretor population (i.e., BH versus NS) and the timescale on which they appear will affect the fraction of the population that is strongly beamed \citep[e.g.,][]{Wiktorowicz2019}. In this respect, the form of the \sedulx used in this work is not exhaustive. The key assumptions for the \sedulx model are that the distribution of observed luminosities for the population is described by the empirical metallicity-dependent XLF, that these observed luminosities can be mapped back to an intrinsic mass transfer rate ($\dot{m}$), and that the mass transfer rate itself is tied to the beaming factor and funnel opening angle ($\theta$). The latter two selections are made to reduce the number of pre-determined paramter values. The former selection to adopt a particular functional form for the XLF, including using best-fit parameter values from \citet{Lehmer2021}, is analogous to making a selection for the form of the IMF, which often includes selection of parameters such as slope and upper mass limit cutoff.

Different assumptions about typical values for ULX parameters such as $\theta$ and $\dot{m}$, or even the form of the XLF for ULXs, will therefore propagate to the intensities of high-ionization nebular emission lines. In practice, this could work to either increase or decrease the line intensities, or perhaps even alter the strict age-dependence for particular line ratios given a model that is extreme enough in the EUV. A more detailed investigation of the effect of these parameters on the resultant nebular emission is therefore beyond the scope of the present work.

Despite these caveats about model assumptions, we demonstrate the applicability of our models to observations via a comparison of simulated and observed \ion{He}{II}~$\lambda$4686/H$\beta$ as a function of \lx/SFR \citep[a typical observable for HMXB and ULX populations; e.g.,][]{Mineo2012,Kovlakas2020}. For the observational comparison set, we use a sample of star-forming dwarf galaxies for which there are spectra from SDSS, \HST/COS, and MMT, and X-ray coverage from \chandra \citep{Senchyna2020}. Given the aforementioned caveats to X-ray detectability, comparison of our simulations with this observed sample rests on the assumption that the observed \lx corresponds roughly to the \lx from the intrinsic (simulated) population. Additionally, given the discussion in Section~\ref{sec:sxp_nebAge}, we must consider whether the SFH assumed for the simulated results is appropriate to the observed sample (i.e., line intensities, SFRs, and \lx are time-averaged in a similar way). For the observed sample, line intensities are derived from the spectroscopic data, and the corresponding SFRs and values for \lx (or upper limits) are determined from the same spectroscopic aperture, or within 1.4$\arcsec$ of the aperture for the X-ray data. SFRs are measured using calibration constants derived from BPASS appropriate to the effective age and metallicity of the stellar population assuming continuous star formation.

\begin{figure}[t!]
    \centering
    \includegraphics[width=9cm]{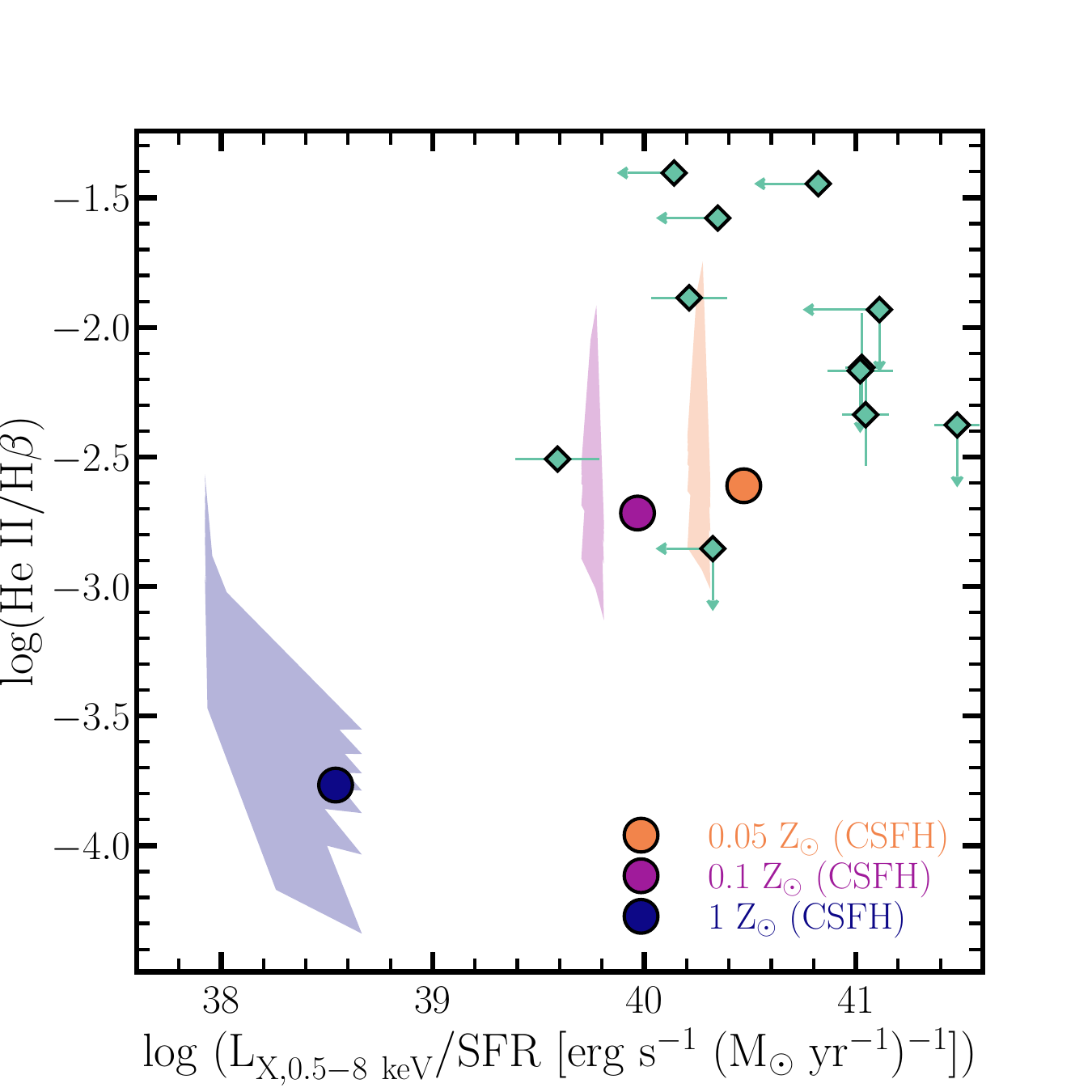}
    \caption{The emergent \lx(0.5--8~keV)/SFR relative to the strength of \ion{He}{II}~$\lambda$4686/H$\beta$ from photoionization simulations with \sxpulx contribution for both continuous star formation (circles) and instantaneous bursts (shaded regions). As before, metallicities are color-coded and labeled. Observed values for a sample of star-forming dwarfs from \citet{Senchyna2020} are shown as green diamonds. The low metallicity simulations are consistent with all but the most extreme observed values and upper limits for the line ratios and \lx/SFR; however, this requires instantaneous bursts where $t_{\rm burst} >$ 10~Myr. Models with continuous SFHs produce \lx/SFR in the range of the observed galaxies and upper limits, but much lower ratios of \ion{He}{II}~$\lambda$4686/H$\beta$. 
    }
    \label{fig:lxM_he2hb}
\end{figure}

To produce a simulated comparison set, we consider both burst and continuous SFH cases. The burst properties are easily recoverable from the simulations, as shown in figures throughout. To recover the continuous SFH case, we integrate the simulated line intensities and \sxpulx \lx (both in terms of stellar mass) assuming a constant SFR of 1~\msunyr. The results are shown in Figure~\ref{fig:lxM_he2hb}, with the continuous SFH models as circles and the burst models as shaded regions, where the extent of the region encompasses the range of simulated \logU and $t_{\rm burst}$. The most extreme line ratios are achieved for high \logU and older $t_{\rm burst}$. The simulations can reproduce much of the observed range of \ion{He}{II}~$\lambda$/H$\beta$ (green diamonds) for reasonable \lx/SFR (i.e., overlap with detections, or are below upper limits); however, this is true only for instantaneous bursts of star formation where $t_{\rm burst} > $ 10~Myr. For the continuous SFH case, the models produce much more modest \ion{He}{II}~$\lambda$/H$\beta$ and there is much less overlap with the observed sample. This demonstrates that the assumed SFH is of critical importance in such comparisons, and that, as discussed in Section~\ref{sec:sxp_nebAge}, proxies for age are integral to determining the importance of \sxpulx high-energy ionizing contribution.

As some of the observed galaxies in Figure~\ref{fig:lxM_he2hb} have values of \lx/SFR in {\it excess} of the simulated results, we briefly comment on the effects of sampling in making such comparisons. In general, testing for agreement between the simulated results for \lx and an observed population is best achieved when the observed population has a well-sampled XLF. This typically holds in the high SFR regime ($\gtrsim$ 3 \msunyr), where the effects of stochastic sampling of the XLF are on the order of, or smaller than, the model uncertainties on \lx, and where the strict linear scaling of \lx with SFR holds \citep{Grimm2003, Justham2012,Lehmer2021}. In the low SFR regime, on the other hand, integrated \lx(obs) can be subject to large ($\sim$ 0.7 dex) scatter due to stochastic sampling of an XLF that, for HMXBs and ULXs, flattens at the bright end \citep[e.g.,][]{Lehmer2021}. Similar but far less severe considerations about statistical sampling apply when determining and comparing SFRs. Empirical or model-derived calibration constants for different SFR indicators are based on the assumption of a fully populated IMF and SFR indicators which have equilibrated. In other words, typical calibration constants are appropriate for regions physically large enough for the IMF to be well-sampled {\it and} for which the SFH has been continuous long enough for a given SFR indicator to reach equilibrium \citep[e.g., 5--100 Myr][]{Kennicutt1998, Kennicutt2012}. The galaxies in Figure~\ref{fig:lxM_he2hb} that are X-ray detected with the highest \lx/SFR ($\gtrsim$ 10$^{41}$ \ergs) are some of the most distant galaxies ($D >$ 200~Mpc) in the sample, such that the spectroscopic apertures encompass large areas on the sky, making it unlikely that statistical sampling affects the measured SFRs. These galaxies have measured SFRs in the range 0.003--6~\msunyr, and therefore may be subject to significant sampling effects at the low SFR end; however, stochastic sampling is likely not enough to explain their high X-ray production efficiencies, even at the very low SFR end, making them $\sim$ 2$\sigma$ outliers relative to empirical scaling relations at low metallicities \citep{Lehmer2021}.  These galaxies may therefore represent cases where alternative high-energy ionizing sources with \lx/SFR well in excess of average XRB scaling relations contribute to the observed line ratios (e.g., Section~\ref{sec:discuss_altIon_IMBH}). 

Taken together, the discussion presented here implies that in order to determine whether \lx from a given \sxpulx is consistent with observations requires modeling the effects of inclination, both on the {\it observable} spectral components and their {\it apparent} luminosities \citep[e.g.,][]{Kovlakas2022}, as well as consideration of the effects of sampling. The former is a critical consideration in the context of these simulations, as it implies an \sxpulx may be intrinsically very luminous, and therefore efficient at producing high-energy ionizing photons, while potentially being difficult to detect in X-rays. The latter is important because all line intensities (either for bursts or calculated for continuous SFH) are simulated assuing a population average \lx. For highly star-forming galaxies, the population average \lx may be a good approximation; however, if these conditions do not hold for an observed sample, comparison with simulation results may result in erroneous inferences about population age, metallicity, and/or source(s) of ionization.  

\subsection{Comparison with Recent Literature}\label{sec:discuss_compLit}

There have been a number of works in the literature exploring the efficacy of X-ray sources in the production of high-energy ionizing photons, with a particular recent emphasis on the X-ray contribution to production of \ion{He}{II}~$\lambda$4686 in high-redshift galaxies and their analogs \citep[e.g.,][]{Kehrig2018,Kehrig2021,schaerer2019, Senchyna2020, Simmonds2021, oskinova2022, Kovlakas2022, umeda2022,Ramambason2022}. The most recent investigations fall broadly into two categories: (1) full photoionization simulations of the ionizing effect of an X-ray source coupled to a stellar population; and (2) analytic approximations for the X-ray source contribution to select line intensities, such as \ion{He}{II}~$\lambda$4686. We focus on the former here for parity with our own implementation.

\begin{figure}[t!]
    \centering
    \includegraphics[width=9cm]{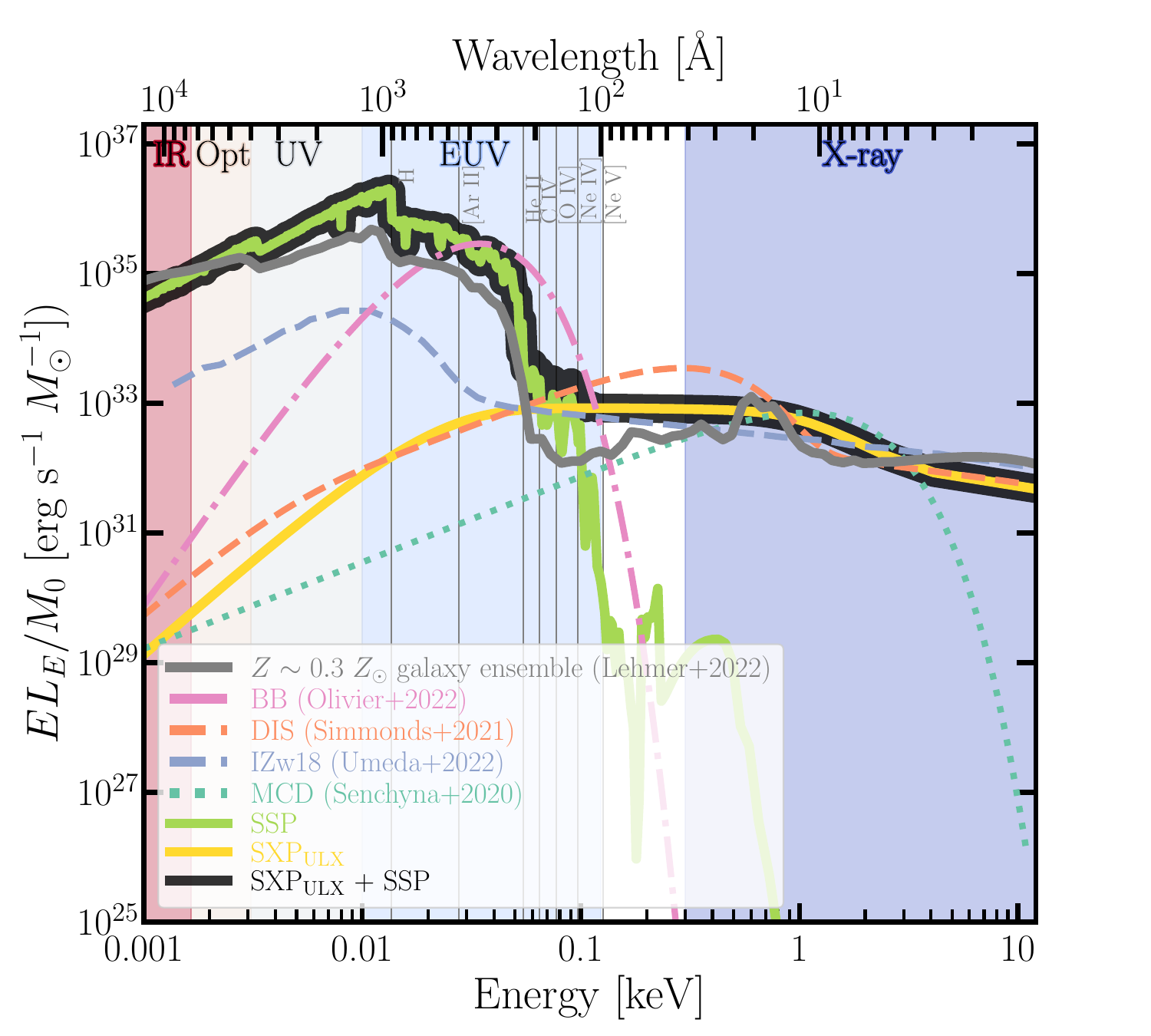}
     \caption{The intrinsic SED for the composite model (\sxpulx + SSP, solid black line) with $t_{\rm burst}$ = 5~Myr and $Z$ = 0.1 \Zsun, normalized to 1~\msun~stellar mass formed. The stellar population and \sedulx components of the composite are shown in green and yellow, respectively. Various SEDs from the literature that have previously been used in photoionization simulations are shown in different colors and linestyles for comparison, all normalized to the same \lx(0.5--8~keV) (or, in the case of the ``BB", $L_{\rm EUV}$) as the \sedulx. As a grey solid line, we also show the intrinsic SED, including contributions from stars, hot gas, and HMXBs, from recent modeling of an ensemble of low-metallicity galaxies \citep{Lehmer2022}. All SEDs have been normalized to the same 0.5--8~keV luminosity for plotting purposes. The grey vertical lines mark ionization potentials for select lines, as labeled.}
    \label{fig:SEDcomp}
\end{figure}

In Figure~\ref{fig:SEDcomp} we show in yellow the \sedulx for a 5~Myr instantaneous burst of star formation at 0.1 \Zsun. We select the 0.1 \Zsun, 5~Myr grid point for this comparison, as it has the highest ionizing photon production efficiency ($\xi_{\rm ion}$) for any grid point with \sxpulx contribution (log($\xi_{\rm ion}$) $\sim$ 25.4 $[{\rm Hz~erg^{-1}}]$), similar to the range of ionizing photon production efficiencies for EELGs \citep[e.g,][]{Olivier2022}. Here the \sedulx has been normalized to 1 \msun~stellar mass formed, corresponding to \lx(0.5--8~keV) $\sim$ 1 $\times$ 10$^{33}$ \ergs. The corresponding SSP is shown in green, while the composite of these two components is shown in black. The additional labeled SEDs (``MCD", ``DIS", ``BB", ``I Zw18", and ``$Z \sim 0.3$ \Zsun~galaxy ensemble") are selections from the literature that have previously been used in photoionization simulations or targeted as high redshift analogs for SED studies. Below, we briefly summarize salient points from recent investigations using these SEDs, and compare to our findings based on the \sxpulx framework.

The photoionization simulations performed in \citet{Senchyna2020} explored the effect of HMXBs on production of \ion{He}{II}~$\lambda$4686 and [\ion{Ne}{V}]~$\lambda$3426. The authors couple an HMXB model, represented via a multi-color disk SED (``MCD", green dotted line in Figure~\ref{fig:SEDcomp}) at different BH masses ($\sim$ 10--100 \msun) to BPASS stellar populations ($Z$ = 0.001--0.020) assuming continuous star formation. The HMXB component is scaled to the stellar population for a range of X-ray production efficiencies (\lx/SFR = 10$^{40} - 10^{44}$ \ergs (\msunyr)$^{-1}$). Much like our approach, they used a modified version of \cloudyfsps~to initialize the \cloudy~input grid, employing a similar range in \logU. With these simulation inputs, the authors conclude that the observed range of \ion{He}{II}~$\lambda$4686/H$\beta$ can only be reproduced for extremely high X-ray production efficiencies (\lx(0.5--8~keV)/SFR $>$ 10$^{42}$ \ergs (\msunyr)$^{-1}$). Such extreme X-ray production efficiencies are not observed, even in nearby extremely metal-poor galaxies, and are likely excluded by reionization-era constraints \citep{Lehmer2021,hera2023}. In Figure~\ref{fig:lxM_he2hb} we compare our model line predictions with the observational sample from \citet{Senchyna2020} which they used, in part, to draw the conclusion that HMXB ionization is insufficient to explain the observed range of \ion{He}{II}~$\lambda$4686/H$\beta$. As already discussed in Section~\ref{sec:discuss_bestDetect}, our simulations are able to reproduce more of the observed range of \ion{He}{II}~$\lambda$4686/H$\beta$ within the observed range of \lx/SFR, but primarily for instantaneous bursts with $t_{\rm burst} >$ 10~Myr at low metallicities. Under the assumption of a continuous SFH, as in \citet{Senchyna2020}, our models are not able to reproduce the more extreme values of \ion{He}{II}~$\lambda$4686/H$\beta$ at the simulated range of \lx/SFR. This underscores the importance of the assumed SFH in setting line ratios; however, even for similar assumptions about the SFH, our models produce stronger \ion{He}{II}~$\lambda$4686/H$\beta$ at a given \lx/SFR as compared with \citet{Senchyna2020}, which we attribute to differences in the assumed SED shape. Our \sedulx is much flatter through the EUV than the multi-color disk model, and therefore produces more ionizing EUV photons per \lx.

The importance of the form of the SED to the production of high-energy ionizing photons was highlighted in the \cloudy~photoionization simulations from \citet{Simmonds2021}. These authors explored the photoionizing effect of ULXs using several empirical models for the ULX SED coupled to a BPASS SSP (1~Myr instantaneous burst with $Z$ = 0.005\footnote{These authors list the metallicity of the BPASS SSP as 12 + log(O/H) = 8.1. For their adopted abundances from \citet{Grevesse2010}, this corresponds to $\sim$ 0.3 \Zsun, which we assume corresponds to the BPASS $Z$ = 0.005 model.}) again for a similar range in \logU~as employed here. In their simulations, the ULX contribution is scaled relative to the BPASS SSP for a range of \lx/SFR, motivated by empirical scalings and scatter therein. These authors find they can reproduce some of the observed range of \ion{He}{II}~$\lambda$4686/H$\beta$ for reasonable X-ray production efficiencies (\lx/SFR $\simeq$ 10$^{40}-10^{41}$ \ergs~(\msunyr)$^{-1}$), roughly in agreement with our findings for instantaneous bursts and the \sedulx model. 

In general, the simulations results presented in \citet{Simmonds2021} also achieve higher values of \ion{He}{II}~$\lambda$4686/H$\beta$ than our grids with \sxpulx contribution, at least for their most optimistic SED model (``DIS" from \citealt{Berghea2012}, as shown via the orange dashed line in Figure~\ref{fig:SEDcomp}). We attribute differences in high-energy ionizing photon production efficiency between simulations, ostensibly both for ULX-type sources, in part to differences in adopted SED shape, and in part to treatment of how the ULX component is scaled relative to the stellar population as a function of metallicity and burst age. As is evident from Figure~\ref{fig:SEDcomp}, the ``DIS" model has a stronger soft X-ray excess than our \sedulx, leading to a different shape when extended into the EUV, particularly around 54~eV where \ion{He}{II}~$\lambda$4686 has its ionization potential. While our \sxpulx simulations are run coupled to a range of SSPs with \lx/\mstar scaled as a function of burst age and metallicity, the simulations in \citet{Simmonds2021} are performed by scaling the ULX component to a single BPASS SSP (1~Myr, $\sim$ 0.3 \Zsun) using a range of \lx/SFR. This amounts to relaxing the assumptions used here for scaling of the \sxpulx formation efficiency with SSP metallicity and age, and the requirement for a delay time for ULX formation. In this way, the approach in \citet{Simmonds2021} complements the average efficiencies employed in the \sxpulx simulations: using a broad range of \lx/SFR for a given $Z$ and $t_{\rm burst}$ is akin to simulating a large degree of scatter in the X-ray production efficiency for that SSP. 

The investigation from \citet{Lehmer2022} did not perform photoionization modeling, but targeted a sample of star-forming, low-metallicity galaxies with excellent multi-wavelength coverage, including X-ray detections from \chandra. This multi-wavelength coverage enabled direct measurements of the high-energy emission, including contributions from HMXBs and hot gas, relative to stellar emission from high-redshift analog galaxies. In this work, the authors constrain the stellar component of the galaxies in their sample using broadband photometry from \Galex, SDSS, PanSTARRS, 2MASS, and WISE, and measure the hot gas and HMXB contributions using data from \chandra. The empirically-constrained intrinsic SED from the full ensemble of galaxies (grey line in Figure~\ref{fig:SEDcomp}) is produced from the physically motivated models for the stars, hot gas, and HMXBs that describe the observed multi-wavelength data, after removal of attentuation and nebular and dust emission. This SED is qualitatively similar to the \sxpulx + SSP model used in this work (black line in Figure~\ref{fig:SEDcomp}), albeit with some differences in the EUV range, particularly around 54~eV. This is the regime where the SED from \citet{Lehmer2022} is dominated by emission due to hot gas and stars, with an HMXB contribution that is a factor of $\approx$ 2--10$\times$ lower than that of the hot gas emission. In our models, there is no hot gas component, but the ULX contribution remains appreciable relative to the stellar ionizing continuum across the EUV range, which explains the difference between the two models. However, it is important to note that the contribution from HMXBs or ULXs relative to any hot gas component is subject to large fluctuations owing to stochastic sampling of the XLF, especially at low SFRs as discussed in Section~\ref{sec:discuss_bestDetect}. As such, the HMXB-to-hot gas normalization could differ by an order of magnitude from the value used for the grey model in Figure~\ref{fig:SEDcomp}, depending on the host galaxy SFR and metallicity. Comprehensive accounting of high-energy ionizing photon production from a given star-forming galaxy will therefore benefit from consideration of stellar and XRB components, as presented here, as well as hot gas or shock components, as presented in \citet{Lehmer2022} and \citet{oskinova2022}, and discussed in Section~\ref{sec:discuss_altIon_Shock}.

While the simulations and observed samples in \citet{Senchyna2020}, \citet{Simmonds2021}, \citet{Lehmer2022}, and this work explore HMXBs and ULXs specifically, there are other recent approaches to the high-energy ionizing photon production question that take a more agnostic approach to source type. In \citet{umeda2022}, the authors attempt to parametrically constrain the EUV shape of the intrinsic ionizing spectra for a sample of extremely metal-poor galaxies. To do so, they construct SEDs consisting of parametrized powerlaw and blackbody components which they run through \cloudy. They then use MCMC to determine the parameters of these SED components that best describe a set of observed high-ionization emission lines from the sample of extremely metal-poor galaxies. Their best-performing models for each galaxy are then compared with more physical SEDs, in this case BPASS SSPs coupled to different ionizing sources such as ULX or AGN, to infer physical properties of the ionizing population. In Figure~\ref{fig:SEDcomp}, we show the results of this approach for one of their galaxies, I Zw18 (blue dashed line), for which there exist excellent broad-band multi-wavelength coverage (i.e., X-ray, optical, and high-ionization nebular emission lines). We note that the results are remarkably consistent with our \sedulx component, despite very different approaches in (re)constructing the intrinsic ionizing spectrum.\footnote{The ionizing spectra appear more discrepant in the UV--to--IR range, however, this regime was not the focus of the modeling in \citet{umeda2022}, and therefore had the smallest contribution (in terms of data sets used) to constraining the best-performing model.} In general, using their full galaxy sample, \citet{umeda2022} show that the intensity of the observed high-ionization emission lines requires different combinations of BPASS SSPs (in terms of burst age) and ULX-like components with various scaling factors relative to the SSPs. Our \compSP grid is constructed with these various scaling factors determined from theoretical binary population synthesis models, while in \citet{umeda2022} the characteristic burst ages and necessary scaling factors for the high-energy ionizing photon production are inferred from the best parametric fits after comparison with physical models. The two approaches are therefore quite complementary, and highlight that determining the sources that contribute to high-energy ionizing photon production may require consideration of not only different sources of ionizing photons, but also their relative contributions as a function of stellar population age and metallicity. 

As a final point of comparison, we highlight the results from \citet{Olivier2022}, where the authors fit the high-ionization emission lines in a set of nearby EELGs with very young ($\lesssim$ 10 Myr) and metal-poor ($\sim$ 0.1 \Zsun) stellar populations. These authors find that BPASS SSPs alone cannot reproduce the observed strengths of some of the highest ionization potential species, such as \ion{He}{II} and [\ion{O}{IV}]. To reproduce the strengths of such lines, they opt to add an 80,000~K blackbody to the BPASS SSPs scaled to a relatively large fractional contribution (45--55\%) to the total luminosity. We show such a blackbody component in Figure~\ref{fig:SEDcomp} as the pink dash-dot line (``BB"). As already discussed in Section~\ref{sec:discuss_bestDetect}, the \sedulx includes a pseudo-blackbody component, albeit at slightly different temperature and normalization than the blackbody employed in \citet{Olivier2022}. However, the temperature of the roughly blackbody component of the \sedulx and its relative contribution to the total composite SED depends on the specific assumptions about the accretion flow geometry, accretor type and mass, and mass supply rate, and our \sedulx model is certainly not exhaustive in these respects. Thus, an EUV blackbody source that is not easily detectable at X-ray wavelengths as employed in \citet{Olivier2022} is not necessarily incompatible with an SXP origin (e.g., Figure~\ref{fig:varseds}). 

These results highlight a few crucial considerations for simulating photoionization due to an SXP-like source. Namely, the results depend heavily on the assumed SED shape for this additional ionizing component, and on how the component is scaled relative to the stellar population. Unfortunately, we still do not have tight empirical constraints on either of these ingredients for the case of ULXs specifically, or XRBs more broadly. However, as these photoionization simulation results suggest, select high-ionization emission lines may offer an additional lever arm for inferring the shape of unseen portions of the SED. The approaches presented in \citet{umeda2022} and \citet{Olivier2022}, for example, have the potential to be very powerful, but the recovered best-performing models depend critically on the suite of lines available for fitting as well as the set of parametric models considered. Whether using parametric or physically-motivated models as inputs to the photoionization simulations, larger grids are likely necessary going forward. We consider the recent approaches highlighted here important first steps in this process. 

\subsection{Alternative Ionizing Sources}\label{sec:discuss_altIon}

For the \sxpulx model used in this work, we have assumed the accretor is a stellar mass BH ($\sim$ 10--100 \msun) and have constructed the \sedulx accordingly. In this section, we discuss the effects of changing these assumptions about the accretor (both mass and  type), and additional processes capable of producing high-energy ionizing photons. 

\subsubsection{Neutron Star ULXs}\label{sec:discuss_altIon_NS}

While the simulations presented here consider only the case of stellar mass BH accretor for ULXs, there is now direct and growing observational evidence for NS accretors in ULXs, via the detection of pulsations \citep[e.g.][]{Bachetti2014,Furst2016,Israel2017,Carpano2018}. Swapping the BH for a NS accretor in the \sxpulx would result in a few fundamental differences, both in terms of longevity of the ionizing output relative to the stellar population and potentially the SED shape.

For the SSPs considered in this work ($t_{\rm burst}$ $\leq$ 20 Myr) BHs should dominate the accretor demographics, while NS likely become the main accretor population at later times \citep[e.g.,][]{Wiktorowicz2019}. The {\it addition} of NS ULXs to the \sxpulx would therefore further prolong the ionizing output of a population, beyond that of the most massive stars and binaries, and the BH \sxpulx presented here. Furthermore, because the stellar ionizing contribution begins to wane considerably on timescales $>$ 20~Myr after the massive star population is depleted, any additional source of high-energy photons present on these timescales will provide a larger fractional contribution to the overall ionizing photon budget. However, the contribution of different accretor demographics to the total \lx/\mstar will be both SFH- and metallicity-dependent, and such a population break-down for the scaling relations is not yet well-constrained theoretically or observationally. 

In addition to timescale differences, the spectral shape and upper limit for the luminosity may well be different for NS accretors relative to stellar mass BHs in the supercritical regime. This is primarily due to the fact, in the case of a NS, the Eddington limit and accretion flow geometry may be modified by the presence of a strong magnetic field. In calculating the Eddington limit, the electron scattering cross section is typically given as the Thomson scattering cross section; however, in the presence of a strong magnetic field, this cross section may be reduced, effectively changing the radiation pressure term in Equation~\ref{eq:ledd}, and increasing the Eddington limit for a given mass \citep{Paczynski1992}. Indeed, there is recent evidence for cyclotron resonance scattering features in the spectra of both a ULX and a hyper-luminous X-ray source (\lx $>$ 10$^{41}$ \ergs), which implies the presence of an accreting NS with very strong magnetic field \citep{Brightman2018,Brightman2022}. It is important to note, however, that an increase in Eddington limit due to a NS with a strong magnetic field is not mutually exclusive with a strongly super-Eddington mass supply rate \citep[e.g.,][]{Bachetti2022}, which may also lead to driving an outflow \citep[e.g.,][]{King2016,King2017}. 

A strong dipole field also funnels accreted material onto the poles of the NS (i.e., creates an accretion column) and possibly further modifies the picture of the supercritical accretion flow presented in Section~\ref{sec:sxp_sedAcc}, depending on how the extent of the Alfv\'en radius ($R_{\rm M}$) compares to the spherization radius ($R_{\rm sph}$). \citet{King2023} present a very detailed review and discussion of the current landscape of models for supercritically accreting NS, both pulsing and non-pulsing. Despite this attention to modeling, there exists no, to our knowledge, energetically self-consistent model in e.g., {\tt XSPEC} for a NS ULX that allows for modeling the effects of a strong magnetic field as well as inclination-dependence on the resultant SED. We consider further investigation of the NS ULX contribution to the \sxpulx a promising avenue for future study, but outside the scope of the present work. 

\subsubsection{Intermediate Mass Black Holes}\label{sec:discuss_altIon_IMBH}

\begin{figure}[t!]
    \centering
    \includegraphics[width=9cm]{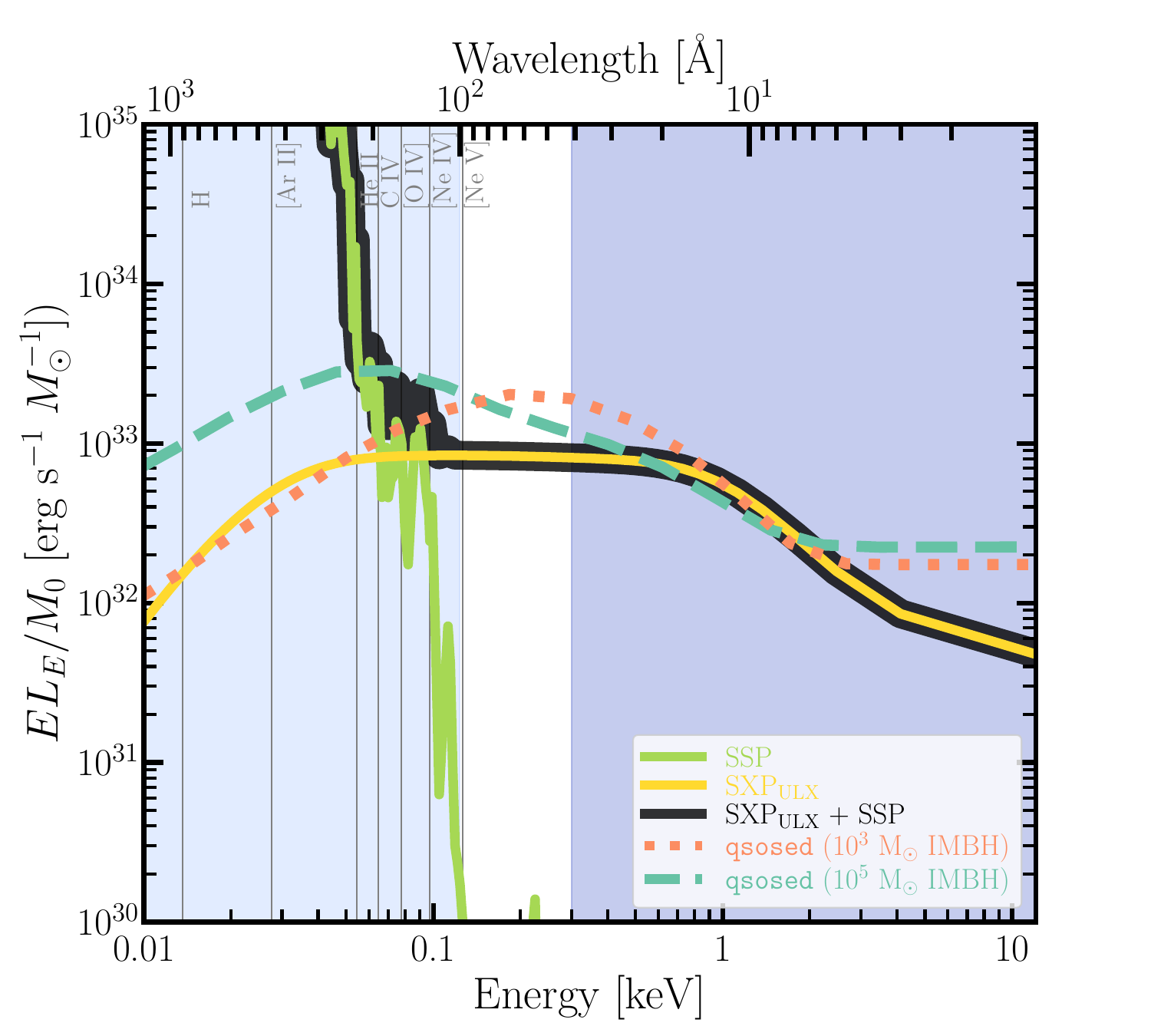}
     \caption{Similar to Figure~\ref{fig:SEDcomp}, but for a comparison with 10$^{3}$~\msun~and 10$^{5}$~\msun~IMBH SEDs (orange dotted and green dashed lines, respectively), as employed in the photoionization simulations from \citet{Richardson2022}. All SEDs are normalized to the same \lx(0.5--8~keV) for plotting purposes. In the absence of broad-band X-ray coverage, it can be difficult to distinguish between the \sxpulx and IMBH SEDs, particularly the sources are similar in bolometric luminosity and sub-dominant relative to the stellar population. However, differences in the intrinsic SEDs in the EUV--to--soft X-ray regime may result in some diagnostic power for distinguishing between these different ionizing sources using high-ionization nebular emission lines. Ionization potentials for a selection of lines are labeled and marked by grey vertical lines.}
    \label{fig:SEDIMBH}
\end{figure}

Moving up in the mass scale, IMBHs (100~\msun~$\leq$ $m$ $\leq$ 10$^{6}$~\msun) have been proposed as an additional source of ionizing photons relative to stellar populations \citep[e.g.,][]{Hatano2023}. As relatively massive BHs with moderate to low mass transfer rates relative to Eddington, IMBHs may also have SEDs with a substantial EUV component. In Figure~\ref{fig:SEDIMBH}, we compare the \sedulx with IMBH SEDs for a 10$^{3}$~\msun~BH and 10$^{5}$~\msun~BH based on the {\tt qsosed} model \citep{Kubota2018}, all normalized to the same \lx(0.5--8~keV) for plotting purposes. As Figure~\ref{fig:SEDIMBH} shows, it can be difficult to distinguish between a ULX and IMBH for sources of similar bolometric luminosity that are sub-dominant with respect to the stellar population, particularly in the absence of broad-band X-ray coverage. However, differences in the SED shapes in the EUV imply there is diagnostic power with high-ionization lines for discerning ionization by accreting BHs of different masses. We provide a preliminary exploration of this via a comparison with select results from the IMBH photoionization simulations presented in \citet{Richardson2022}. 

The full set of IMBH photoionization simulations from \citet{Richardson2022} is expansive, encompassing different BH masses, prescriptions for the IMBH SED, mixing methodologies between the IMBH and stellar population, and a relatively fine gridding in stellar metallicity and \logU. We compare only with the subset of their results most closely matching our simulation inputs and set-up. Namely, we show only their simulations for a 10$^{3}$~\msun~IMBH ({\tt qsosed} model) combined with a 20~Myr BPASS SSP assuming coincident mixing and a closed cloud geometry for grid points with -2 $\leq$ log($Z$) $\leq$ 0 and -4 $\leq$ \logU $\leq$ 1. In Figures~\ref{fig:ir_bpt} and \ref{fig:ne5ne3ir} we plot their simulation results for IMBHs with fractional contributions $\leq$ 16\% relative to the 20~Myr SSP (bottom right panels). 

\begin{figure}
    \centering
    \includegraphics[width=9cm]{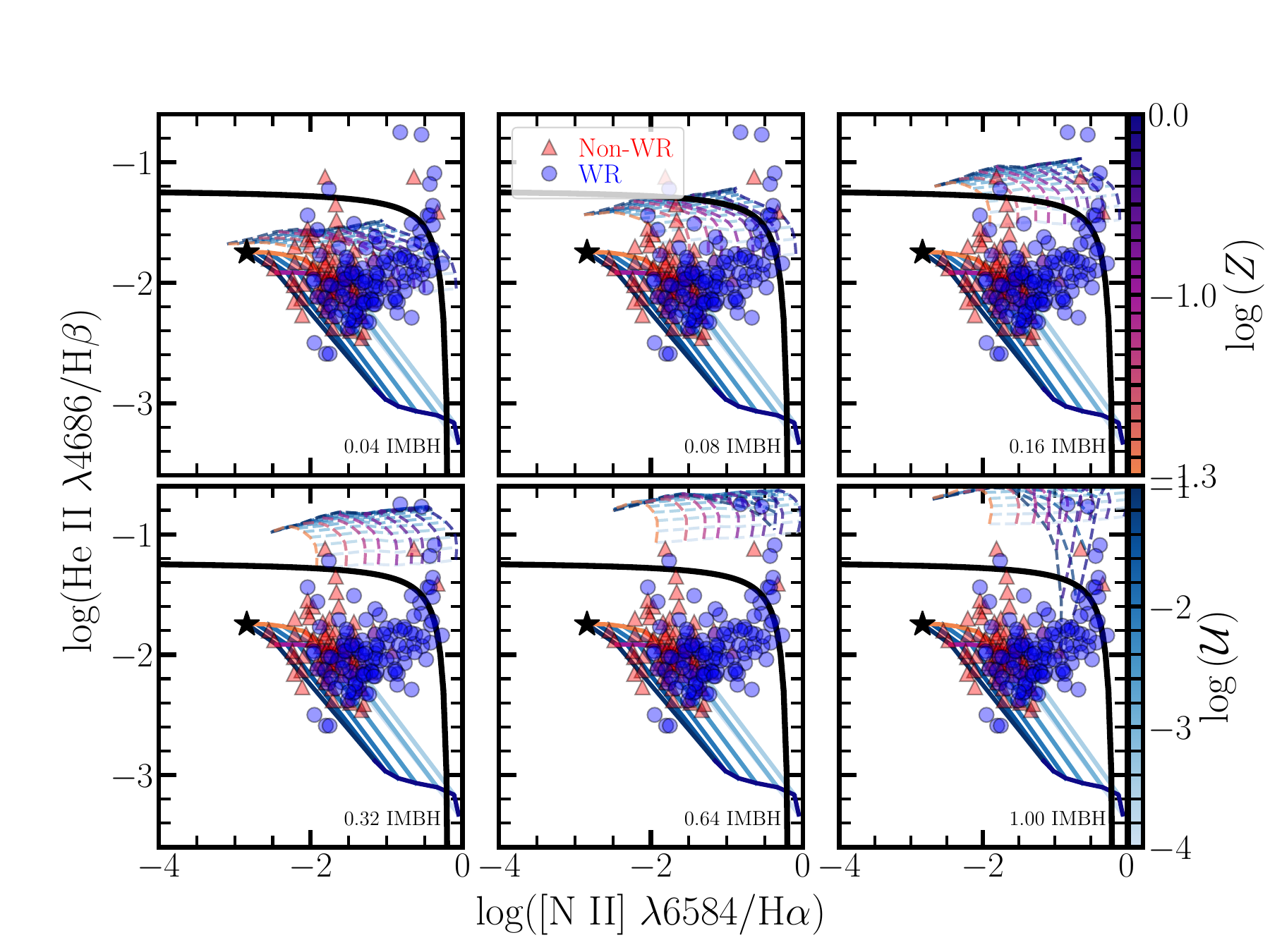}
    \caption{Similar to Figure~\ref{fig:n2ha_v_he2hb}, but with each panel showing only the 20~Myr \compSPU grid (solid lines), relative to 
    grids covering similar \logU~and~$Z$ with different fractional IMBH contributions (dashed lines) from \citet{Richardson2022}, where the IMBH contribution is annotated in the lower right of each panel. Emission line diagnostics including high-ionization species such as \ion{He}{II}~$\lambda$4686 can be useful for distinguishing between ionization due to SF versus accreting black holes across the mass scale (i.e., \sxpulx, IMBH, AGN); however, classification regions depend on the fractional contribution of the additional ionizing source (e.g., \sxpulx or IMBH) relative to the stellar population, and whether this contribution scales strongly as a function of metallicity and burst age.} 
    \label{fig:n2ha_v_he2hb_IMBHfrac}
\end{figure}

The comparisons in Figures~\ref{fig:ir_bpt} and \ref{fig:ne5ne3ir} illustrate that there is strong diagnostic potential with very high-ionization line species for distinguishing between the ionizing impact of BHs across orders of magnitude in mass \citep[e.g.,][]{Cann2018,Richardson2022}. Offsets between the IMBH grids and grid with \sxpulx contribution in these emission line diagnostics are driven primarily by SED shape, suggesting these particular line ratios are sensitive to BH mass. However, this implies the diagnostic power hinges on assumed SED shape for a given BH mass, as well as typical fractional contribution of the BH ionizing component relative to the stellar population.  

To more clearly show the differences due to changing IMBH ionizing fraction, in Figure~\ref{fig:n2ha_v_he2hb_IMBHfrac} we instead plot different fractional contributions of the IMBH (from 4--100\%) relative to the SSP in each panel, as compared only with the 20~Myr \compSPU grid from this work. As before, the black solid line represents the empirically derived star formation/AGN demarcation, which the grid with \sxpulx contribution never exceeds, and the IMBH grids only exceed for ionizing contribution $\geq$ 16\%. In addition to the fractional contribution relative to the stellar population, differences between the grids in this diagnostic space also depend critically on how the additional ionizing component is scaled with respect to the metallicity of the stellar population. For example, the IMBH simulations that we compare to do not include a metallicity-dependence for the scaling of the IMBH component relative to the SSP. In this way, the IMBH grids are produced in a fashion similar to our ``no metallicity-dependence" model (e.g., Figure~\ref{fig:n2ha_v_he2hb_nometal}), and can therefore achieve more extreme line ratios across a range in metallicities from solar to sub-solar. If instead the typical IMBH mass scales with stellar population metallicity, some regions of the IMBH grids as presented here would potentially be excluded. Harnessing the full power of such emission line diagnostics for investigating BH ionizing impact across the mass scale will therefore likely require careful consideration of typical host galaxy environments for ULXs and IMBHs, and how this affects their scaling and mixing with stellar populations \citep[e.g.,][]{Polimera2022,Richardson2022}. 

\subsubsection{Shock Ionization}\label{sec:discuss_altIon_Shock}

While the grid with \sxpulx contribution presented here is capable of reproducing some of the range of observed line ratios for high-ionization emission lines, the models fail elsewhere. This implies the need either for tweaks to the \sxpulx implementation, or consideration of some alternative energetic process. 

Here we explore the ionizing impact of fast radiative shocks, which have been considered by many others as a potential source of the requisite high-energy ionizing photons to explain the observed line intensities in select EELGs \citep[e.g.,][]{thuan2005,izotov2012,Plat2019,izotov2021}. However, instead of suggesting fast shocks as an {\it alternative} source of ionizing photons relative to the \sxpulx, we consider them as an {\it additional} source related to the \sxpulx itself. 

The accretion flow model prescribed for the \sxpulx in this work includes a quasi-spherical outflow component, which will likely interact with the surrounding medium. The interaction of ULX outflows with the interstellar medium is fairly well motivated by observations. A number of nearby ($\lesssim$ 10 Mpc) ULXs have been observed to sit inside large ($>$ 100~pc) bubbles. Such nebulae may be radiatively or mechanically inflated \citep[e.g.,][]{Cseh2012,King2023}. Several very bright ULXs show more direct evidence for the presence of fast outflows via the detection of blue-shifted absorption lines in high-resolution X-ray spectra \citep{Poutanen2007,Pinto2016,Kosec2018a,Kosec2018b}. Very recently, a detailed optical spectroscopic study of a ULX with direct evidence for fast outflows, as well as a bubble detection, confirmed the ULX wind or jet likely drives bubble expansion \citep{Gurpide2022}. And indeed, investigation of the emission line diagnostics for such ULX bubbles reveal signatures of shock and photoionization \citep[e.g.,][]{Abolmasov2007,Lopez2019, Gurpide2022}. Therefore, in at least {\it some} cases, ULX feedback is sufficient to inflate a bubble and drive both shock and photoionization. 

The \sedulx model employed here, by virtue of including the outflow component, is therefore capable of driving shock ionization as well as photoionization, though we model only the latter in this work. Detailed shock and precursor ionization model grids are available from MAPPINGS \citep{Allen2008}, which may in principle be matched to a corresponding \sxpulx to simulate the combined signature of shock and photoionization. However, including the effects of shocks self-consistently alongside \sxpulx photoionization, assuming the shock is driven by the \sxpulx itself, will likely require more than the addition of separate grids. In all simulations presented here we assume the gas is uniformly distributed in the cloud, but the presence of shocks related to the \sxpulx (or indeed the stellar population) could preferentially displace gas from certain regions of the cloud. This may lead to shock ionization occurring in regions with different density than where the primary photoionization is occurring \citep[e.g.,][]{izotov2021}. Such a non-uniform density may also alter the intensities of typical strong lines such as [\ion{O}{III}], which is produced in the inner regions of the cloud. If gas is displaced from the central regions, this would amount to effectively lowering the ionization parameter for species such as [\ion{O}{III}], relative to species produced elsewhere in the cloud \citep[e.g.,][]{kewley2013}. It is clear a more careful treatment of joint consideration of \sxpulx shock and photoionization is warranted in terms of timescales, cloud properties, and mixing methodologies to understand the overall effects on typical line diagnostics. Given the modeling complexities, we defer consideration of joint \sxpulx shock and photoionization to a future related work. 

As a final note, we remark that the presence of relatively ubiquitous ULX outflows implies the potential for feedback that can strongly shape the interstellar medium, akin to feedback from stellar winds and supernovae. For example, a 10$^{7}$~\msun~instantaneous burst of star formation yields $\sim$ 10$^{40}$~\ergs~mechanical luminosity over the course of 10~Myr as the most massive stars evolve and explode as supernovae \citep[e.g.,][]{starburst99,Prestwich2015}. From Figure~\ref{fig:lxm_age}, a $\lesssim$ 0.1~\Zsun \sxpulx provides $\sim$ 10$^{39}$--10$^{40}$~\ergs~mechanical luminosity for the same instantaneous burst, assuming ULX mechanical luminosity on par with radiative output. For at least some observed ULXs, it appears mechanical power from the outflow actually exceeds radiative power, suggesting the above estimate may even be a lower limit for \sxpulx feedback \citep[e.g.][]{Justham2012,Pinto2016,Gurpide2022}. Feedback from an \sxpulx could therefore contribute to favorable conditions for ionizing photon escape \citep[e.g., similar to the scenario for stellar populations from][]{jaskot2013}. Indeed, resolved studies of Lyman continuum leakers at X-ray wavelengths suggest ULXs could be an important source of mechanical feedback, facilitating the escape of ionizing photons by carving channels in the surrounding medium \citep{Prestwich2015,Kaaret2017}. An \sxpulx with strong outflow component could therefore be important for (1) producing high-energy ionizing photons via both shock and photoionziation; (2) facilitating ionizing photon escape; and (3) producing the soft ($<$ 2~keV) photons that can more efficiently heat the intergalactic medium prior to the epoch of reionization \citep[e.g.,][]{Das2017,hera2023}. As such, continued attention to modeling \sxpulx-like sources is extremely relevant to high redshift studies. 

\section{Summary}\label{sec:summary}

In this work, we have presented a framework for combining a ``simple X-ray population" alongside the corresponding simple stellar population in a physically consistent and meaningful manner in order to simulate and explore their combined ionizing impact as a function of (1) instantaneous burst age and (2) metallicity. Using our combined SED as input to the photoionization code \cloudy, we have produced a nebular line and continuum emission grid, which we make publicly available. Our principal findings can be summarized as follows:

\begin{itemize}

    \item The addition of the \sxpulx prolongs the ionizing output relative to single massive stars or the products of binary evolution already included in BPASS. This is a consequence of the physically robust combination of an \sxpulx to the corresponding BPASS SSP with relative scaling factors determined as a function of burst age and stellar metallicity. Namely, the \sxpulx is scaled with an increasing fractional contribution as a function of decreasing stellar metallicity, in line with empirical constraints, and combined only with an SSP of the corresponding age, following the expectation that the \sxpulx evolves from the parent stellar population.
    \item For the burst timescales modeled here ($t_{\rm burst}$ $\leq$ 20~Myr) the ionizing photons due to the \sxpulx contribute no more than $\sim$ 5\% to the total nebular line and continuum emission. Broad-band FIR--to--UV colors are therefore not strongly affected by the addition of this component, relative to the case of a stellar-only ionizing component. 
    \item Given that the \sxpulx has an SED which is relatively flat through the EUV, the addition of this component increases the intensity lines with ionization potentials $\gtrsim 54$ eV such as \ion{He}{II}~$\lambda$1640,4686 and [\ion{O}{IV}]~25.9$\mu$m by a factor of at least two relative to the BPASS SSPs alone. Emission line diagnostics including such lines can therefore be used to infer the presence of \sxpulx ionization.
    \item The intensity of very high-ionization lines (ionization potentials $>$ 90 eV) such as [\ion{Ne}{V}]~$\lambda$3426, 14.3$\mu$m are strongly enhanced ($>$ 10$^{5}\times$) due to the addition of the \sxpulx; however, these lines are typically very weak in the simulations presented here (e.g., log($\widetilde{f}_{\rm line}/\widetilde{f}_{{\rm Pa}\beta}$) $\sim$ -6 for coronal lines in the IR). This is due in large part to the fixed \sxpulx normalization relative to the SSP as a function of burst age and metallicity, where the maximum relative scaling is achieved for the lowest stellar metallicities. Because gas-phase and stellar metallicities are coupled in all simulations, the regime in which the \sxpulx reaches maximum ionizing contribution relative to the SSP is also the one where many elements have their lowest absolute abundances in the cloud, which can be the dominant factor in setting line intensity.
    \item \sxpulx photoionization is capable of reproducing the observed strengths of high-ionization emission line ratios (e.g., \ion{He}{II}~$\lambda$4686/H$\beta$), but primarily for instantaneous burst models with $Z \lesssim$ 0.1~\Zsun and $t_{\rm burst} >$ 10~Myr. This is again a consequence of the coupling of the \sxpulx to the SSP as a fixed function of stellar metallicity and burst age. On timescales $<$ 10~Myr, the stellar ionizing continuum is still substantial, effectively diluting the \sxpulx ionizing output, particularly for high-ionization lines as measured relative to strong nebular lines for which there is little \sxpulx ionizing contribution. As a result, the \sxpulx is unlikely to contribute significantly to producing strong high-ionization emission lines in galaxies with very recent ($\lesssim$ 10~Myr) bursts of star formation.
    \item Comparison of the simulation outputs to observations in terms of emission line ratios and/or typical X-ray observables (e.g., \lx/SFR) should be performed with careful consideration of the underlying assumptions, including \sxpulx SED shape, cloud abundance patterns, assumed star formation history, and \sxpulx detectability as a function of viewing angle and variability. 
    \item Though the X-ray detectability of the \sxpulx depends on viewing angle and duty cycle, the \sxpulx production of high-ionization nebular emission lines is ubiquitous for the \sedulx modeled here. This is due to the fact that the EUV and soft X-ray photons from the \sxpulx emanate from the outflow component, which is assumed to be relatively isotropic. In this way, high-ionization emission lines can be indirect tracers of a transient or unfavorably oriented \sxpulx. 
    \item The outflow component of the \sxpulx may additionally drive fast shocks and therefore shock ionization. We discuss complexities in joint modeling of shock and photoionization due to an \sxpulx, and consider it a promising avenue for future study. This is particularly true given that \sxpulx mechanical feedback due to a disk wind or outflow could be similar in magntiude, if not timescale, to feedback from stellar winds and supernovae. As such, the \sxpulx could contribute to conditions in the interstellar medium conducive to ionizing photon escape, and additionally provide a substantial number of soft (0.5--2~keV) photons that could directly contribute to heating the intergalactic medium prior to the epoch of reionization. 
\end{itemize}

\begin{acknowledgments}
K.G.'s research was supported by an appointment to the NASA Postdoctoral Program at NASA Goddard Space Flight Center, administered by Oak Ridge Associated Universities under contract with NASA. We gratefully acknowledge support under NASA award 80GSFC21M0002 (A.R.B., P.T.), NASA award 80NSSC22K0407 (K.G., A.R.B., P.T., A.H.), and \chandra grant No. GO0-2JO76A (B.D.L., A.R.B.) C.R. acknowledges the support of the Elon University Japheth E. Rawls Professorship. K.G. would like to thank Jenna Cann and Anna Ogorza\l{}ek for helpful conversations during the preparation of this manuscript, and give a special thanks to Nell Byler for inspiration. 
\end{acknowledgments}

\vspace{5mm}

\software{
    \cloudy~\citep{cloudy17}, \cloudyfsps~\citep{cloudyfsps}, \fsps~\citep{Conroy2009,Conroy2010}, {\tt Python-FSPS}~\citep{newpyfsps}, {\tt XSPEC}~\citep{xspec}, {\tt HEASoft}~\citep{heasoft}, {\tt astropy}~\citep{astropya,astropyb,astropyc}, {\tt scipy}~\citep{scipy}, {\tt matplotlib}~\citep{mpl}, {\tt astroquery}~\citep{astroquery}, {\tt numpy}~\citep{np}
          }

\appendix

\section*{}
\refstepcounter{section}

\counterwithin{figure}{section}
\counterwithin{table}{section}

\renewcommand{\thefigure}{A.\arabic{figure}}
\setcounter{figure}{0}

\renewcommand{\thetable}{A.\arabic{table}}
\setcounter{table}{0}

\startlongtable
\begin{deluxetable}{ccc}
\centering
\tablecaption{Line list used in photoionization simulations, including vacuum wavelength, line identifier, and \cloudy~specific identifier. Bolded lines are those available in \fsps. The full version of this table is available online. \label{tab:line_list}}
\tablehead{
\colhead{Vacuum Wavelength (\AA)} & \colhead{Line ID} & \colhead{Cloudy ID} \\
\colhead{(1)} & \colhead{(2)} & \colhead{(3)}
}
\startdata 
\hline
917.473 & O I 917.473 & {\tt O  1 917.473A} \\
917.726 & O I 917.726 & {\tt O  1 917.726A} \\
917.970 & O I 917.97 & {\tt O  1 917.970A} \\
918.147 & P III 918.147 & {\tt P  3 918.147A} \\
918.493 & Ar I 918.493 & {\tt Ar 1 918.493A} \\
\ldots   & \ldots                  & \ldots   \\
1486.50 & {\bf N IV] 1486.5}\tablenotemark{{\rm \textbar\textbar}} & {\tt N  4 1486.50A} \\
1533.43 & Si II 1533.43 & {\tt Si 2 1533.43A} \\
1548.19 & {\bf C IV 1548.19}\tablenotemark{{\rm \textdagger}} & {\tt C  4 1548.19A} \\
1550.78 & {\bf C IV 1550.78}\tablenotemark{{\rm \textdagger}} & {\tt C  4 1550.78A} \\
1561.33 & C I 1561.33 & {\tt C  1 1561.33A} \\
1577.10 & C III 1577.1 & {\tt C  3 1577.10A} \\
1640.43 & {\bf He II 1640.43}\tablenotemark{{\rm *}} & {\tt He 2 1640.43A} \\
\ldots   & \ldots                  & \ldots   \\
242135. & {\bf [Ne V] 24.2135$\mu$m}\tablenotemark{{\rm \textdaggerdbl}} & {\tt Ne 5 24.2065m} \\
245191. & Fe II 24.5191$\mu$m & {\tt Fe 2 24.5120m} \\
258906. & {\bf O IV 25.8906$\mu$m}\tablenotemark{{\rm *}} & {\tt O  4 25.8832m} \\
\ldots   & \ldots                  & \ldots   \\
1455370. & {\bf O I 145.537$\mu$m} & {\tt O  1 145.495m} \\
1576810. & {\bf C II 157.681$\mu$m} & {\tt C  2 157.636m} \\
2053030. & {\bf N II 205.303$\mu$m} & {\tt N  2 205.244m} \\
3703750. & {\bf [C I] 370.375$\mu$m} & {\tt C  1 370.269m} \\
6097650. & {\bf [C I] 609.765$\mu$m} & {\tt C  1 609.590m} \\
\hline 
\enddata
\tablenotetext{{\rm *}}{Line that is strong relative to H$\beta$ or Pa$\beta$, and strongly enhanced by the addition of the \sxpulx, relative to SSP-only models.}
\tablenotetext{$\textdagger$}{Line that is strong relative to H$\beta$ or Pa$\beta$, but not strongly enhanced relative to SSP-only models.}
\tablenotetext{$\textdaggerdbl$}{Line that is weak relative to H$\beta$ or Pa$\beta$, but strongly enhanced relative to SSP-only models.}
\tablenotetext{$\textbar\textbar$}{Line that is weak relative to H$\beta$ or Pa$\beta$ (or has zero flux), and is not strongly enhanced relative to SSP-only models.}
\tablecomments{Criteria for strength relative to H$\beta$ or Pa$\beta$, or enhancement relative to the SSP-only models are described in Section~\ref{sec:sxp_neb}. Only lines with ionization potentials $\geq$ 54~eV are marked in the table according to the above criteria.} 
\end{deluxetable}

\begin{figure}[h]
    \centering
    \includegraphics[width=9cm]{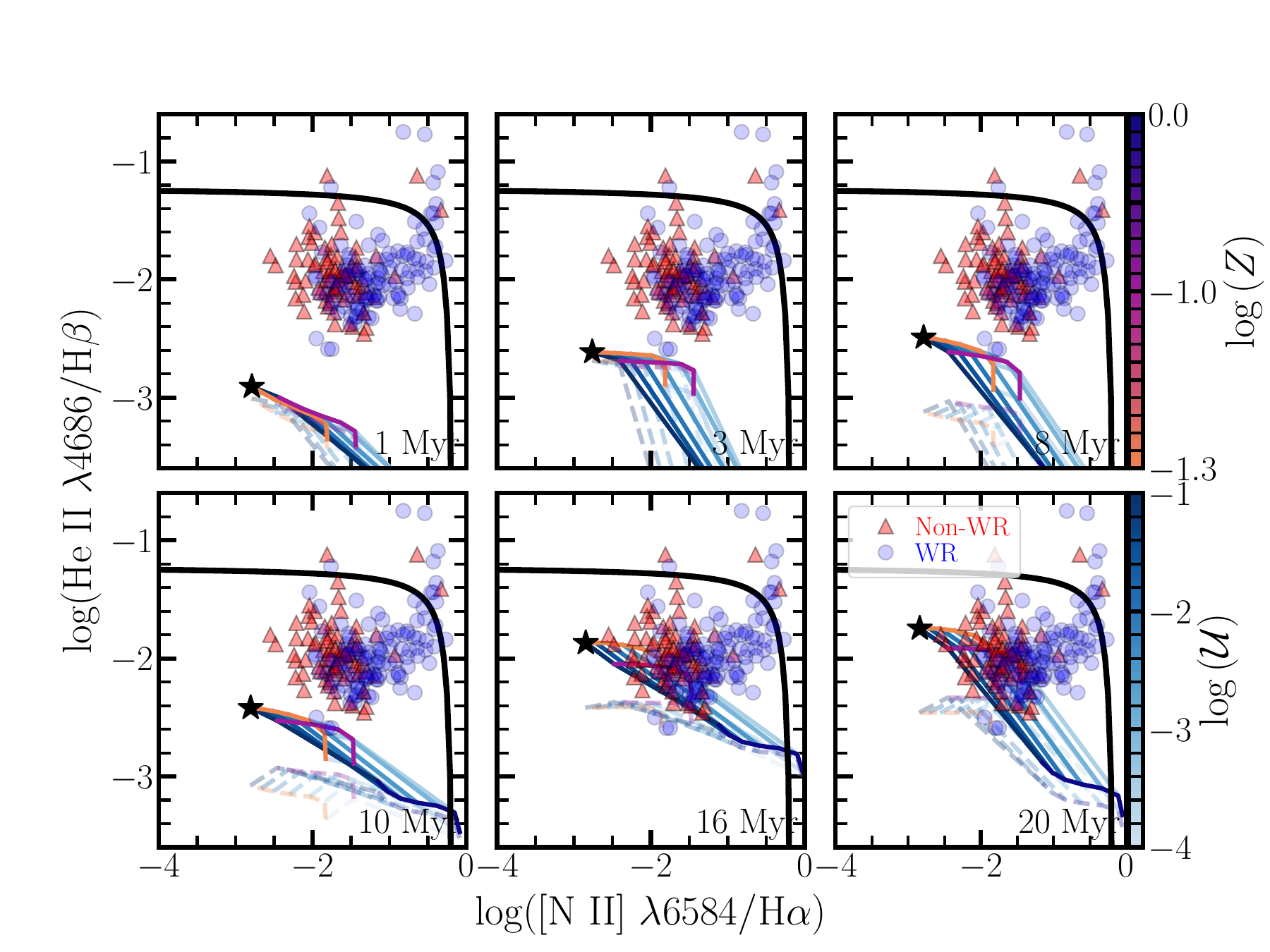}
    \caption{Same as Figure~\ref{fig:n2ha_v_he2hb}, but with the strict delay-time dependence for the \sxpulx removed. Here we show $t_{\rm burst}$ = 1--3~Myr in the top left and center panels to illustrate the effect of allowing the \sxpulx to form immediately with the corresponding SSP.}
    \label{fig:n2ha_v_he2hb_nodelay}
\end{figure}

\begin{figure}[h]
    \centering
    \includegraphics[width=9cm]{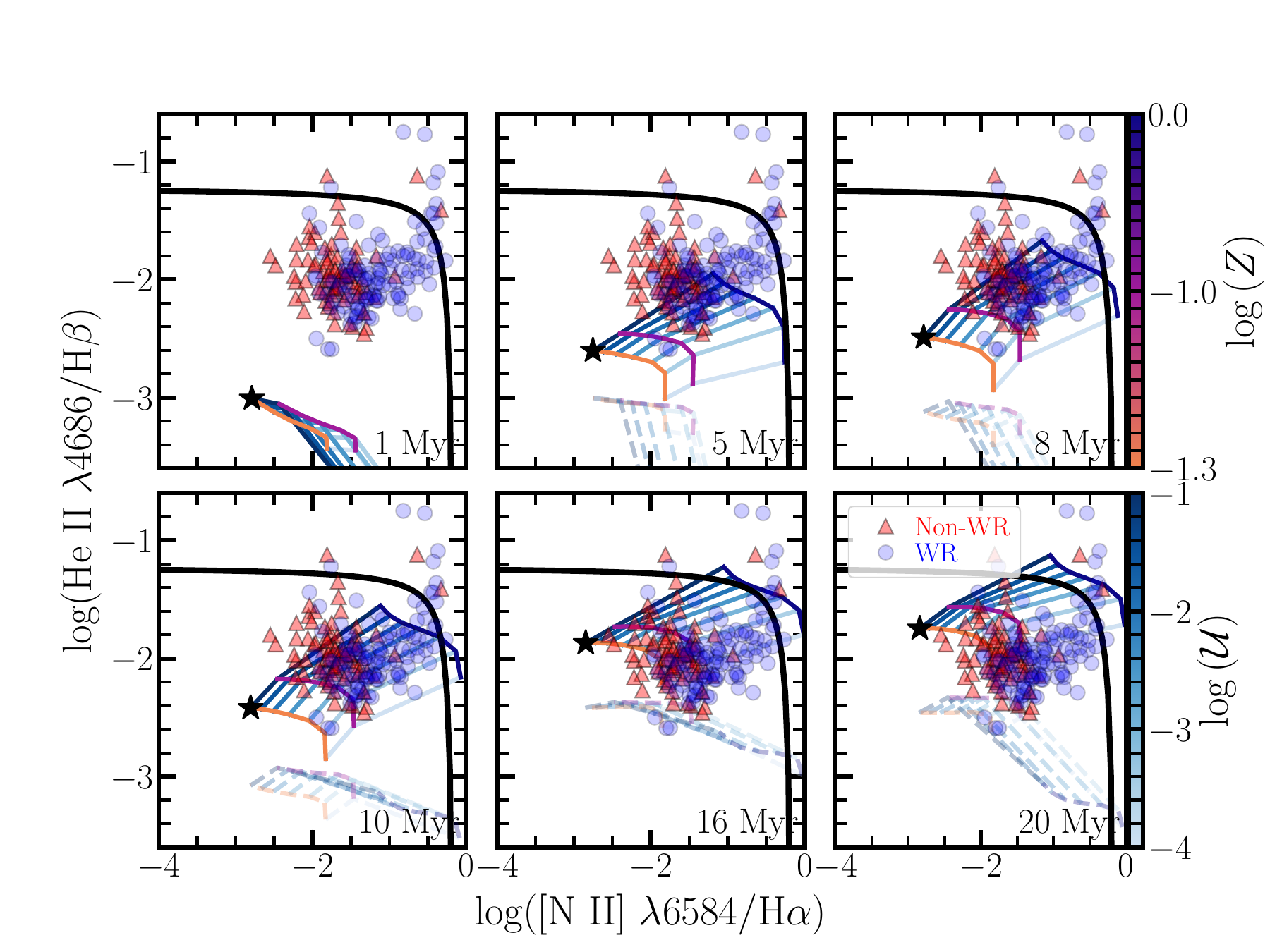}
    \caption{Same as Figure~\ref{fig:n2ha_v_he2hb}, but with the strict correspondence of the metallicity dependence of the \sxpulx relative to the SSP removed. In this case, the grid with \sxpulx contribution overlaps with nearly the entire range of observed line ratios, even at very extreme \ion{He}{II}~$\lambda$4686/H$\beta$, but requires extremely high, and therefore likely unphysical X-ray production efficiencies (i.e., \lx/SFR) for ULXs at high metallicity in order to do so.}
    \label{fig:n2ha_v_he2hb_nometal}
\end{figure}

\end{document}